\documentclass[11pt]{article}
\pdfoutput=1

\usepackage{epsfig,multicol,bbm,simplewick}

\usepackage[usenames]{color}
\usepackage{fix-cm}

\usepackage{booktabs}

\newcommand{\Feyn}[1]{#1\kern-0.45em/}

\usepackage[english]{babel}
\usepackage{amsmath,amssymb,amsbsy,amstext}
\usepackage{graphicx}
\usepackage{amsfonts}
\usepackage{amssymb}
\usepackage{subfigure}
\usepackage{color}
\usepackage{float}
\usepackage[small]{caption}

\numberwithin{equation}{section}

\def\beq{\begin{equation}}
\def\eeq{\end{equation}}
\def\bea{\begin{eqnarray}}
\def\eea{\end{eqnarray}}
\def\ba{\begin{align}}
\def\ea{\end{align}}

\def\d{{\rm d}}

\def\a{{\rm a}}

\def\H{{\cal H}}

\def\k{{\boldsymbol{k}}}
\def\q{{\boldsymbol{q}}}
\def\p{{\boldsymbol{p}}}
\def\v{{\boldsymbol{v}}}
\def\x{{\boldsymbol{x}}}
\def\r{{\boldsymbol{r}}}

\DeclareRobustCommand{\SkipTocEntry}[4]{}

\begin{document}

\begin{titlepage}

\setcounter{page}{1} \baselineskip=15.5pt \thispagestyle{empty}


\bigskip\
\begin{center}
{\fontsize{17.5}{30}\selectfont  \bf Cosmological Non-Linearities as an Effective Fluid}
\end{center}

\vspace{0.5cm}
\begin{center}
{\fontsize{14}{30}\selectfont  Daniel Baumann,$^{1,2}$ Alberto Nicolis,$^{3}$}
\vskip 6pt
{\fontsize{14}{30}\selectfont  Leonardo Senatore,$^{1}$ and Matias Zaldarriaga$^{1}$}
\end{center}


\begin{center}
\vskip 8pt
\textsl{${}^1$ School of Natural Sciences,
 Institute for Advanced Study,
Princeton, NJ 08540}

\vskip 4pt
\textsl{${}^2$ Department of Physics, Harvard University, Cambridge, MA 02138}


\vskip 4pt
\textsl{${}^3$ Department of Physics and ISCAP, Columbia University, New York, NY 10027}

\end{center} 


\vspace{1.2cm}
\hrule \vspace{0.3cm}
{ \noindent \textbf{Abstract} \\[0.2cm]
\noindent
The universe is smooth on large scales but very inhomogeneous on small scales. Why is the spacetime on large scales modeled to a good approximation by the Friedmann equations? Are we sure that small-scale non-linearities do not induce a large backreaction? Related to this, what is the effective theory that describes the universe on large scales?
In this paper we make progress in addressing these questions. 
We show that the effective theory for the long-wavelength universe
behaves as a viscous fluid coupled to gravity: integrating out short-wavelength perturbations renormalizes the homogeneous background and introduces dissipative dynamics into the evolution of long-wavelength perturbations.
The effective fluid has small perturbations and is characterized  by a few parameters like an equation of state, 
a sound speed and a viscosity parameter. 
These parameters 
can be matched to numerical simulations or fitted from observations. 
We find that the backreaction of small-scale non-linearities is very small, being suppressed by the large hierarchy between the scale of non-linearities and the horizon scale.
The effective pressure of the fluid is always positive and much too small to significantly affect the background evolution.
Moreover, we prove that virialized scales decouple completely from the large-scale dynamics, at all orders in the post-Newtonian expansion.
We propose that our effective theory
be used to formulate a well-defined and controlled alternative to conventional perturbation theory, and we discuss possible observational applications.  Finally, our way of reformulating results in second-order perturbation theory in terms of a long-wavelength effective fluid provides the opportunity to understand non-linear effects in a simple and physically intuitive way.}
 \vspace{0.3cm}
 \hrule

\vspace{0.6cm}

\end{titlepage}

\newpage
\tableofcontents

\newpage
\section{Introduction and Summary}
\label{sec:intro}

The formation of large-scale structures in the universe 
is characterized by two fundamental scales:\footnote{In a universe filled with a combination of different matter components there are additional scales associated with the horizon at the time of transition between the different eras; {\it e.g.}~in our universe the horizon at matter-radiation equality defines the scale $k_{\rm eq}^{-1} \equiv {\cal H}^{-1}(\eta_{\rm eq})$. This complication is of little consequence for the arguments made in this section, but will of course be taken into account in the remainder of the paper.}
\begin{enumerate}
\item[i)] the (comoving) {\it Hubble scale} ${\cal H}^{-1}(\eta)$ defines the extent of the observable 
universe at any given time $\eta$ and limits the range over which
interactions can influence the evolution of large-scale perturbations;
\item[ii)] the {\it non-linear scale} $k_{\rm NL}^{-1}(\eta)$ describes the size of structures whose density contrast $\delta(\eta, \x) \equiv [\rho(\eta, \x)/\bar \rho(\eta) - 1]$ exceeds unity.
\end{enumerate}
The large hierarchy between these two scales,
$\varepsilon \equiv k^{-1}_{\rm NL}/{{\cal H}^{-1}} \ll 1$, is of fundamental importance for cosmology.
It is this hierarchy of scales that is responsible for the success of linear perturbation theory:
the most important features of the anisotropies of the cosmic microwave background (CMB) and the large-scale structure (LSS) observed in galaxy surveys are accurately described by
linear perturbations around a homogeneous Friedmann-Robertson-Walker (FRW) background. However, with the advance of observations the study of small non-linear corrections to the long-wavelength dynamics is becoming more and more relevant.

\vskip 6pt
\noindent
{\sl UV-IR coupling in cosmology.} \hskip 8pt
While for linearized perturbations different Fourier modes evolve independently, at the non-linear level two short-wavelength (UV) perturbations can couple to produce a long-wavelength (IR) perturbation, {\it i.e.}~starting at quadratic order Fourier modes don't evolve independently.
Beyond linear perturbation theory short-wavelength perturbations can therefore, in principle, affect the evolution of the long-distance universe.
In particular, small-scale, non-linear fluctuations can give subtle backreaction effects both on the evolution of the background spacetime and the dynamics of long-wavelength perturbations.
In fact, it has been claimed that the renormalization of the background from small-scale structures can be large enough to explain the acceleration of the universe without the need for dark energy~\cite{Kolb:2005da}.  In addition, the effect of small-scale non-linearities on the evolution of superhorizon perturbations from inflation provides an interesting case study~\cite{Boubekeur:2008kn}. These effects and others will be explored in this paper.
 
We expect corrections to the linear evolution of scales with wavelengths comparable to (or larger than) the Hubble scale to be suppressed by the aforementioned hierarchy between the non-linear scale and the Hubble scale. This decoupling of short-wavelength (high-energy) fluctuations from the long-wavelength (low-energy) theory is of course a common feature of effective field theories~\cite{Appelquist:1974tg} (see {\it e.g.}~\cite{Weinberg:1996kr,Goldberger:2007hy, Burgess:2007pt, Pich:1998xt} for recent reviews).
It should therefore be possible to derive a 
long-wavelength effective theory in which corrections to the linear evolution appear systematically suppressed by powers of $\varepsilon^2$ (corrections linear in $\varepsilon$ are forbidden by the isotropy of the background).
In this paper we aim to formalize the effective field theory approach as applied to cosmological perturbations. Specifically, we are interested in an effective description of the long-wavelength universe obtained by `integrating out' short-wavelength modes.\footnote{We define this procedure in detail in \S\ref{sec:PerfectFluid}. There we explain that {\it integrating out short-wavelength fluctuations} amounts
to smoothing the equations of motion 
and taking expectation values of the short-wavelength modes in the presence of long-wavelength perturbations, so that one is left with equations in terms only of the long-wavelength modes.}
We will derive the small corrections to the effective theory of long-wavelength perturbations arising from
the UV-IR coupling imposed by the non-linearities of the Einstein equations and the matter sources.

\vskip 6pt
\noindent
{\sl Matter fluctuations and perturbation theory.} \hskip 8pt
When following this logic, one may worry that the density contrast $\delta$ becomes large below the non-linear scale and that small scales therefore lead to large backreaction effects on the long-wavelength modes.
However, while the density contrast indeed becomes large, the spacetime perturbations and the particle velocities remain small (at least outside of black holes). The system is therefore still amenable to perturbation theory if organized in terms of the gravitational potential $\Phi$ and the average particle velocity $v$ rather than the density perturbation $\delta$. 
Such an analysis reveals that very small scales in fact decouple from the large-scale evolution.
We believe that this decoupling of short-wavelength non-linearities should even apply in the extreme case that the universe is filled with a gas of black holes. In this case, our perturbative scheme breaks down, but the effective theory can be matched continuously  to the effective theory for the dynamics of black holes by Goldberger and Rothstein~\cite{Goldberger:2004jt}, making a large backreaction of gravitational non-linearities even in this case unlikely.

Below the virial scale
$k_{\rm vir}^{-1}$ there exist definitive relations between the gravitational potentials $\Phi$ and the velocities $v$, with large cancellations between the potential and kinetic energies.
One of the main results of our work will be a proof that virial scales indeed decouple completely -- {\it i.e.}~they don't even lead to $\varepsilon^2$ suppressed contributions to the effective pressure, although they of course lead to a small renormalization of the background density. 
In other words, the non-linear source terms for the evolution of large-scale modes can be expressed in a form that vanishes identically in the virial limit. We stress that this is more than standard effective field theory decoupling: indeed, according to the latter, the leading long-distance effect of short-distance physics is just a renormalization of the parameters of the long-distance effective theory \cite{Appelquist:1974tg}. Here instead, we claim that one such parameter -- the effective pressure -- does not even get renormalized in the virial limit. Therefore, our result goes more properly under the name of a `non-renormalization theorem'. This significantly constrains our expectations for backreaction effects of small-scale non-linearities on the long-distance cosmological dynamics.
We stress that the decoupling of virialized structures holds at all orders in the post-Newtonian expansion. Therefore it applies equally well to {\em relativistic} virialized systems, like for instance those containing black holes.

\vskip 6pt
\noindent
{\sl The effective theory.} \hskip 8pt
Ultimately, our theory will be formulated as an FRW universe with small (quasi-linear) long-wavelength perturbations evolving in the presence of an effective fluid whose properties are determined by non-linear short-wavelength modes.
We present the details of the effective theory in \S\ref{sec:ImperfectFluid}. 
The key element of the theory is the effective stress-energy tensor $\tau_{\mu \nu}$ induced by the short-wavelength modes.  Given its importance\footnote{The non-linear scalar source terms captured by $\tau_{\mu \nu}$ have the following effects:
\begin{enumerate}
\item the generation of vector perturbations $\omega_i^{(2)}$ \cite{vectors1, vectors2, matta};
\item the generation of tensor perturbations $\chi_{ij}^{(2)}$ \cite{matta, wands, baum, wands2};
\item the superhorizon evolution of scalar perturbations $\dot \Phi^{(2)} \ne 0$\ \cite{Boubekeur:2008kn} (see \S\ref{sec:evolution});
\item the viscous damping of density perturbations (see \S\ref{sec:BAO}).
\end{enumerate}
All of these effects are absent in linear perturbation theory.
}, we will derive $\tau_{\mu \nu}$ in two different ways:

\vskip 6pt
\noindent
{\sl Effective stress-energy via (post-)Newton -- constructive approach.} \hskip 8pt We aim at understanding the effects of the short-scale non-linearities on the background Hubble expansion and on the long-wavelength fluctuations.
In \S\ref{sec:PT} we will study non-linear cosmological perturbations in a general-relativistic framework. However, we will encounter a number of technical complications that hide to some extent the physical relevance and intuitive nature of our findings.
Fortunately, there are two important points emerging from the former discussion that simplify the problem considerably, allowing us to give a quicker yet rigorous derivation. 
First, the scale at which non-linearities in the perturbation equations become relevant is much smaller than the horizon scale. This allows us to concentrate on subhorizon distances and neglect the general-relativistic effects associated with the {\em background} FRW expansion. Second, for non-relativistic structures like clusters or galaxies, these non-linearities involve the matter sector only. That is, the short-scale gravitational dynamics is Newtonian to a very good approximation. Of course, the effect that these non-linear  structures then have on the long-scale perturbations is {\em post}-Newtonian in nature -- since it involves the coupling of gravity to itself -- but given the above considerations the following simplified approach suggests itself: short-scale perturbations evolve according to flat-space Newtonian equations, where all gravitational fields (short-scale as well as long-scale) are encoded in the Newtonian potential $\Phi$. The total stress-energy tensor $\tau^{\mu\nu}$ of this system is conserved, in the ordinary sense, {\it i.e.}~$\partial_\mu \tau^{\mu \nu} =0$. 
Indeed, the existence of a conserved stress-energy tensor follows from locality and from invariance under space-time translations, regardless of Lorentz-invariance.
The tensor $\tau^{\mu\nu}$ has a gravitational contribution, of order $\rho \Phi$, because the gravitational potential energy obviously participates in the stress-energy conservation already in Newtonian physics. We then smooth this stress-energy tensor over some scale $\Lambda^{-1}$ larger than the typical inhomogeneity scale, and define an effective long-scale stress-energy tensor. The post-Newtonian leap is now to {\em declare} that the effective stress-energy tensor thus defined is what couples to long-wavelength gravitational fields. However, there is not much freedom in this assumption. Long-wavelength gravitational fields must couple to anything (including themselves) through a locally conserved symmetric tensor~\cite{Weinberg:1964ew, Weinberg:1965rz}. For any given system, the only tensor with such properties is the (symmetrized) stress-energy tensor, which is unique up to total derivative terms of the form \cite{Weinberg:1965rz}
\beq \label{Tmn_ambiguity}
\partial _\alpha \partial _\beta \Sigma^{[\alpha \mu] [\beta \nu]} \; .
\eeq
Here, the tensor $\Sigma$ is symmetric under the exchange of the two index pairs, and antisymmetric within each pair. The addition of a term of the form of (\ref{Tmn_ambiguity}) to a system's stress-energy tensor is identically conserved, does not affect the associated global charges ({\it i.e.}, the total four-momentum), and most importantly for our purposes vanishes like two powers of momentum in the small-momentum limit. This ambiguity thus belongs in the class of `higher-derivative' corrections (which we will discuss below), and can be neglected for long wavelengths.
We work out the details of this approach in \S \ref{sect:Newton}.

\vskip 6pt
\noindent
{\sl Effective stress-energy via Einstein -- deductive approach.} \hskip 8pt
A second equivalent description of the UV-IR coupling of cosmological fluctuations arises from a simple reorganization of the Einstein equations.
First, we decompose the Einstein tensor into a homogeneous background (denoted by overbars) and terms that are linear (L) and non-linear (NL) in the metric perturbations, collectively denoted by $\delta X(\eta,  \x)  \equiv X(\eta, \x) - \bar X(\eta)$.
The Einstein equations can then be written as
\beq
\bar G_{\mu \nu}[\bar X] + (G_{\mu \nu})^{\rm L}[\delta X] + (G_{\mu \nu})^{\rm NL}[\delta X^2] = 8 \pi G \, T_{\mu \nu}\, .
\eeq
The background equations, $\bar G_{\mu \nu} = 8\pi G \, \bar T_{\mu \nu}$,
and the linearized Einstein equations, $(G_{\mu \nu})^{\rm L} = 8\pi G\, (T_{\mu \nu})^{\rm L}$,
are then defined in the standard way.
The non-linear Einstein equations can be written in a form that is very similar to the linear equations,
\beq
\label{equ:nonlinear}
(G_{\mu \nu})^{\rm L} = 8\pi G\, (\tau_{\mu \nu} - \bar T_{\mu \nu})\, ,
\eeq
where we defined the {\it effective stress-energy pseudo-tensor}
\beq
\tau_{\mu \nu} \ \equiv \  T_{\mu \nu} - \frac{(G_{\mu \nu})^{\rm NL}}{8\pi G} \, .
\eeq
The stress-energy pseudo-tensor has a long history in studies of General Relativity
({\it e.g.}~as an approach to studying gravitational raditation; see the books by Weinberg~\cite{Weinberg} or Landau and Lifshitz~\cite{Landau}) and will play a key role in this paper.

\vskip 6pt
\noindent
{\sl Properties of the effective fluid.} \hskip 8pt
Given the form of $\tau_{\mu \nu}$ in terms of short-scale (high-momentum) fields -- $\Phi_s$, $v_s$ -- we can analyze its effects on the large-scale modes -- $\delta_\ell$, $\Phi_\ell$, $v_\ell$ -- and on the homogeneous background.
For the benefit of the impatient (curious) reader, we will now state some of the highlights of that analysis, leaving detailed derivations and explanations to the main text and the appendices:
\begin{enumerate}
\item In the absence of long-wavelength perturbations or on very large (superhorizon) scales, the gravitational small-scale (subhorizon) non-linearities mimic 
an isotropic fluid whose effective density and pressure simply renormalize the background by terms of order of the velocity dispersion, $\langle v^2_s \rangle$. 
The effective pressure of the fluid is always positive and much too small to significantly affect the background evolution. 
Moreover, the spatial part of the stress-energy tensor is equal to the second time-derivative of the moment of inertia tensor
\beq
[\tau_{ij} ]_\Lambda = \frac{1}{2} \frac{d^2 I_{ij}}{d \eta^2}\, ,
\eeq
where $[\dots]_\Lambda$ denotes spatial averaging over a region of size $\Lambda^{-1}$, and $I_{ij}$ is the moment of inertia associated with the same region.
This shows that virialized structures decouple completely from the effective theory at large scales. The backreaction effects that we capture in our effective treatment therefore {\it cannot} explain the acceleration of the universe.
Finally, the small induced pressure and the associated renormalization of the background explain the apparent superhorizon evolution of primordial curvature perturbation $\zeta$ \cite{Boubekeur:2008kn}. After the renormalization of the background is taken into account, $\zeta$ is indeed constant on superhorizon scales.

\item The fluid is an imperfect fluid in the sense that the small-scale non-linearities induce dissipative terms and non-negligible anisotropic stress into the evolution of long-wavelength perturbations.
At long wavelengths the fluid is characterized by only a few parameters like an equation of state, a sound speed and a viscosity parameter.\footnote{We point out that the effective viscosity of the fluid leads to a damping in the non-linear evolution of density fluctuations $\delta_\ell$ and suggest that this intuitively explains the non-linear broadening of the peak of baryon acoustic oscillations (see \S\ref{sec:Application}).}
For instance, to leading order, the source term in the Euler equation may be written as
\beq
\frac{k_i k_j}{k^2} \frac{\langle [\tau_{ij}]_\Lambda \rangle}{\bar \rho} = c_s^2 \delta_\ell - c_{\rm vis}^2 \frac{k_i v^i_\ell}{{\cal H}}\, , 
\eeq
where $c_s$ and $c_{\rm vis}$ are time-dependent coefficients.
These parameters can be calibrated by computing the small-scale dynamics with numerical $N$-body simulations. 
Alternatively, the fluid parameters may simply be retained as free parameters to be measured by fitting predictions of the effective theory to observations.
This kind of matching calculation is of course common in effective field theory.

\item The short-wavelength fluctuations provide a source of noise to the dynamics of the long-wavelength perturbations. Although this statistical noise has a negligible effect on the evolution of the background cosmology, it is not irrelevant in all contexts. For example, the stochastic contribution to the pressure fluctuations can be comparable to the pressure fluctuations induced by long-wavelength density fluctuations $\delta_\ell$ for a wide range of scales. However, these fluctuations are uncorrelated at leading order with the amplitude of the long-wavelength modes, which results in a suppression of their importance in averaged quantities such as the power spectrum.

\end{enumerate}

\begin{figure}[h!]
    \centering
        \includegraphics[width=0.55\textwidth]{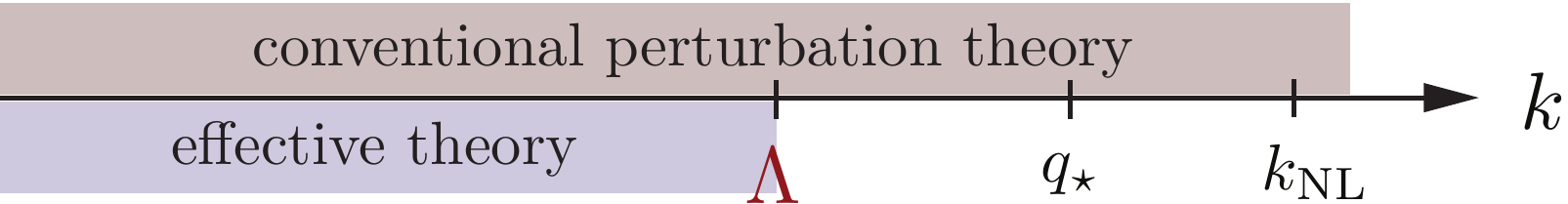}
    \caption{\sl Hierarchy of scales and perturbative expansions. In the effective theory loop integrals only contain modes with $k < \Lambda$, while conventional perturbation theory contains modes with $k \sim k_{\rm NL}$ where the perturbative expansion is known to break down (see \S\ref{sec:Application}).}
    \label{fig:scales0}
\end{figure}

\vskip 6pt
\noindent
{\sl An alternative to conventional perturbation theory.} \hskip 8pt
After the fluid parameters are determined from $N$-body simulations of scales with high momenta, $k > \Lambda$, the effective fluid has small expansion parameters -- $\{ \delta_\ell , \Phi_\ell, v_\ell\} \ll 1$ -- allowing for a controlled perturbative expansion at low momenta, $k < \Lambda$ (see Fig.~\ref{fig:scales0}). 
Conceptually, our approach 
therefore offers a well-defined and controlled treatment of the effects of short-distance non-linearities on the long-wavelength universe.
This is to be contrasted with the failure of many cosmological perturbation theory techniques to include the effects of gravitational non-linearities~\cite{Carlson:2009it}.

\vskip 8pt
\noindent
\newpage
{\sl Outline}: \hskip 8pt The outline of the paper is as follows:
In Section~\ref{sect:Newton} give a bottom-up construction of the effective stress-energy tensor using the Newtonian approximation on subhorizon scales.
Alternatively, in Section~\ref{sec:PT} we explain our basic perturbative approach within General Relativity and present second-order results for metric and matter perturbations in Poisson gauge.
In Section~\ref{sec:PerfectFluid} we derive the gravitational stress-energy pseudo-tensor and define an effective fluid by taking its long-wavelength limit.
We show that on very large scales the fluid behaves as 
an isotropic fluid and gravitational non-linearities only renormalize the background density and pressure.
Furthermore, we prove that virial scales decouple completely from the long-wavelength theory.
In Section~\ref{sec:ImperfectFluid} we show that on scales comparable to and smaller than the horizon scale the fluid anisotropic stress is non-negligible and important for the evolution of perturbations. We characterize the properties of this imperfect fluid in detail.
In Section~\ref{sec:Application} we suggest possible applications of our effective theory.
Finally, we present our conclusions in Section~\ref{sec:conclusions}.

A number of appendices contain technical details:
Appendix~\ref{sec:Euler} gives an alternative derivation of the effective fluid properties starting from the Newtonian conservation equations.
This derivation provides considerable intuition for the physical origin of the effective fluid.
Appendix~\ref{sec:Einstein} presents more details of second-order cosmological perturbation theory.
We collect results in Poisson gauge, cite second-order gauge transformations and discuss the long-wavelength limit of the spacetime.
In Appendix~\ref{sec:FluidReview} we review key elements of dissipative fluid dynamics that are used in Section~\ref{sec:ImperfectFluid}.
Finally, in Appendix~\ref{sec:estimates} we illustrate some of our ideas by presenting example calculations in perturbation theory.

\vskip 6pt
\small
\hrule
\vskip 1pt
\hrule
\vskip 4pt
\small
Short {\bf Proofs} and {\bf Examples} are separated from the main text by horizontal lines. These parts can be omitted without loss of continuity, but they often illuminate the underlying physics.
\vskip 4pt
\hrule
\vskip 1pt
\hrule
\vskip 6pt
\normalsize 

Our work builds on a long history of studies in non-linear perturbation theory \cite{Peebles, Bernardeau:2001qr, Matarrese:1997ay,  Crocce:2005xy, Jeong:2006xd, Shoji:2009gg, Pietroni:2008jx, Matarrese:2007wc, Matsubara:2008wx, Boubekeur:2008kn}, numerical simulations \cite{Siegel:2005xu, Heitmann:2008eq, Heitmann:2009cu, Widrow:2009ru}, and effective field theory \cite{Goldberger:2007hy, Burgess:2007pt, Goldberger:2004jt, Cheung:2007st}.
Related ideas have appeared recently in Refs.~\cite{Dominguez:2000dt, Ishibashi:2005sj, Gruzinov:2006nk, Pueblas:2008uv, McDonald:2009hs, Peebles:2009hw, Shoji:2010hm}.

\vskip 8pt
\noindent
{\sl Notation and conventions}: \hskip 8pt
Except for the Newtonian analysis of \S \ref{sect:Newton}, we will work exclusively with conformal time $\eta$ and in units where the speed of light is set to unity, $c \equiv 1$.
Greek indices $\mu, \nu = 0,1,2,3$ are used for four-dimensional spacetime coordinates, while Latin indices $i, j = 1,2,3$ are reserved for spatial coordinates. 
We will use overdots to indicate derivatives with respect to conformal time.
We use commas to denote partial derivaties and semi-colons for covariant derivatives, {\it i.e.}
\beq
(\dots)_{,\mu} =\frac{\partial}{\partial x^\mu} (\dots) \equiv \partial_\mu (\dots) \qquad {\rm and} \qquad  (\dots)_{;\mu} = \nabla_\mu (\dots) \, .
\eeq
Our Fourier convention will be
\beq
\phi_{\k} =  \int_{\x} e^{-i \k \cdot \x } \phi(\x)\, ,  \qquad \phi(\x) = \int_{\k}  e^{i \k \cdot \x } \phi_{\k}\, ,
\eeq
where we used the notation
\beq
\int_\x \equiv \int d^3 \x \, , \qquad \int_\k \equiv  \int \frac{d^3 \k}{(2\pi)^3}\, .
\eeq
Reality of $\phi(\x)$ demands that $\phi^*_\k = \phi_{-\k}$.
The product of two functions in real space is a convolution in Fourier space
\beq
\int_\x e^{-i \k \cdot \x} \phi(\x) \psi(\x) = \int_\q \phi_{\q} \psi_{\k - \q} \, .
\eeq
We define two forms of the power spectrum,
\beq
\langle \phi_{\k}(\eta) \phi_{{\k}'}(\eta) \rangle = (2\pi)^3 \delta_{\rm D}({\k} + {\k}') P_\phi(k, \eta)\, ,
\eeq
and
\beq
\Delta^2_\phi(k, \eta) \equiv \frac{k^3}{2\pi^2} P_\phi(k, \eta)\, ,
\eeq
such that $\langle \phi^2 \rangle = \int d \ln k\, \Delta_{\phi}^2(k)$.
The time-evolution of the gravitational potential on subhorizon scales during the radiation era leads to a momentum-dependent transfer function $T_\phi(k, \eta)$, where
\beq
\phi_{\k}(\eta) = T_\phi(k, \eta) \phi_\k(0)\, .
\eeq
We take the initial conditions to be scale-invariant, with an amplitude fixed by cosmic microwave background observations~\cite{Komatsu:2010fb},
$\Delta_\phi^2(k, 0) \approx 10^{-9}$.

\section{Bottom-Up Construction of the Effective Stress-Energy}\label{sect:Newton}

As we stressed in the previous section, the large hierarchy between the non-linear scales and the horizon size, and the smallness of typical velocities and gravitational potentials for the former, make the problem of `backreaction' amenable to a Newtonian analysis.
To make use of the fact that the expansion of the universe is negligible at subhorizon scales, it is particularly convenient to use coordinates for which locally around any given comoving point the unperturbed FRW geometry is manifestly a small perturbation of Minkowski space~\cite{MM, Nicolis:2008in}.  That this is possible is a straightforward consequence of the equivalence principle. However, what is not that straightforward is that we can do so 
for an entire comoving geodesic---{\it i.e.}~the metric is approximately flat at subhorizon distances from the geodesic, {\em for all times}.\footnote{Local coordinates with these properties are called Fermi coordinates \cite{MM}.} This is important because we do not want to limit the validity of our approximation to times shorter than $H^{-1}$ (since $H^{-1}$ is the time-scale we will eventually be interested in).
Taking $\x =0$ as the preferred comoving point, the background FRW metric locally becomes \cite{Nicolis:2008in}
\beq \label{FRWlocal}
\d s^2_{\rm FRW} \simeq - \Big[ 1 - \big( \dot H + H^2 \big)\x ^2  \Big] \d t^2 +  \Big[ 1 - \frac12 H^2 \x ^2  \Big] \d \x^2 = \big( \eta_{\mu\nu} + h^{\rm FRW}_{\mu\nu} \big) \, \d x^\mu \d x^\nu  \; ,
\eeq
where $\x$ and $t$ are suitably defined `physical' (as opposed to comoving) coordinates, and $H$ and $\dot H$ are evaluated at $t$.
The actual relation between $(t, \x)$ and the standard comoving FRW coordinates $(t_c, \x_c)$ is
\beq \label{FRWcomoving}
t_c = t - {\textstyle \frac12} H(t) \x^2 \; , \qquad \x_c = \frac{\x}{a(t)}\big[ 1+  {\textstyle \frac14} H^2(t) \x^2\big] \; .
\eeq
In these coordinates the Hubble flow corresponds to $\v \simeq H \x$. Corrections to the above metric are suppressed by further powers of $\sim H^2 \x^2$. We thus see that when studying a non-relativistic system of subhorizon size, we can encode the Hubble expansion into a Newtonian potential\footnote{For subhorizon {\em relativistic} systems we would also need the second Newtonian potential $\Psi_{\rm FRW}=  \frac{1}{4}H^2 \x ^2$. In this section we are using the standard convention for the metric in the Newtonian approximation: $\d s^2=-(1+2\Phi)\d t^2+(1-2\Psi)\d\x^2$.} 
\beq \label{PhiFRW}
\Phi_{\rm FRW} =- \frac12 \big( \dot H + H^2 \big)\x ^2 
\eeq
and a background velocity field
\beq \label{vFRW}
\v_{\rm FRW} = H \x \;.
\eeq
Perturbations of the background homogeneity appear as further contributions to the total $\Phi$ and $\v$ fields.

For simplicity, let us imagine the universe to be filled with a pressureless fluid.\footnote{We discuss the limitations of this approximation in Appendix~\ref{sec:Euler}.} In the Newtonian approximation the dynamics of the fluid coupled to gravity is governed by the continuity and Euler equations, supplemented with Poisson's equation for the Newtonian potential, 
\begin{align}
\dot \rho_m + \boldsymbol{\nabla} \cdot  (\rho_m \v) &= 0\, , \label{continuity} \\
\dot \v + (\v \cdot \boldsymbol{\nabla}\big) \v &= - \boldsymbol{\nabla} \Phi\, , 
\label{euler} \\
\label{poisson}
\nabla^2 \Phi & =  4\pi G \, \rho_m\, .
\end{align}
For instance, the background fields (\ref{PhiFRW}) and (\ref{vFRW}) obey these equations exactly, for homogeneous $\rho_m$, provided that $H^2$ and $\dot H$ satisfy the usual Friedmann equations.
We stress that here $\rho_m$ is the fluid's Newtonian {\em mass} density. We use this notation to be consistent with the rest of the paper, where we stick to the standard relativistic convention of reserving $\rho$ for the fluid's  energy density as measured in its local rest-frame. Our aim is to construct a symmetric stress-energy tensor $\tau^{\mu\nu}$ that is conserved by virtue of the above equations of motion. We are guided by knowing the conserved total Newtonian energy and momentum of the system
\bea
E & = & \int d^3x \Big[ \rho_m + \frac12 \rho_m v^2 + \frac12 \rho_m \Phi \Big]  \equiv \int d^3x \, \tau^{00} \, ,\label{total_energy} \\
P^i & = & \int d^3x \,  \rho_m v^i \equiv \int d^3x \, \tau^{0i} \; .
\eea
Of course, these integral expressions make sense only for finite systems---for infinite systems like the universe, they are divergent. We use them here just as a guide to formulate an ansatz for the local stress-energy tensor. Once we explicitly check that the stress-energy tensor thus guessed is conserved, we have no need to define global quantities like the total energy and momentum.
Let us thus start by postulating
\beq \label{t0i}
\tau^{0i} =  \rho_m v^i
\eeq
and by imposing the $i$-component of the conservation equation
\begin{align}
0 = \partial_{\mu} \tau^{\mu i} & =  \dot \rho_m v^i + \rho_m \dot v^i + \partial_{j} \tau^{j i}  \\
					    & =  - \partial_j \big( \rho_m \, v^i v^j\big) - \rho_m \partial_i \Phi + \partial_{j} \tau^{j i} \; ,
\end{align}
where we used the continuity and Euler equations.
If we are able the rewrite the second term also as a divergence, we have an expression for $\tau^{ij}$. This is straightforward to do upon using Poisson's equation
\beq
\rho_m \, \partial_i \Phi = \frac1{4 \pi G} \nabla^2 \Phi  \, \partial_i \Phi 
= \frac1{4 \pi G} \partial_j \Big[ \partial_i \Phi \partial_j \Phi - \frac12 \delta_{ij} \big( \nabla \Phi \big)^2 \Big]  \; .
\eeq
We thus have a natural candidate for $\tau^{ij}$:
\beq \label{tij}
\tau^{ij} = \rho_m v^i v^j + \frac1{8 \pi G} \Bigl[ 2\partial_i \Phi \partial_j \Phi -  \delta_{ij} \big( \nabla \Phi \big)^2 \Bigr]  \; .
\eeq

For $\tau^{00}$ instead, we have several natural choices. This is because, as usual, the gravitational contribution to the energy density in (\ref{total_energy}) can be written as $\rho_m \Phi$ or as $(\nabla \Phi)^2$ or as any linear combination thereof---these possibilities all coincide upon using Poisson's equation and integration by parts. However, this ambiguity corresponds precisely to higher-derivative terms of the form~(\ref{Tmn_ambiguity}). Indeed
\beq
(\nabla \Phi)^2 \ \to \ - 4 \pi G \, \rho_m  \Phi + \frac 12 \nabla^2 \Phi^2 \; .
\eeq
The associated $\Sigma$ tensor---{\it cf.}~Eqn.~(\ref{Tmn_ambiguity})---is thus
\beq
\Sigma^{[i0][j0]} \propto  \Phi^2 \delta^{ij} \; ,
\eeq
and zero for all other components not related to these by symmetry. We find it more convenient to use the $(\nabla \Phi)^2$ representation for the gravitational potential energy. Our ansatz for $\tau^{00}$ therefore is
\beq \label{t00}
\tau^{00} = \rho_m + \frac12 \rho_m v^2 - \frac{1}{8\pi G}\big( \nabla \Phi \big)^2 \; .
\eeq
We now impose that this obeys the $0$-component of stress-energy conservation
\beq
0 = \partial_ \mu \tau^{\mu0} = \partial_0 \tau^{00} + \partial_i \tau^{i0} \; .
\eeq
We do {\em not} assume that $\tau^{i0}$ is the same as $\tau^{0i}$ defined in (\ref{t0i}). As we will see in a moment this is an interesting point.
Taking the time-derivative of (\ref{t00}) and using repeatedly the continuity, Euler, and Poisson equations, we get
\beq
\partial_0 \tau^{00}  = - \partial_i \Big[  \rho_m v^i \Big( 1+ \frac12 v^2 + \Phi \Big)  + \frac{1}{4\pi G} \Phi \partial_i \dot \Phi \Big] \; .
\eeq
This is consistent with the local conservation law for 
\beq \label{ti0}
\tau^{i0} = \rho_m v^i \Big( 1+ \frac12 v^2 + \Phi \Big)  + \frac{1}{4\pi G} \Phi \partial_i \dot \Phi \; .
\eeq

Now, as we have just shown, the stress-energy tensor defined by Eqns.~(\ref{t0i}), (\ref{tij}), (\ref{t00}), and (\ref{ti0}) is exactly conserved for our Netwonian system and yields the usual total energy and momentum of Netwonian mechanics. However  $\tau^{0i} \neq \tau^{i0}$, so before using $\tau^{\mu\nu}$ as a source for long-wavelength gravitational fields, we should make it symmetric. Symmetrization of the stress-tensor is always possible (see {\it e.g.}~Ref.~\cite{Weinberg:1995mt}), but it requires Lorentz-invariance because it explicitly involves the use of the  Lorentz algebra generators. This means that we cannot make the stress-energy tensor symmetric within our non-relativistic approximation. This is hardly a problem however, since the mismatch between $\tau^{i0}$ and $\tau^{0i}$ is negligible at non-relativistic speeds and gravitational fields. This is evident for the terms in parentheses in Eqn.~(\ref{ti0}), whereas for the last term we have
\beq
\frac{1}{4\pi G} \Phi \partial_i \dot \Phi \ \sim \ \frac1{4\pi G} \Phi \,  v  \, \partial_i \partial_j \Phi \ \sim \ \rho_m v \,  \Phi \ll \rho_m v \; .
\eeq
This means that in our non-relativistic Newtonian approximation we can use our $\tau^{\mu\nu}$ with $\tau^{i0}$ set equal to $\tau^{0i} = \rho_m v^i$, as the post-Newtonian source for gravity.
Notice in passing that we never required inhomogeneities in $\rho_m$  to be small. In fact, they can be huge, as long as velocities and gravitational fields are non-relativistic.

A few remarks are in order: For a Newtonian system governed by Eqns.~(\ref{continuity})--(\ref{poisson}) there is another locally conserved and symmetric tensor $\tilde \tau^{\mu\nu}$,
\beq
\tilde \tau {}^{00} = \rho_m \; , \qquad \tilde \tau {}^{0i} = \tilde \tau {}^{i0} = \rho_m v^i \; , 
\eeq
and $\tilde \tau {}^{ij} = \tau ^{ij}$ as given in Eqn.~(\ref{tij}). The $0$-component of its conservation equation is precisely the continuity equation, which expresses the conservation of {\em mass}. Indeed the global charge associated with $\tilde \tau {}^{00}$ is the total mass of the system
\beq
M = \int d^3 x \,\rho_m \; .
\eeq
Now, in Newtonian physics mass and energy (kinetic plus potential) are separately conserved. However, while the latter is conserved  thanks to the dynamics, the conservation of the former is a trivial kinematical constraint---in the sense that the mass of a system is part of its definition rather than being a dynamical quantity whose value depends on the system's configuration.\footnote{The formal quantum-mechanical counterpart of this statement is that the mass appears as a central charge in the algebra of Galilean transformations and translations \cite{Weinberg:1995mt}. As a consequence, states of different mass belong to different super-selection sectors of the theory.}
We know that upon inclusion of relativistic corrections this degeneracy is gone, and only one form of energy is conserved, namely that which in the Newtonian limit reduces to (\ref{total_energy}). This guarantees that the stress-energy tensor that sources the gravitational field is indeed $\tau^{\mu\nu}$ rather than
$\tilde \tau^{\mu\nu}$ (or any linear combination of the two). 
This conclusion will be manifest in the general-relativistic treatment of \S\ref{sec:PT}.

Next, we want to consider the contribution of the FRW background fields (\ref{PhiFRW}) and (\ref{vFRW})
to the effective stress-energy tensor. First, in the absence of fluctuations we schematically have
\beq
\tau^{00}_{\rm FRW} \sim \rho_m \big( 1+ H^2  x^2\big) \; , \qquad \tau^{0i}_{\rm FRW} \sim \rho_m Hx \; ,
\qquad \tau^{ij}_{\rm FRW} \sim \rho_m H^2 x^2 \; .
\eeq
The explicit $x$-dependence is clearly related to our choice of coordinates, which is not particularly convenient to address questions beyond leading order in $H x$. Of course, the FRW geometry {\em is} invariant under properly defined translations, and so should be any suitably defined physical quantity, but this is not manifest at all in our coordinate system.
Likewise, when we have fluctuations in $\Phi$ and $\v$,
\beq
\Phi = \Phi_{\rm FRW} + \delta \Phi \; , \qquad \v = \v_{\rm FRW} + \delta \v \; ,
\eeq
the mixed background-fluctuation terms  in $\tau^{\mu\nu}$  are weighed by $H x$. This yields a stress-energy tensor for the fluctuations that depends explicitly on position. Given the background physical homogeneity, we would like to get rid of this fake explicit position-dependence and be left just with the physical one implicitly contained in the fluctuation fields. We want to stress that as long as we stay sufficiently close to the origin so that special- and general-relativistic effects are negligible, our expression for $\tau^{\mu\nu}$ is certainly the correct one in the coordinates we have been using, even in the presence of an FRW background. Only, such coordinates hide the nice symmetries of the cosmological background and are thus not suitable for cosmological applications. Of course, in the rest of the paper we will use standard FRW coordinates---so how do we translate our $\tau^{\mu\nu}$ to FRW comoving coordinates?
One possibility is to perform the actual coordinate transformation, but the gravitational part of the stress-energy tensor, infamously, does {\em not} transform as a tensor under generic coordinate transformations. Indeed, it can always be made to vanish at any given point by choosing a locally inertial frame \cite{Landau}. Another possibility, which we will adopt, is to evaluate our $\tau^{\mu\nu}$ at $\x =0 $. That is, given the homogeneity of the background, any comoving point $P$ is as good as any other to play the role of $\x =0 $. Therefore, if we are interested in the components of $\tau^{\mu\nu}$ at $P$ in FRW coordinates we can proceed as follows: We identify the origin of our Netwonian coordinates (\ref{FRWlocal}) with $P$. We compute $\tau^{\mu\nu}$ at $\x = 0$ according to Eqns.~(\ref{t0i}), (\ref{tij}), and (\ref{t00}). The background fields  (\ref{PhiFRW}) and (\ref{vFRW}), given their $\x$-dependence, give no contribution at $\x = 0$. That is, at the origin only the {\em fluctuations} in $\Phi$ and $\v$ contribute to $\tau^{\mu\nu}$. Finally, we perform the coordinate change (\ref{FRWcomoving}) which brings us back to comoving FRW coordinates, and which at the origin is exceedingly simple. There is a factor of $a(t)$ for each spatial index, and more importantly, the problematic non-tensor part of the transformation law for $\tau_{\mu\nu}$ is gone, because the point $\x = 0 $ is at rest for all times in {\em both} coordinate systems---its state of motion is untouched by the coordinate change.
In conclusion, when we go back to standard FRW coordinates, we can use our expressions for $\tau^{\mu\nu}$ directly in terms of the Newtonian-gauge potential and peculiar velocity, modulo obvious factors of $a(t)$.

\vskip 6pt
Finally, it is instructive to present the derivation of our effective stress-tensor $\tau_{\mu\nu}$ in the Newtonian context in yet another way. As we mentioned in the Introduction, we will later define the effective theory for long-wavelength fluctuations by smoothing the stress-energy tensor $\tau_{\mu\nu}$ on a scale $\Lambda$ and declaring that long-wavelength gravitational fields are coupled to it. It is particularly illuminating to see how $\tau_{\mu\nu}$ arises in we perform the smoothing immediately at the level of the Euler and Poisson equations (\ref{continuity}) and (\ref{euler}).  We apply a filter on scales of order $\Lambda^{-1}$ to the Euler equation 
\beq
\label{equ:NLEulerIntro}
\int d^3 \x' \ W_\Lambda(|\x-\x'|) \cdot \left\{\ \rho_m \left[ \dot v^i + v^j \nabla_j v^i \right] + \rho \nabla_i \Phi\ \right\} = 0  \, .
\eeq
We define smoothed quantities of all fields $X \equiv \{\rho_m, \Phi, \rho_m \v\}$ as
\beq\label{eq:smoothingleo}
X_\ell \equiv [X]_\Lambda(\x) = \int d^3 \x' \, W_\Lambda(|\x-\x'|) X(\x')\, ,
\eeq
and split the fields into short-wavelength and long-wavelength fluctuations $X \equiv X_\ell+X_s$. Straightforward algebra then shows (see Appendix~\ref{sec:Euler}) that the Euler equation can be recast in the following way
\beq
\label{equ:SmoothEulerIntro}
\rho_\ell \left[ \dot v^i_\ell + v^j_\ell \nabla_j v^i_\ell \right] + \rho_\ell \nabla_i \Phi_\ell = - \nabla_j \bigl[ \tau^j_{\ i}\bigr]^{s} ,
\eeq
where 
\beq
\label{equ:TauEffIntroN}
\bigl[\tau_{ij}\bigr]^{s} \equiv \left[\rho_m v_i^s v_j^s\right]_\Lambda +\frac{1}{8\pi G}\left[2 \partial_i\Phi_s\partial_j\Phi_s-\delta_{ij}(\nabla\Phi_s)^2\right]_\Lambda \, .
\eeq
We see that the long-wavelength fluctuations obey an Euler equation in which the stress tensor $\tau_{ij}$ receives contributions from the short-wavelength fluctuations.
Eqn.~(\ref{equ:TauEffIntroN}) is exactly of the form of the effective $\tau_{ij}$ derived in Eqn.~(\ref{tij}), 
and it shows explicitly how the effective long-wavelength fluid is different from the pressureless fluid we started with in the continuity and Euler equations (\ref{continuity}) and (\ref{euler}).

\vskip 6pt
The skeptical reader will find comfort in the analysis of the next section, where we derive the effective stress-energy tensor directly in FRW by more traditional means.\footnote{In comparing the results of \S2 and \S3 the reader should be warned that in \S3 we will switch the convention for the metric fluctuations $\Phi$ and $\Psi$ (see Eqn.~(\ref{eq:preliminary-newtnian-gauge-metric})) in order to follow the standard convention for Newtonian gauge (to be distinguished from the convention used in the Newtonian approximation in \S2). The precise correspondence is: $\delta\Phi_{\S2}=\Psi_{\S3}\,,\ \delta\Psi_{\S2}=\Phi_{\S3}$.  In comparing the results for the stress tensor $\tau_{\mu \nu}$ at second order in $\Phi$ and $\Psi$ one should also use that in our approximations the first-order perturbations are equal: $\Phi_{\S3}^{(1)}=\Psi_{\S3}^{(1)}$.} Those already convinced that our $\tau^{\mu\nu}$ is the correct post-Newtonian source for gravity, can instead skip directly to \S \ref{sec:PerfectFluid} without substantial loss of continuity. For these readers, the starting point should be Eqns.~(\ref{equ:first}) and  (\ref{equ:second}). There, $\tau^{\mu\nu}$ is expressed in terms of $\rho$, the energy density as measured by an observer comoving with the fluid. This  differs from our $\rho_m$ by terms of order $\rho v^2$ and $\rho \Phi$. Also, the $\tau^{00}$ given in Eqn.~(\ref{equ:first}) differs from that of this section by a different choice for the $\rho \Phi$ {\em vs.}~$(\nabla \Phi)^2$ ambiguity in the potential energy. As we argued different choices correspond to different higher-derivative terms of the form (\ref{Tmn_ambiguity}), which are irrelevant at long wavelengths.

\section{Perturbation Theory and Velocity Expansion}
\label{sec:PT}

We now turn to the more standard approach to cosmological perturbation theory, where one expands in perturbations around the FRW metric. However, contrary to the standard way of organizing the expansion in fluctuations, we will show that perturbation theory may be formulated systematically as an expansion in perturbed velocities, 
in analogy with the post-Newtonian approach to General Relativity.
This has certain advantages when trying to capture non-linear perturbations.
In \S\ref{sec:general} we describe the general philosophy of this approach.
We introduce the concept of the effective stress-energy tensor arising from short-wavelength perturbations.
The treatment will be schematic with details postponed to the following sections.
In \S\ref{sec:second} we give the equations of second-order perturbation theory in Poisson gauge (see also Appendix~\ref{sec:Einstein}). This is used in \S\ref{sec:effectiveX}
 to derive the effective stress-energy tensor.
 
\subsection{Generalities}
\label{sec:general}

Consider the perturbed spacetime\footnote{For the purposes of this introductory section we neglect vector and tensor perturbations and we don't worry about technical details like gauge-fixing. These deficiencies will be cured momentarily.}
\beq\label{eq:preliminary-newtnian-gauge-metric}
\d s^2 = a^2(\eta) \left[ - e^{2\Psi} \d \eta^2 + e^{-2\Phi} \d {\x}^2 \right]\, ,
\eeq
sourced by matter perturbations $\delta \rho$, $\delta p$ and $v_i$.
In standard cosmological perturbation theory the Einstein equations are expanded in metric perturbations 
and matter perturbations.
Perturbation theory is then said to break down when the magnitude of density fluctuations $\delta \rho(\eta, {\x})$ becomes comparable to the background value $\bar \rho(\eta)$.
However, this gives up too early.
It is well-known that metric perturbations and matter velocities remain small even if the density perturbations become large, $\delta \rho > \rho$. 
Indeed, the density contrast $\delta \equiv \delta \rho/\rho$ is $\sim 10^2$, $10^{10}$ and $10^{30}$ on cluster scales, galaxy scales and solar system scales, respectively, while the spacetime on these scales is still well-approximated by a homogeneous spacetime with small perturbations $\Phi \lesssim 10^{-5}$.

The following alternative to conventional cosmological perturbation theory therefore suggests itself:
the Einstein equations are expanded in metric perturbations $\Phi$, $\Psi$ and matter velocities $v^i$, but {\it not} in density perturbations $\delta \rho$.
In perturbation theory the linear peculiar velocities are related to the Newtonian potential and the density contrast as follows
\beq
v^2 \  \sim \ \Phi\, \delta\, .
\eeq
When both $\Phi$ and $\delta$ are much smaller than unity, $v^2$ is therefore considered a second-order quantity. However, when entering the quasi-nonlinear regime, $\delta_{\rm NL} \sim 1$, the fluid kinetic energy (per unit mass) effectively becomes a `first-order' contribution, $v^2 \sim \Phi$.
Similarly, for virialized objects, $\delta_{\rm vir} \sim 10^2$, one also finds that kinetic and gravitational energies are of the same order, $v^2 \sim \Phi$. 

Furthermore, on small scales, gradients of the gravitational potential can change the power-counting of  
standard perturbation theory.
For instance, quadratic terms with two spatial derivatives satisfy the following scaling relations
\beq
\frac{(\nabla \Phi)^2}{{\cal H}^2}\ \sim \ \frac{\Phi \nabla^2 \Phi }{{\cal H}^2}\ \sim \ \Phi \, \delta \ \sim \ v^2\, ,
\eeq 
where ${\cal H} \equiv \dot a/a$ is the comoving Hubble parameter and
 we have used the Poisson equation (for a pure matter universe), $\nabla^2 \Phi = \frac{3}{2} \H^2 \, \delta$.
For small-scale perturbations the terms $(\nabla \Phi)^2$ and $\Phi \nabla^2 \Phi$ are therefore enhanced relative to terms without at least two gradients\footnote{In General Relativity, the equations of motion have at most two derivatives. This implies that the perturbation expansion naturally truncates at second order and third-order terms are suppressed by an extra power of $\Phi$, {\it e.g.}~$\Phi^2 \nabla^2 \Phi \ll \Phi \nabla^2 \Phi$.}, like $\H^2 \Phi^2$.
For a (scale-invariant) spectrum of perturbations with different wavelengths, terms with the maximal number of gradients will hence give the leading effect.
We will capture this by taking $v^2$ as our expansion parameter.
At the non-linear scale $\Phi^2$ is order $v^4$ while each gradient (rendered dimensionless by the Hubble scale ${\cal H}$) reduces the order in $v$ by one.
This is reminisent of the post-Newtonian approach to General Relativity.

Of course, our statement about expanding in velocities is not gauge-invariant: one could even choose a comoving gauge where velocities are zero, in which case clearly our expansion would not be well-defined. However, this is not a source of worry for the following reason: the fluctuations that become non-linear are those related to the matter degrees of freedom, while the metric fluctuations are not large even on very short scales (except close to black holes\footnote{The case of order one metric fluctuations near black holes can also be taken into account quite straightforwardly. We have briefly commented on this in the Introduction.}). Of course, one can choose a gauge in which the matter fluctuations vanish, and all the fluctuations are `eaten' by the metric.
In this gauge metric fluctuations will be large. In this case, the large fluctuations of the metric simply are related to a contorted slicing of the spacetime, 
which is, in fact, never very different from FRW. This particular fact implies that there exist gauges where metric fluctuations are small and matter fluctuations can be large. This is the case for Newtonian gauge, or for any gauge where the matter fluctuations are not made explicitly small by the gauge condition. Our expansion scheme applies in those gauges. A gauge where instead metric fluctuations are as large as the matter fluctuations is the so called $\zeta$-gauge, or unitary gauge. In this case, matter fluctuations are actually taken to be zero, and our expansion is not applicable. 

The reason why there are universes that at subhorizon distances feature large matter fluctuations but small metric perturbations can be stated in a language that is familiar from particle physics where is goes by the name of the equivalence theorem for massive vector fields.  
A non-trivial matter background density spontaneously breaks some of the spacetime symmetries, which are gauged in General Relativity. As a consequence the gravitational field gets a mass of order Hubble. Matter fluctuations can be considered as Goldstone bosons of these symmetries, which according to the equivalence theorem, are the most strongly-coupled degrees of freedom at scales smaller than the graviton inverse Compton wavelength, that is at subhorizon scales. This is because their self-couplings involve more derivatives than those they have with the metric fluctuations or than the metric fluctuations' self-couplings. As a consequence, on small distances matter fluctuations are more strongly coupled, and therefore non-linear, than the metric ones. This hierarchy is completely obscure in the gauge in which matter fluctuations are set to zero---in which they are `eaten' by the metric---which is called unitary gauge in the particle physics context. This logic is expanded upon in Ref.~\cite{Creminelli:2006xe, Cheung:2007st}.

As we just remarked, our approach of expanding in velocities is valid for all gauges in which there is an hierarchy between matter fluctuations and metric fluctuations. One can consistently change gauges within this set (see Appendix~\ref{sec:Einstein}). Our approach will lead us to neglect terms quadratic in the metric perturbations of order $\Phi^2\sim 10^{-10}$. Because of this, our approach is not gauge-invariant at this level. Furthermore, at this order it even becomes unclear how to split the metric into long-wavelength and short-wavelength fluctuations (see \S\ref{sec:evolution}), as we did in Eqn.~(\ref{eq:smoothingleo}) and will do in the rest of the paper. Luckily, there are many non-linear effects in the matter sector whose size is larger than $10^{-10}$. It is those effects that can be addressed unambiguously.  It is unclear to us if effects of order $ \Phi^2\sim 10^{-10}$ will ever be relevant from the observational point of view in the context of LSS surveys. In any case, describing those effects would require a qualitative improvement of our treatment that goes beyond the scope of the current paper.

We are now ready to return to the main objective in this work, which is to develop an effective description for the evolution of long-wavelength perturbations in the presence of short-wavelength non-linearities valid for effects larger than order $\Phi^2\sim 10^{-10}$.
To achieve this, we decompose the Einstein tensor into a homogeneous background (denoted by overbars) and terms that are linear (L) and non-linear (NL) in the perturbations $\delta X(\eta, {\x})  \equiv X(\eta, {\x}) - \bar X(\eta)$.
The Einstein equations can then be written as
\beq
\bar G_{\mu \nu}[\bar X] + (G_{\mu \nu})^{\rm L}[\delta X] + (G_{\mu \nu})^{\rm NL}[\delta X^2] = 8 \pi G \, T_{\mu \nu}\, .
\eeq
The background equations, $\bar G_{\mu \nu} = 8\pi G \, \bar T_{\mu \nu}$,
and the linearized Einstein equations, $(G_{\mu \nu})^{\rm L} = 8\pi G\, (T_{\mu \nu})^{\rm L}$,
are defined in the standard way. 
The non-linear Einstein equations can be written in a form that is very similar to the linear equations,
\beq
\label{equ:nonlinear}
(G_{\mu \nu})^{\rm L} = 8\pi G\, (\tau_{\mu \nu} - \bar T_{\mu \nu})\, ,
\eeq
where we defined the {\it effective stress-energy pseudo-tensor}
\beq
\tau_{\mu \nu} \ \equiv \  T_{\mu \nu} - \frac{(G_{\mu \nu})^{\rm NL}}{8\pi G} \, .
\eeq
In the following we compute the effective stress-energy pseudo-tensor 
associated with the short-wavelength non-linearities at quadratic order and as an expansion in the matter velocities.

\subsection{Second-Order Perturbation Theory}
\label{sec:second}

We now start to define things more precisely.
We begin by collecting standard results in second-order perturbation theory in Poisson gauge, a generalization of Newtonian gauge to second order. 

\vskip 6pt
\noindent
{\sl Metric perturbations.} \hskip 8pt
The metric in Poisson gauge is given by
\beq
\label{equ:PoissonMetric0}
\d s^2 =
 a^2(\eta) \left[ - e^{2\Psi} \d \eta^2 + 2 \omega_i \d x^i \d \eta + (e^{-2\Phi} \delta_{ij} + \chi_{ij}) \d x^i \d x^j\right]\, ,
\eeq
with $\chi_{ii} = 0$---{\it i.e.}~the trace of $g_{ij}$ is absorbed into $\Phi$---and
\beq
\label{equ:gauge}
\omega_{i,i} = \chi_{ij, i} = 0\, .
\eeq
The gauge condition (\ref{equ:gauge}) eliminates one scalar degree of freedom from $g_{0i}$ and one scalar and one transverse vector degree of freedom from $g_{ij}$.  Thus, $\omega_i$ is a transverse vector, while $\chi_{ij}$ is a transverse-traceless tensor.
The metric determinant is $\sqrt{-g} = e^{\Psi-3\Phi} a^4$ up to second order in perturbations.
Expressions for the Christoffel symbols, Riemann and Einstein tensors corresponding to the spacetime (\ref{equ:PoissonMetric0}) are given in Appendix~\ref{sec:Einstein} (see also Refs.~\cite{Senatore:2008vi,Bartolo:2005kv, Malik:2008im}).

Perturbations are formally split into first-order and second-order terms
\beq
\Psi = \Psi^{(1)} + \frac{1}{2} \Psi^{(2)}\, , \qquad \Phi = \Phi^{(1)} + \frac{1}{2} \Phi^{(2)}\, , \qquad \omega_i = \omega_i^{(2)}\, , \qquad \chi_{ij} = \chi_{ij}^{(2)}\, . \label{equ:pert}
\eeq
Here, we have ignored first-order vector and tensor perturbations, {\it i.e.}~$\omega_i^{(1)}= \chi_{ij}^{(1)} \equiv 0$.  Many models of inflation indeed produce initial conditions where vector modes are zero and tensor modes are negligibly small \cite{Baumann:2009ds, Baumann:2008aq}.
Furthermore, at first order, subhorizon vectors and tensors decay as the universe expands.
The absence of first-order vectors and tensors has the following important consequence: second-order scalar modes are only sourced by first-order scalars. In fact, although scalars, vectors and tensors mix beyond linear order in perturbation theory, the second-order parts of scalars, vectors and tensors do {\it not} mix, {\it e.g.}~second-order scalars are only sourced by quadratic contributions from the first-order vectors and tensors. 

The non-linear parts of the components of the Einstein tensor are (see Appendix~\ref{sec:Einstein}),
\bea
\label{equ:GnlFirst}
- a^2 (G^0_{\ 0})^{\rm NL} &=& 12 \H^2 \Psi^2 + 12 \H \dot \Phi \Psi + 3 \dot \Phi^2 - \Phi_{,k} \Phi_{,k} + 4 \Phi \Phi_{,kk}\, ,\\
 \frac{a^2}{2} (G^i_{\ 0})^{\rm NL} &=& 2\Phi [\dot \Phi + \H \Psi]_{,i} - \dot \Phi \Psi_{,i} \, ,\\
 a^2(G^i_{\ j})^{\rm NL} &=&  \Bigl[  -4 ({\cal H}^2 + 2 \dot{\cal H}) \Psi^2 - 2 \dot \Phi \dot \Psi - 3 \dot \Phi^2 - 4 {\cal H} (2\dot \Phi + \dot \Psi) \Psi - 4 \Psi \ddot \Phi \nonumber \\
&&\ \ \ \ \ \ + \ \Psi_{,k} \Psi_{,k} - 2 \Phi [\Phi - \Psi]_{,kk}  \Bigr]\delta^i_j \nonumber \\
&& \     + \
2 \Phi [\Phi - \Psi]_{,ij} + \Phi_{,i} \Phi_{,j}  -\Psi_{,i} \Psi_{,j}  - \Psi_{,i} \Phi_{,j}  - \Phi_{,i} \Psi_{,j} \, .
\label{equ:GnlLast}
\eea

\vskip 6pt
\noindent
{\sl Matter perturbations.} \hskip 8pt
To define matter perturbations we introduce the timelike velocity four-vector 
\beq
u^\mu \equiv \frac{d x^\mu}{d \tau}\, ,
\eeq
where $\tau$ is the proper time comoving with the fluid, so that $g_{\mu \nu} u^\mu u^\nu = -1$.
We define the tensor $\gamma_{\mu \nu} \equiv g_{\mu \nu} + u_\mu u_\nu$ which projects tensors orthogonal to the four-velocity into the fluid's instantaneous rest space at each event.
The energy-momentum tensor of a general (imperfect) fluid decomposes, at leading order in derivatives, into irreducible parts as
\beq
\label{equ:pf}
T_{\mu \nu} = \rho u_\mu u_\nu + (p-\zeta\theta) \gamma_{\mu \nu} + \Sigma_{\mu \nu}\, ,
\eeq
where $\rho = T_{\mu \nu} u^\mu u^\nu$ is the matter energy density, $p$ is the isotropic pressure, $\zeta$ is the bulk viscosity, $\theta = \partial_i u_i$ is the velocity divergence and $\Sigma_{\mu \nu} = \gamma_{\langle \mu}^{\ \alpha} \gamma_{\nu \rangle}^{\ \beta} T_{\alpha \beta}$ is the symmetric and trace-free anisotropic stress tensor.\footnote{Here we use the notation $t_{\langle \mu \nu \rangle} = \gamma_{(\mu}^{\ \alpha} \gamma_{\nu)}^{\ \beta} t_{\alpha \beta} - \frac{1}{3} \gamma^{\alpha \beta} t_{\alpha \beta} \gamma_{\mu \nu}$ and $t_{(\mu \nu)} = \frac{1}{2}(t_{\mu \nu}+t_{\nu \mu})$.} $\Sigma_{\mu\nu}$ and $\zeta$ vanish for
a perfect fluid, 
 $\zeta=\Sigma_{\mu \nu} = 0$. It holds that $p-\zeta\theta= \frac{1}{3} T_{\mu \nu} \gamma^{\mu \nu}$.\footnote{It is worth pointing out that the viscosity is a higher-derivative term that locally in space and time is indistinguishable from the pressure. This explains why we cannot isolate the pressure through a tensorial contraction.}
 
 
For cold dark matter we furthermore impose that the background pressure vanishes. Our starting point is therefore the stress-energy tensor,\footnote{On very small scales the fluid description is expected to break down. In that limit we will treat the dark matter as a collection of point particles that interact only via gravity~\cite{Weinberg} (see Appendix~\ref{sec:Einstein}). For notational simplicity we will often use the perfect fluid description of the dark matter, {\it i.e.}~the continuum limit of the particle model and the $\Sigma_{\mu \nu} = p=0$ limit of Eqn.~(\ref{equ:pf}). However, all our results are easily generalized to the background of particles and sometimes we will find it more convenient to think of the dark matter in terms of particles ({\it e.g.}~when discussing the effects of tidal forces from long-wavelength perturbations in \S\ref{sec:tidal}). When 
relating our approach to cosmological $N$-body simulations of dark matter (see \S\ref{sec:Application}) the effective stress-energy is that of non-relativistic point particles.}
\beq
\label{equ:Tmunu}
T^\mu_{\ \nu} = \rho\, u^\mu u_\nu \, .
\eeq

To second order in metric and three-velocity perturbations the components of the fluid four-velocity are
\beq
u^0 = a^{-1} e^{-\Psi} \gamma(v) \, , \qquad u^i = a^{-1} e^\Phi v^i\, ,
\eeq
where we defined the Lorentz factor
\beq
\gamma(v) \equiv \frac{1}{\sqrt{1-v^2}} \approx 1 + \frac{1}{2} v^2\, .
\eeq
At order $v^2$ the components of the stress-energy tensor during matter-domination therefore are
\beq
T^0_{\ 0} = - \gamma^2 \rho = - \rho(1 + v^2)\, ,\qquad
T^i_{\ 0} = -  e^{\Psi + \Phi} \rho v^i \, ,\qquad
T^i_{\ j} = \rho  v^i v_j\, .
\eeq
We emphasize again that we do {\it not} expand in density fluctuations. We may still write
\beq
\rho(\eta, {\x}) = \bar \rho(\eta) [1+ \delta(\eta, {\x})]\, , \qquad {\rm where} \quad \bar \rho = \frac{3 \H^2}{8\pi G a^2}\, ,
\eeq
but we do {\it not} assume that $\delta < 1$ (unless we are specifically referring to long-wavelength density perturbations).

\vskip 6pt
\noindent
{\sl Einstein equations.} \hskip 8pt
The metric and matter perturbations 
are coupled to each other by the Einstein equations.
As we have argued above, the non-linear Einstein equations can be written in a form that is very similar to the linear equations if we replace the stress-energy energy tensor $T^\mu_{\ \nu}$ by the stress-energy pseudo-tensor $\tau^\mu_{\ \nu}$:
\beq
(G^\mu_{\ \nu})^{\rm L} = 8\pi G (\tau^\mu_{\ \nu} - \bar T^\mu_{\ \nu})\, ,
\eeq
or explicitly for scalar fluctuations,
\bea
\nabla^2 \Phi - 3 \H (\dot \Phi + \H \Psi) & =& - 4 \pi G a^2 \, (\tau^0_{\ 0} - \bar \tau^0_{\ 0}) \, ,\label{equ:NL1_0}\\
\bigl[ \dot \Phi + \H \Psi \bigr]_{,i}  & = & 4 \pi G a^2 \,\tau^i_{\ 0} \, , \\
\ddot \Phi + \H (2\dot \Phi + \dot \Psi) + (\H^2 +2 \dot \H) \Psi - \frac{2}{3} \nabla^2 (\Phi-\Psi) &=&  \frac{4\pi G a^2}{3}\, (\tau^i_{\ i} - \bar \tau^i_{\ i}) \, ,\label{equ:NLT_0}\\
\partial_i \partial_j \left[ (\Phi - \Psi)_{,ij} - \frac{1}{3} \delta_{ij} \nabla^2 (\Phi - \Psi)\right] & =& 8\pi G a^2 \,\partial_i \partial_j \left[ \tau^i_{\ j} - \frac{1}{3} \delta^i_j \tau^k_{\ k} \right]\, . \label{equ:NL2_0}
\eea
This defines the dynamics of long-wavelength scalar fluctuations sourced by products of short-wavelength fluctuations.
All gravitational non-linearities have been moved to the r.h.s.~of the Einstein equations to define the effective stress-energy tensor $\tau^\mu_{\ \nu}$.
The tensor $\tau^\mu_{\ \nu}$ is conserved by virtue of the linearized Bianchi identity---see Eqn.~(\ref{equ:BFRW3})).

\subsection{Effective Stress-Energy}
\label{sec:effectiveX}

From the expressions for the second-order Einstein tensor, Eqns.~(\ref{equ:GnlFirst})--(\ref{equ:GnlLast}), we read off the components of the stress-energy pseudo-tensor 
\bea
\label{equ:first} \tau^0_{\ 0} &=& - \rho (1+ v_{k} v_{k}) - \frac{\phi_{,k} \phi_{,k} -4 \phi \phi_{,kk}}{8\pi G a^2}\, ,\\
\tau^i_{\ j} &=& \rho  v_{i} v_{j} - \frac{\phi_{,k} \phi_{,k} \delta^i_j - 2 \phi_{,i} \phi_{,j}}{8\pi G a^2}\, . \label{equ:second}\eea
Here we have used the fact that $\Psi^{(1)} = \Phi^{(1)} \equiv \phi$.
Corrections to $\tau^0_{\ i}$ are of order $v^3$ and hence not shown.
We define the trace $\tau \equiv \tau^i_{\ i}$ and the traceless part $\hat \tau^i_{\ j} \equiv \tau^i_{\ j} - \frac{1}{3} \delta^i_j \tau^k_{\ k}$.
The divergence of Eqn.~(\ref{equ:second}) is
\beq
\label{equ:diver}
\nabla_i \tau^i_{\ j} = \nabla_i (\rho v_i v_j) + \frac{\nabla^2 \phi \nabla_j \phi}{4\pi G a^2} \approx  \nabla_i (\rho v_i v_j) + \rho \nabla_j \phi\, .
\eeq
Eqns.~(\ref{equ:first}) and (\ref{equ:second}) form the basis of our exploration of gravitational non-linearities as an effective fluid.

\section{Non-Linear Gravity as an Effective Fluid}
\label{sec:PerfectFluid}

Having derived the effective stress-tensor associated with non-linear fluctuations in two different ways ({\it cf.}~Sections~\ref{sect:Newton}, \ref{sec:PT} and Appendix~\ref{sec:Euler}), we are now in the position to discuss in detail the effects of short-wavelength fluctuations on the long-wavelength universe.

In \S\ref{sec:integratingout2} we explain the procedure of integrating out short-wavelength fluctuations to arrive at the long-wavelength limit of the effective stress-energy pseudo-tensor $\tau_{\mu \nu}$.
In \S\ref{sec:Boltzmann} we digress to explain why higher-order moments in the Boltzmann hierarchy are suppressed for dark matter particles in a FRW universe. This truncation of the Boltzmann hierarchy allows us to describe the system at long wavelengths as a fluid with density, pressure and anisotropic stress.
We describe the properties of $\tau_{\mu \nu}$ in terms of an effective fluid in \S\ref{sec:tau}.
In \S\ref{sec:super} we show that anisotropic stress is negligible on very large scales and that integrating out small scales is therefore captured simply by a renormalization of the density and pressure of the background.
However, on scales smaller than the horizon, anisotropic stress becomes important for the evolution of long-wavelength perturbations.
We will describe the parameters of this imperfect fluid in the next section, \S\ref{sec:ImperfectFluid}.

\subsection{Integrating out Small Scales}
\label{sec:integratingout2}

The leading contributions to the non-linear gravitational source terms, Eqns.~(\ref{equ:first}) and (\ref{equ:second}), contain two spatial derivatives. For scale-invariant perturbations in the Newtonian potential the effects are hence dominated by small scales with characteristic momentum $q_\star$.
Since we are interested in the theory at scales $k$ much larger than the scale of non-linearities $q_\star$, we define an effective long-wavelength theory by `integrating out' short-wavelength modes below a scale $\Lambda \ll q_\star$.
Here, integrating out short-wavelength fluctuations amounts to smoothing the equations of motion and taking expectation values of the short-wavelength modes in the presence of long-wavelength perturbations, so that one is left with equations in terms of only the long-wavelength modes.\footnote{Taking the expectation value of the short modes isn't strictly necessary.
The backreaction of the short-wavelength modes on the long-wavelength modes could also be dealt with also on a realization by realization basis.}
In real space, the smoothing of perturbations corresponds to a convolution of all fields $X \equiv \{\rho, \Phi, \Psi, \rho \v\}$ with a window function $W_\Lambda$,
\beq
\label{equ:smoothX}
X_\ell \equiv [X]_\Lambda(\x) = \int d^3 \x' \, W_\Lambda(|\x-\x'|) X(\x')\, .
\eeq
Essentially, Eqn.~(\ref{equ:smoothX}) amounts to averaging the fields over domains of size $\Lambda^{-1}$.
Note that we smooth the momentum density, ${\boldsymbol{j}} = \rho \v$, rather than the velocity $\v$. From a
dynamical point of view, it is more natural to average the momentum rather than the velocity, since the former
is an additive quantity for a system of particles. 
We then split all fields into long and short modes,
\beq
X = X_\ell + X_s\, . \label{equ:XX}
\eeq

Next, we consider the
smoothing of the Einstein equations (\ref{equ:NL1_0}) -- (\ref{equ:NL2_0}),
\bea
\int d^3 \x' \, W_\Lambda(|\x-\x'|)  \cdot G_{\mu \nu}^{\rm L}(\x') &=& \int d^3 \x' \, W_\Lambda(|\x-\x'|)  \cdot \{\ \tau_{\mu \nu} - \bar T_{\mu \nu} \ \} \label{equ:EinsteinSmooth}\\
 &\equiv& [\tau_{\mu \nu}]_\Lambda - \bar T_{\mu \nu}\, .
\eea
Being linear in metric fluctuations $\Phi$ and $\Psi$, the l.h.s.~of Eqn.~(\ref{equ:EinsteinSmooth}) extracts the long-wavelength metric perturbations $\Phi_\ell$ and $\Psi_\ell$.
To compute the non-linear source terms for $\Phi_\ell$ and $\Psi_\ell$, we substitute the fields in Eqn.~(\ref{equ:XX}) into the r.h.s.~of Eqn.~(\ref{equ:EinsteinSmooth}), using Eqns.~(\ref{equ:first}) and (\ref{equ:second}) for the effective stress-energy tensor $\tau_{\mu \nu}$.
We find, 
\beq
\label{equ:effTau}
[\tau_{\mu \nu}]_\Lambda = [\tau_{\mu \nu}]^\ell + [\tau_{\mu \nu}]^s +  [\tau_{\mu \nu}]^{\partial^2} \, ,
\eeq
where $ [\tau_{\mu \nu}]^\ell$ depends only on long-wavelength modes, $X_\ell$, and $ [\tau_{\mu \nu}]^s$ is quadratic in short-wavelength modes, $X_s$.
Mixed terms with one long-wavelength mode and one short-wave\-length mode vanish up to higher-derivative terms which are suppressed by powers of $k/\Lambda$. 
Collectively we denote all higher-derivative corrections by $[\tau_{\mu \nu}]^{\partial^2}$.
Typically, these higher-derivative terms can be dropped unless the aim is very high precision.
Non-linear metric contributions in $ [\tau_{\mu \nu}]^\ell$ may be moved back to the l.h.s.~of the Einstein equations to define the non-linear completion of the Einstein tensor for long-wavelength modes,  $[G_{\mu \nu}]^\ell =[G_{\mu \nu}^{\rm L}]^\ell + [G_{\mu \nu}^{\rm NL}]^\ell$. These modes are sourced by $ [\tau_{\mu \nu}]^s$.

After a computation completely analogous to that performed explicitly in Appendix~\ref{sec:Euler} (see also the {\bf Example} below) we find:
\begin{itemize}
\item {\it long-wavelength modes}
\beq
 [\tau^0_{\ 0}]^\ell = - \rho_\ell  (1+v_\ell^k v_\ell^k) - \frac{\phi^\ell_{,k} \phi^\ell_{,k} - 4 \phi^\ell \phi^\ell_{,kk}}{8\pi G a^2}
 \, , \eeq
 \beq
 [\tau^i_{\ j}]^\ell = \rho_\ell  v_{i}^\ell v_{j}^\ell - \frac{\phi_{,k}^\ell \phi_{,k}^\ell \delta^i_j - 2 \phi_{,i}^\ell \phi_{,j}^\ell}{8\pi G a^2}\, .
\eeq
\item {\it short-wavelength non-linearities}
\beq
\label{equ:tauShort1}
 [\tau^0_{\ 0}]^s = - [\rho v_s^k v_s^k]_\Lambda - \frac{[\phi^s_{,k} \phi^s_{,k}]_\Lambda - 4 [\phi^s \phi^s_{,kk}]_\Lambda}{8\pi G a^2}
 \, ,
\eeq
\beq
\label{equ:tauShort2}
 [\tau^i_{\ j}]^s =[\rho  v_{i}^s v_{j}^s]_\Lambda - \frac{[\phi_{,k}^s \phi_{,k}^s]_\Lambda \delta^i_j - 2 [\phi_{,i}^s \phi_{,j}^s]_\Lambda}{8\pi G a^2}\, .
\eeq
\item {\it higher-derivative terms}
\beq\label{eq:higher_derivatives}
 [\tau^0_{\ 0}]^{\partial^2} = - \rho_\ell \frac{\nabla v^k_\ell \cdot \nabla v^k_\ell}{\Lambda^2}  - \frac{\nabla \phi_{,k}^\ell \cdot \nabla \phi_{,k}^\ell  - 4 \nabla \phi^\ell\cdot  \nabla \phi_{,kk}^\ell}{8\pi G a^2 \cdot \Lambda^2}\, ,
\eeq
\beq
 [\tau^i_{\ j}]^{\partial^2} = \rho_\ell \frac{\nabla v^i_\ell \cdot \nabla v^j_\ell}{\Lambda^2}  + \frac{\nabla \phi_{,k}^\ell \cdot \nabla \phi_{,k}^\ell \, \delta^i_j - 2 \nabla \phi_{,i}^\ell\cdot  \nabla \phi_{,j}^\ell}{8\pi G a^2 \cdot \Lambda^2}\, .
 \label{equ:effTauLast}
\eeq
\end{itemize}
\vspace{0.5cm}
 \hrule
 \vskip 1pt
 \hrule \vspace{0.3cm}
 \small
\noindent 
{\bf Example}:\vskip 4pt

The derivation of Eqns.~(\ref{equ:effTau}) -- (\ref{equ:effTauLast}) is tedious, but instructive.
In order to provide some intuition for the physical origin of the different terms we now give a sample computation, which applies to all bilinear terms in $\tau_{\mu\nu}$.
All other terms ({\it i.e.}~the trilinear ones $\sim \rho v v$) work in a similar way and are shown in Appendix~\ref{sec:Euler}.

Consider smoothing of a bilinear quantity $f g$,
\beq
\label{equ:star}
[ fg]_\Lambda =  \int_{\x'} W_\Lambda\, f(\x') g(\x') \, . 
\eeq
We split the fields $f$, $g$ into long-wavelength modes $f_\ell$, $g_\ell$ and short-wavelength modes $f_s$, $g_s$.
We get
\beq \label{fg}
[ fg]_\Lambda  = [f_\ell g_\ell]_\Lambda  + [f_s g_s ]_\Lambda + [ f_\ell g_s]_\Lambda + [ f_s g_\ell]_\Lambda\, .
\eeq
The second term is already in the short-short form of Eqns.~(\ref{equ:tauShort1}) and (\ref{equ:tauShort2}). We now massage the other terms. In the first term, given their mild dependence on $\x'$, we can expand the long modes  $f_\ell(\x')$ and $g_\ell(\x')$ in a Taylor series about $\x$:
\bea \label{flgl}
[f_\ell g_\ell]_\Lambda (\x) & = & \int_{\x'} W_\Lambda\, f_\ell(\x') g_\ell(\x') \\
				& = & f_\ell g_\ell + \frac{1}{\Lambda^2} \Big( \nabla f_\ell \cdot \nabla g_\ell + \frac12 f_\ell \nabla^2 g_\ell +  \frac12 g_\ell \nabla^2 f_\ell \Big) + \dots \; ,
\eea
where the dots stand for even higher derivative terms. We used the normalization condition for the window function, $\int W_\Lambda = 1$, and the characterization of $1/\Lambda^2$ as the real-space variance of $W_\Lambda$:
\beq
\int_{\x'} W_\Lambda \,  (\x - \x')^i (\x - \x')^j = \frac{1}{\Lambda^2} \delta^{ij}  \; .
\eeq
In fact, we take this as the {\em definition} of what we mean by the smoothing scale $\Lambda^{-1}$. This way the expansion (\ref{flgl}) does not depend on the specific window function we adopt (we are of course assuming isotropy of $W_\Lambda$).

To simplify the mixed long-short terms in Eqn.~(\ref{fg}) we have to work a little harder. For instance, for the third term we
rewrite
\beq
[ f_\ell g_s]_\Lambda = [ f_\ell g]_\Lambda - [ f_\ell g_\ell]_\Lambda  \; .
\eeq
Here, the second term we already dealt with, so we have to simplify the first term. However, to proceed we need further assumptions on $W_\Lambda$ -- {\it i.e.}~the result will depend explicitly on the form of the window function. A particularly convenient choice is a Gaussian, because then we have useful identities:
\begin{align}
 \label{equ:Gaussian1x}
\partial_{j'} W_\Lambda &= - \partial_j W_\Lambda = \Lambda^2 (\x-\x')^j W_\Lambda \, , \\
\partial_{i'} \partial_{j'} W_\Lambda &= \partial_i \partial_j W_\Lambda = - \Lambda^2 \delta_{ij} W_\Lambda + \Lambda^4 (\x-\x')^i (\x-\x')^j W_\Lambda\, ,  \label{equ:Gaussian2x} \, .
\end{align}
We can then expand $f_\ell$ in a Taylor series about $\x$ like before and get
\bea
[ f_\ell g]_\Lambda & = & f_\ell g_\ell - \nabla f_\ell \cdot [(\x - \x') g(\x')]_\Lambda + \frac12 \nabla_i \nabla_j f_\ell \cdot [  (\x - \x')^i (\x -\x')^j \, g(\x') ] _\Lambda + \dots \\
& = &   f_\ell g_\ell + \frac1{\Lambda^2} \Big(  \nabla f_\ell \cdot \nabla g_\ell + \frac12 g_\ell \nabla^2 f_\ell \Big) + \dots 
\eea
In conclusion, our smoothed bilinear is
\beq
[fg]_\Lambda = f_\ell g_\ell + [f_s g_s]_\Lambda +  \frac1{\Lambda^2}\nabla f_\ell \cdot \nabla g_\ell + \dots \; ,
\eeq
which has exactly the structure of Eqns.~(\ref{equ:effTau}) -- (\ref{equ:effTauLast}).
\vspace{0.2cm} 
 \hrule
 \vskip 1pt
 \hrule
 \vspace{0.5cm}
 
 \normalsize
For notational simplicity we will often drop the index `$s$' denoting short-wavelength quantities, but it is important to keep in mind that the fields entering the effective stress-energy 
for long-wavelength perturbations
are the short-wavelength fields $v_s$ and $\phi_s$. 
In the remainder of the paper $\tau_{\mu \nu}$ will always refer to $ [\tau_{\mu \nu}]^s + [\tau_{\mu \nu}]^{\partial^2} \approx  [\tau_{\mu \nu}]^s$.

When considering the effect on scales with $k \ll \Lambda$ we will often replace the Fourier modes inside the small domains defined by the scale $\Lambda$ by their {\it ensemble averages},
\beq
\label{equ:avetau}
\langle [\tau_{\mu \nu}]_\Lambda \rangle(\x)\, .
\eeq
The ensemble average is used to determine what the spatial average would be in a {\it typical domain}.
The effect of random statistical fluctuations is captured by the variance,
\beq
{\rm Var}([\tau_{\mu \nu}]_\Lambda) \equiv \langle [\tau_{\mu \nu}]_\Lambda^2  \rangle -  \langle [\tau_{\mu \nu}]_\Lambda \rangle^2\, ,
\eeq
{\it i.e.}~the variance quantifies the expected statistical variation of the spatial average between different domains.
This goes to zero as the domain size becomes large, but can be important in some applications, see \S\ref{sec:Stochastic}.

In the following we will assume this smoothing, $[...]_\Lambda$, and ensemble averaging, $\langle ... \rangle$, in the long-wavelength limit, but we will often simplify our notation to 
\beq
\langle [\tau_{\mu \nu}]_\Lambda \rangle \ \to \ \langle \tau_{\mu \nu} \rangle \ \to \ \tau_{\mu \nu} \, .
\eeq
The meaning of this notation should be clear from the context.

Finally, we remark that the spatial dependence of Eqn.~(\ref{equ:avetau}) can arise from two different effects:
\begin{enumerate}
\item {\it Long-wavelength fluctuations} (Section~\ref{sec:ImperfectFluid}): Long-wavelength density fluctuations correlate small-scale fluctuations in spatially separated regions.
\item {\it Random statistical fluctuations} (Appendix~\ref{sec:estimates}): Stochastic fluctuations arise from quantities in a given realization of the universe being different from their ensemble averages.
\end{enumerate}

\subsection{Truncation of the Boltzmann Hierarchy}
 \label{sec:Boltzmann}
 
 In the next section we will interpret the coarse-grained effective stress-energy tensor $\langle \tau_{\mu \nu} \rangle$ as an effective fluid with density, pressure and anisotropic stress. This fluid description relies on a truncation of the hierarchy of moments of the distribution function for dark matter particles. We therefore digress briefly to argue that this truncation naturally arises for collisionless dark matter particles in an FRW universe. Our discussion is based on the classic treatment of the Boltzmann hierarchy by Ma and Bertschinger \cite{Ma:1995ey}.
 
 A collisionless gas of non-relativistic  particles is described by the particle phase space density $f(\x, \p, \eta)$, where $f d^3 x d^3 p$ is the number of particles in an infinitesimal phase space volume. 
 Here, $\x$ is the comoving spatial coordinate of the particle and $\p \equiv a m \dot \x$ its conjugate momentum.
 Phase-space conservation leads to the collisionless Boltzmann equation (or Vlasov equation) \cite{Ma:1995ey}
 \beq
 \label{equ:Boltz}
 \frac{\partial f}{\partial \eta} + \frac{\p}{am} \cdot \nabla f - a m \nabla \Phi \cdot \frac{\partial f}{\partial \p} = 0\, .
 \eeq
 Moments of the distribution function $f$ define the particle density, $\rho$, the particle momentum flow, $\rho \v$, as well as isotropic and anisotropic contributions to the pressure, $p$ and $\sigma$ (see Appendix~\ref{sec:Euler}).
 Inhomogeneities 
and anisotropies are described perturbatively 
 \beq
 f(\x, \p, \eta ) = f_0(p) \left[ 1+ \delta_f(\x, p, \hat \p, \eta)\right]\, .
 \eeq
Let us assume that we are in the quasi-linear regime, and expand the Fourier modes of the perturbation $\delta_f$ in terms of Legendre polynomials $P_n$,
\beq
\delta_f(\k, p, \hat \p, \eta) \equiv \sum_{n=0}^\infty (-i)^n (2n+1) \delta_{f}^{[n]}(\k, p, \eta) P_n(\hat \k \cdot \hat \p)\, ,
\eeq
where $\hat \k$ and $\hat \p$ are vectors of unit norm.
The first few moments of the perturbation $f_0 \delta_f$ then give the perturbed energy density, pressure, energy flux, and shear stress
\bea
\delta \rho &=& m a^{-3} \int 4\pi p^2 d p\,  f_{0}(p) \delta_f^{[0]}  \, , \\
\delta p &=& \frac{1}{3} m a^{-3} \int 4\pi p^2 d p \, \frac{p^2}{a^2 m^2} f_0(p)  \delta_{f}^{[0]} \, , \\
(\bar \rho + \bar p) \theta &=&  m a^{-3}\, k \int 4\pi p^2 d p\, \frac{p}{am} f_0(p)  \delta_{f}^{[1]} \, , \\
(\bar \rho + \bar p) \sigma &=& - \frac{2}{3} m a^{-3} \int 4\pi p^2 d p\, \frac{p^2}{a^2 m^2} f_0(p)  \delta_{f}^{[2]}\, .
\eea
We now argue that higher moments are systematically suppressed for scales larger than the non-linear scale, {\it i.e.}
\beq
 \delta_{f}^{[n \ge 3]}\ \ll\  \delta_{f}^{[2]}\, , \qquad {\rm for} \quad k \gg k_{\rm NL}\, .
\eeq
First, we note that the Boltzmann equation (\ref{equ:Boltz}) implies a hierarchy of evolution equations for the higher moments \cite{Ma:1995ey}
\bea
\dot  \delta_{f}^{[n]} &=& k v_{\rm p} \left[ \frac{n+1}{2n+1}  \delta_{f}^{[n+1]} - \frac{n}{2n+1}  \delta_{f}^{[n-1]}\right]\, , \quad n \ge 2\, , \label{equ:Boltz2}
\eea
where $\v_{\rm p} =\frac{\p}{am}= \dot \x$ is the particle's peculiar velocity which is to be distinguished from the mean peculiar 
velocity entering $\theta = \nabla \cdot \v$.
At long wavelengths, any time evolution is of order the Hubble time $ {\cal H}^{-1}$, so that Eqn.~(\ref{equ:Boltz2}) may be estimated to give
\bea
\label{equ:Boltz3}
 \delta_{f}^{[n]} &\sim&  k v_{\rm p} \H^{-1}  \left[ \frac{n+1}{2n+1}  \delta_{f}{[n+1]} - \frac{n}{2n+1}  \delta_{f}^{[n-1]}\right]\, , \quad n \ge 2\, .
\eea
If $k v_{\rm p} \H^{-1}$ is much smaller than unity, this leads to a natural hierarchy of higher moments
\beq
 \delta_{f}^{[n]} \sim (k v_{\rm p} \H^{-1})^{n-2} \,  \delta_{f}^{[2]}\, .
\eeq
This is similar to what happens in a conventional fluid, for which an equation like Eqn.~(\ref{equ:Boltz3}) applies with prefactor $k v_{\rm p} \tau_c$, where $\tau_c$ is the characteristic collision time. In a sense, the presence of such a hierarchy is really the definition of a fluid.
In our case, $v_{\rm p} \tau_c$ is the mean free path and at distances larger than the mean free path, $k v_{\rm p} \tau_c \ll 1$, we can truncate the hierarchy and describe the system by coarse-grained fluid variables. In the absence of such a hierarchy, our approach for deriving a long-distance effective fluid would be doomed. 

For the dark matter fluid is not obvious that $k v_{\rm p} \H^{-1}$ should be small since $k/\H \gg 1$ inside the horizon.
To see that the truncation of the Boltzmann hierarchy indeed occurs we need to estimate the typical particle velocities.
As a bound on the maximal particle velocities we take the velocity at the non-linear scale
\beq
v_{\rm p} \le \Delta_v(k_{\rm NL})\, ,
\eeq
where (see Appendix~\ref{sec:Einstein})
\beq
\Delta_v^2(k_{\rm NL}) \sim \Delta_\delta^2(k_{\rm NL}) \frac{\H^2}{k_{\rm NL}^2} \sim \frac{\H^2}{k_{\rm NL}^2}\, .
\eeq
This implies that
\beq
k v_{\rm p} \H^{-1} \lesssim \frac{k}{k_{\rm NL}}\, ,
\eeq
which is indeed small for scales larger than the non-linear scale, $k \ll k_{\rm NL}$.
The Boltzmann hierarchy therefore can be truncated beyond $ \delta_{f}^{[2]}$ (the moment responsible for anisotropic stress). This may be understood intuitively: during a Hubble time particles don't move more than a non-linear distance, not because they interact strongly like in a conventional fluid, but because they move slowly and haven't had time to move far.
The hierarchy between the Hubble scale $\H^{-1}$ and the non-linear scale plays a fundamental role in this interpretation.
Gravity induces a finite age and a finite horizon, which allows for the formation of a hierarchy among the multipoles. On the contrary, in the absence of gravity, the system could be infinitely old, and we would be left with no hierarchy and therefore no fluid. It is in this sense that we refer to the result of integrating out short-wavelength non-linearities in a FRW universe as a `gravitational fluid', highlighting the fundamental importance of gravity.

\subsection{The Effective Fluid}
\label{sec:tau}

Let us now begin to discuss the physical content of the stress-energy tensor arising from second-order matter fluctuations at short scales ($k \gg \Lambda$) as described by Eqns.~(\ref{equ:tauShort1}) and (\ref{equ:tauShort2}). We argued in the previous subsection that at long wavelengths ($k \ll \Lambda$) the effective stress-energy tensor can be put into the form of a fluid with density, pressure and anisotropic stress, 
\beq
\rho_{\rm eff}= \langle \tau_{\mu \nu} \rangle U^\mu_\ell U^\nu_\ell \approx - \langle \tau^0_{\ 0} \rangle \, , \qquad 3 p_{\rm eff} =\langle  \tau_{\mu \nu} \rangle \gamma^{\mu \nu}_\ell \approx  \langle \tau^i_{\ i} \rangle  \, , \qquad (\Sigma^i_{\ j})_{\rm eff} \approx \langle  \hat \tau^i_{\ j}  \rangle\, ,
\eeq
where $\hat \tau^i_{\ j}$ is the traceless part of $\tau^i_{\ j}$, and we have absorbed a small correction from bulk viscosity---{\it cf.}~Eqn.~(\ref{equ:pf})---into the definition of $p_{\rm eff}$.
We notice from the defining equations of $\tau_{\mu \nu}$ that our description of the effective fluid requires as the basic
input the following correlation functions of the short-wavelength fluctuations,
\bea
\label{equ:kij}
\kappa_{ij} &\equiv& \frac{1}{2}\langle (1+\delta ) v_i v_j \rangle  \\
\omega_{ij} &\equiv& - \frac{\langle \phi_{,i} \phi_{,j} \rangle}{8\pi G a^2 \bar \rho} \approx \frac{\langle \phi_{,ij} \, \phi \rangle}{8\pi G a^2 \bar \rho} \, . \label{equ:wij}
\eea
In the last passage of Eqn.~(\ref{equ:wij}) we have assumed that $\langle\phi\phi_{,i}\rangle=0$. As we will discuss in \S\ref{sec:tidal}, this quantity is in general non-zero and proportional to a derivative of long-wavelength fluctuations. This term is therefore irrelevant in the long-wavelength limit. For the moment we neglect this effect and concentrate only on the long-wavelength limit. With the above definitions,
the spatially-averaged components of the effective stress-energy tensor, Eqns.~(\ref{equ:tauShort1}) and (\ref{equ:tauShort2}), can be written as
\bea
\langle \tau^0_{\ 0} \rangle &=& - \bar \rho (1+2 \kappa - 5\omega)\, , \label{equ:tau1}\\
\langle \tau^i_{\ j} \rangle &=& \bar \rho (2\kappa_{ij} + \omega \delta_{ij} -2 \omega_{ij})\, ,\label{equ:tau2}
\eea
where $\kappa$ and $\omega$ are
the kinetic energy and the potential energy of a fluid element normalized with respect to the background energy density $\bar \rho$,
\beq
\kappa = \kappa_{ii}= \frac{1}{2} \langle (1+\delta) v^2  \rangle \qquad {\rm and}
 \qquad \omega = \omega_{ii} =\frac{1}{2} \langle \delta\, \phi  \rangle \, < \, 0\, .
\eeq
The spatial averages of the kinetic and potential energies are related by the {\it cosmic energy equation} (or {\it  Irvine-Layzer equation}) \cite{Peebles}
\beq
\label{equ:ce}
\frac{d (\kappa+\omega)}{d \eta} + {\cal H}(2\kappa + \omega) = 0\, .
\eeq

Finally, we note that the properties of the effective stress-energy tensor, Eqns.~(\ref{equ:tau1}) and (\ref{equ:tau2}), are expressed in terms of the power spectra of small-scale density and velocity fluctuations, {\it e.g.}
\beq
\omega_{ij} \sim - \int_\Lambda d \ln q \ \frac{q_i q_j}{\H^2} \, \Delta_\phi^2(q)\, ,
\eeq
where the integral is over modes with $q > \Lambda$.
This information is available through the halo model \cite{Cooray:2002dia} or $N$-body simulations \cite{Heitmann:2008eq, Heitmann:2009cu, Widrow:2009ru}.
Alternatively, the cosmic energy equation (\ref{equ:ce}) relates $\kappa$ and $\omega$ given some initial conditions \cite{Siegel:2005xu}. 
\subsection{Renormalization of the Background}
\label{sec:super}

From the above analysis it is straightforward to see that
integrating out short-wavelength fluctuations leads to a renormalization of the background. We define the new background as the $k \ll \Lambda$ limit of the effective fluid,
\beq
\label{equ:back}
\bar \rho_{\rm eff} \equiv - \lim_{k \ll \Lambda} \langle \tau^0_{\ 0} \rangle \, , \qquad 3 \bar p_{\rm eff} \equiv  \lim_{k\ll \Lambda} \langle \tau^i_{\ i} \rangle \, , \qquad (\bar \Sigma^i_{\ j})_{\rm eff} \equiv \lim_{k \ll \Lambda} \langle  \hat \tau^i_{\ j}  \rangle\, .
\eeq
Eqn.~(\ref{equ:back}) 
describes the fluid on very large scales, where spatial fluctuations are suppressed by $k^2/q_\star^2$, with $q_\star$ the typical scale of non-linearities.
In particular, on superhorizon scales these fluctuations are highly suppressed.
We defer a treatment of fluctuations on subhorizon scales until \S\ref{sec:ImperfectFluid}.

\vskip 6pt
\noindent
{\sl Density.} \hskip 6pt
We find that the {\it effective energy density} receives contributions from the kinetic and potential energies associated with small-scale fluctuations
\beq
\bar \rho_{\rm eff} =  \bar \rho (1+ 2\kappa - 5 \omega)\, .
\eeq
Introducing the `boosted mass density'\footnote{The physical role of $\rho_m$ becomes clear when thinking about dark matter in terms of particles and their conserved number density, $n \propto \rho_m$. A relativistic boost relates the physical number density in the inertial frame where the fluid velocity is $v$ to the number in the fluid rest frame, $n_{\rm rest} \propto \rho$. The factor $\gamma(v) e^{-3\phi}$ in Eqn.~(\ref{equ:boost}) clearly relates the volume element in the rest frame to the physical volume in the moving frame.} (see Appendix~\ref{sec:Einstein})
\bea
\rho_m &\equiv& \gamma(v) e^{-3\phi} \rho\, , \label{equ:boost} \\
&=&
\rho \left(1+ \frac{v^2}{2} - 3 \phi \right) = \rho + \bar \rho (\kappa - 6\omega)\, ,
\eea
the effective density of the fluid is
\bea
\bar \rho_{\rm eff}  &=& \langle \gamma \rho_m \rangle + \frac{1}{2} \langle \rho_m \phi \rangle\, ,\\
&=& \bar \rho_m(1+ \kappa + \omega)\, . 
\eea
This shows that the {\it background energy density is corrected precisely by the total kinetic and potential energies associated with non-linear small-scale structures}, in perfect agreement with our analysis in \S\ref{sect:Newton}.
For virialized scales with $2\kappa+\omega=0$ the cosmic energy equation (\ref{equ:ce})
 implies that the sum of kinetic and potential energies, $\kappa+\omega$, is constant, so that {\it virialized scales only contribute a time-independent renormalization of the background density}.

\vskip 6pt
\noindent
{\sl Pressure.} \hskip 6pt
The {\it effective pressure} of the fluid is
\beq
\label{equ:peffX}
3 \bar p_{\rm eff}  = \bar \rho_m (2\kappa + \omega)\, ,
\eeq
and its {\it equation of state} is
\beq
\label{equ:eos}
\bar w_{\rm eff} \equiv \frac{\bar p_{\rm eff}}{\bar \rho_{\rm eff}} = \frac{1}{3} (2\kappa + \omega)\, .
\eeq
We see that for virialized scales the effective pressure vanishes.
As intuitively expected, {\it a universe filled with virialized objects acts like pressureless dust}. 
(This agrees with the conclusion reached by Peebles in \cite{Peebles:2009hw}.)
Non-virialized structures, however, do have a small effect on the long-wavelength universe, giving corrections to the background of order the velocity dispersion, ${\cal O}(v^2)$.  We will have more to say about this in \S\ref{sec:evolution}.
Furthermore, in Appendix~\ref{sec:estimates} we will show in perturbation theory that $2\kappa+\omega > 0$ ({\it e.g.}~in linear theory $2\kappa_{\rm L} + \omega_{\rm L} = \frac{1}{2} \kappa_{\rm L} >0$ in Einstein-de Sitter). Peebles \cite{Peebles}, using the cosmic energy equation (\ref{equ:ce}), shows that the inequality $2\kappa + \omega > 0$ even extends to the non-perturbative regime.
We have therefore shown that {\it the induced effective pressure is always positive}, $\bar p_{\rm eff} > 0$.

\vskip 6pt
\noindent
{\sl Anisotropic stress.} \hskip 6pt
On very large scales the anisotropic stress  $(\bar \Sigma^i_{\ j})_{\rm eff}$ averages to zero, {\it i.e.}~it has no long-wavelength contribution:
\beq
\lim_{k \ll \Lambda} (\bar \Sigma^i_{\ j})_{\rm eff} \approx 0 \; .
\eeq
This straightforwardly follows from the isotropy of the fluctuation power spectrum.
On very large scales, the gravitationally-induced fluid therefore acts like 
an isotropic fluid; its only effects are small ${\cal O}(v^2)$ corrections to the background density and pressure.
Anisotropic stress, however, does become important when studying the evolution of perturbations on subhorizon scales.
We discuss this limit in \S\ref{sec:ImperfectFluid}.

\subsection{Decoupling and the Virial Theorem}

 In the Einstein-de Sitter universe, $\Omega_m =1$, the Newtonian potential is scale-invariant on all scales and the main contribution to the effective gravitational stress-energy comes from very short-wavelength modes.\footnote{In the Einstein-de Sitter universe the non-linear scale $k_{\rm NL}$ is always a fixed fraction of the horizon scale, $k_{\rm NL}/{\cal H} = {\rm const.}$ In the real universe, the scale of matter-radiation equality is always bigger than the non-linear scale. The effect of non-linear scales therefore receives an additional suppression from the fact that they entered the horizon during radiation-domination and then decayed until matter-domination. In that case the biggest contribution is expected from modes with $k \sim k_{\rm eq}$ \cite{baum}.}
 In principle one may worry that the effect becomes arbitrarily large for modes with $k 
 \gg {\cal H}$.
 In this section we show that modes of very short wavelengths, in fact, give {\it no} effect if they correspond to virial equilibrium.
 
In the previous section we have already seen an example for the decoupling of virial scales in the fact that they do not contribute an effective pressure.
Here, we will demonstrate that this result extends to the full tensorial version of the virial theorem.
Specifically, we will prove that the spatial part of the effective stress-energy tensor, averaged over a domain of size $\Lambda^{-1}$, is equal to the second derivative of the inertia tensor associated with that domain,
\beq
\label{equ:virialR}
[ \tau_{ij} ]_\Lambda= \frac{1}{2}\frac{d^2  I_{ij} }{d \eta^2} \, .
\eeq

\vspace{0.5cm}
\hrule \vskip 1pt
 \hrule \vspace{0.3cm}
 \noindent 
\small
{\sl Virial scales decouple from the long-wavelength dynamics.}
\vskip 2pt
\noindent
{\bf Proof I}:\vskip 4pt
Consider
\bea
 [\tau_{ij}]_\Lambda(\x)  &=& \int_{\x'} W_\Lambda(|\x -\x'|) \,\tau_{ij}(\x') \\
&=&  [\rho v_i v_j]_\Lambda - \frac{1}{8\pi G} [\phi_{,k} \phi_{,k}\delta_{ij} - 2 \phi_{,i} \phi_{,j}]_\Lambda \, . \label{equ:vir}
\eea
Since we are interested in the effect of virial scales deep inside the horizon we ignored the expansion of the universe by setting $a=1$. There is a subtlety associated with this assumption that we will discuss at the end of the next proof.
The first term in Eqn.~(\ref{equ:vir}) may be written in terms of the tensor
\beq
K_{ij} \equiv \frac{1}{2} \int_{\x'} W_\Lambda(|\x-\x'|)\,  \rho v_i v_j(\x')\, ,
\eeq
whose trace, $K = K_{ii}$, is the non-relativistic kinetic energy of the fluid. We now show that the second term can be written as the tensor
\beq
\label{equ:Wij}
W_{ij} \equiv \frac{1}{2} \int_{\x'} W_\Lambda(|\x-\x'|) \rho(\x') \Phi_{ij}(\x') \, , 
\eeq
where
\beq
\label{equ:Phiij}
 \Phi_{ij}(\x') \equiv - G \int_{\r} \rho(\x'+{\r})\frac{r_i r_j}{r^3}\, .
\eeq
The trace of $W_{ij}$ is the gravitational potential energy of the fluid.

Consider the Fourier transform of the term $\phi_{,i} \phi_{,j}$,
\bea
\int_\x e^{-i \k \cdot \x} \phi_{,i} \phi_{,j} &=&-  \int_\q q^i (\k-\q)^j \phi_\q \phi_{\k -\q}\, \\
&\rightarrow& (4\pi G)^2 \int_\q \frac{q^i q^j}{q^4} \rho_\q \, \rho_{\k - \q}\, .
\eea
In the second line we used the Poisson equation in the Newtonian limit, $q^2 \phi_\q = 4\pi G \rho_\q$, and took the limit $k \ll q$.
For the gravitational term in Eqn.~(\ref{equ:vir}) this implies
\beq
\label{equ:intX}
\star \ \equiv \ - \frac{1}{8\pi G} \int_\x e^{-i\k \cdot \x} \left(\phi_{,k} \phi_{,k}\delta_{ij} - 2 \phi_{,i} \phi_{,j} \right) = - \frac{G}{2} \int_\q \frac{4\pi}{q^4} \left[q^2 \delta_{ij} - 2 q_i q_j\right] \rho_\q\, \rho_{\k -\q}\, .
\eeq
Using the identity
\beq
\int_\q e^{ i \q \cdot \x} \frac{q_i q_j}{q^4} = \frac{1}{8\pi |\x|} \left[\delta_{ij} - \frac{x_i x_j}{|\x|^2} \right]\, ,
\eeq
and hence
\beq
\int_\q e^{ i \q \cdot \x} \frac{4\pi}{q^4} \left[q^2 \delta_{ij} - 2 q_i q_j\right] = \frac{x_i x_j}{x^3}\, ,
\eeq
we can write Eqn.~(\ref{equ:intX}) as
\beq
\star \ =\ \frac{1}{2} \int_\q \Phi_{ij}(\q) \, \rho_{\k-\q}\, ,
\eeq
where $\Phi_{ij}(\q)$ is the Fourier transform of Eqn.~(\ref{equ:Phiij}).
Considering the Fourier transform of Eqn.~(\ref{equ:Wij}),
\beq
W_{ij}(\k) = \frac{1}{2} W_\Lambda(k) \cdot \int_\q \Phi_{ij}(\q)\, \rho_{\k - \q}\, ,
\eeq
we find
\beq
 - \frac{1}{8\pi G} [\phi_{,k} \phi_{,k}\delta_{ij} - 2 \phi_{,i} \phi_{,j}]_\Lambda  = W_{ij}\, .
\eeq
Hence,
\beq
[ \tau_{ij}]_\Lambda = 2 K_{ij} + W_{ij}\, .
\eeq
In a background with non-zero pressure $p$ we would have found
\beq
\label{equ:chandra}
[ \tau_{ij}]_\Lambda = 2 K_{ij} + W_{ij} + P \delta_{ij}\, , \qquad P \equiv \int_{\x'} W_\Lambda \, p\, .
\eeq
It is a standard result from hydrodynamics ({\it e.g.}~Ref.~\cite{Chandrasekhar:1990uf})\footnote{In the next proof we will not actually need to rely on standard results from fluid dynamics.} that the r.h.s.~of Eqn.~(\ref{equ:chandra}) equals the second time-derivative of the moment of inertia tensor
\beq
I_{ij} \equiv \int_{\x'} W_\Lambda(|\x-\x'|) \rho x_i' x_j'\, .
\eeq
Hence, we have proven the desired result, Eqn.~(\ref{equ:virialR}).
\hfill QED $\blacksquare$
\vspace{0.2cm} 
\hrule \vskip 1pt \hrule
 \vspace{0.5cm}
 
 \newpage
\hrule \vskip 1pt
 \hrule \vspace{0.2cm}
  \noindent
 {\sl Virial scales decouple from the long-wavelength dynamics.}
\vskip 2pt
\noindent 
{\bf Proof II}:\vskip 4pt
Eqn.~(\ref{equ:virialR}) also follows straightforwardly from conservation of the total stress-energy tensor $\tau_{\mu\nu}$ for matter plus gravity,
\beq \label{conservation}
\partial_\mu \tau^{\mu\nu} = 0\, ,
\eeq
as we now show. For the moment we are ignoring the expansion of universe, since we are interested in scales well inside the horizon.
Consider then the following integral:
\beq
0=\int_{\x'} \partial_k \partial_l \big[ W_\Lambda (|\x - \x'|) \, x^i x^j \, \tau^{kl}\big] \; .
\eeq
This vanishes because the integrand is a total derivative, and it is localized. Expanding the derivatives we get
\beq
0 = \int_{\x'} W_\Lambda (|\x - \x'|) \left[ 2 \tau^{ij} + x^ i x^j \, \partial_k \partial_l \tau^{kl} + 2 \, (x^i \delta^j_l + x^j \delta^i_l) \partial_k \tau^{kl} \right] \; ,
\eeq
where (and henceforth) we neglect derivatives acting on $W_\Lambda$, because we are assuming that the smoothing scale $\Lambda^{-1}$ is much larger than the typical variation scale of $\tau^{\mu\nu}$. 
Upon integration by parts, the last term can be written as minus twice the first term. The second term instead, can be converted into a second {\em time}-derivative by using conservation of $\tau^{\mu\nu}$ twice. We finally get
\beq
0 = \int_{\x'} W_\Lambda (|\x - \x'|) \left[ -2 \tau^{ij} + \partial_0^2 \big(x^ i x^j \,  \tau^{00}\big) \right] \; ,
\eeq
which is exactly Eqn.~(\ref{equ:virialR}).

Notice that this proof holds at all orders in the post-Newtonian expansion, since Eqn.~(\ref{conservation}) does. So, even if we consider a {\em relativistic} virialized system, say a black hole-black hole binary, the total stress tensor $\tau_{ij}$, smoothed over a scale larger than the system's size, obeys Eqn.~(\ref{equ:virialR}).

Now, the expansion of the universe is negligible at subhorizon scales, but can its effect build up over time and give ${\cal O}(1)$ modifications to Eqn.~(\ref{equ:virialR}) on time-scales of order Hubble? To show that this cannot happen, it is particularly convenient to use the Newtonian coordinates of \S \ref{sect:Newton}. The background FRW metric is given in Eqn.~(\ref{FRWlocal}).
Consider then a virialized system localized around $\x = 0$, of size $\ell$ much smaller than Hubble. In these coordinates the effect of the FRW expansion on this system is manifestly a tidal effect suppressed by $H^2 \ell^2$, with nothing explicitly becoming large\footnote{In fact for spherical systems there is no effect at all, due to Birkhoff's theorem. So the effect is really a tidal-coupling between the background expansion and the traceless quadrupole moment of the system.}  for $t \sim H^{-1}$. 
Of course, even such a tiny tidal-coupling does perturb the system to some extent, and it will have an integrated effect over long time-scales. However, for this effect to become of order one, one needs to wait for a time of order $H^{-1} ( H^{-1}/ t_s) \gg H^{-1}$, where $t_s$ is the system's dynamical time.
Notice that this is perfectly compatible with there being the usual secular redshift effects associated with the Hubble expansion. For instance right at $\x = 0$  tidal effects are strictly zero, yet the background $\rho$ does redshift according to the standard FRW equation
\beq
\dot \rho = -3 H (\rho + p)\, ,
\eeq
which yields an order-one variation of $\rho$ over a Hubble time. In our coordinates it is clear that this is not a gravitational effect. Rather, it is enforced by the {\em special}-relativistic continuity equation, applied to the Hubble velocity field $\v \simeq H \x$.

In conclusion, the expansion of the universe does not perturb Eqn.~(\ref{equ:virialR}) appreciably, even on time-scales of order Hubble.
\hfill QED $\blacksquare$
\vspace{-0.2cm}  
\hrule \vskip 1pt \hrule
 \vspace{0.5cm}

\normalsize
Note that for virialized sources the r.h.s.~of Eqn.~(\ref{equ:virialR}) is not just zero. Rather, it is of the same order of magnitude as for non-virialized sources, namely $\sim \rho \, v^2$. The crucial property of Eqn.~(\ref{equ:virialR}), however, is the explicit time-derivative structure. Since we are interested in the effect of short-scale structures on the long-scale and long-time evolution of the universe, we are effectively only sensitive to the average of Eqn.~(\ref{equ:virialR}) over a long time $T$, much longer than the typical short-scale dynamical time $t_s$,
\beq
\Big\langle \frac{d^2 I_{ij}}{d \eta^2} \Big\rangle_T = \frac{1}{T} \frac{d I_{ij}}{d \eta} \bigg|^T_0 \sim \frac1T \rho v \ell   \sim \rho v^2 \Big( \frac{t_s}{T} \Big) \ll \rho v^2 \; ,
\eeq
where $\ell$ is the typical size of virialized structures. Equivalently, in Fourier-space we are interested in the very low frequency ($\omega \sim H$) component of Eqn.~(\ref{equ:virialR}), which is automatically small because of explicit powers of the frequency,
\beq
\frac{d^2 I_{ij}}{d \eta^2} \to - \omega^2 I_{ij} (\omega) \; .
\eeq
Note that this Fourier-space argument does not go through for non-viralized sources, because their motion is peaked at low frequencies. For instance, for a free particle with $\rho = m \, \delta_{\rm D}(\x - \x_0 (t))$  and $\x_0(t) = \v_0 t$ we have
\beq
I_{ij} (t) = m \, v_i v_j \, t^2 \delta_{\rm D}(\x-\x_0(t))\;, 
\eeq
whose second time-derivative is unsuppressed, and in fact supported exclusively, at low frequencies.

This decoupling (or non-renormalization) result proves that we shouldn't expect large effects from small-scale non-linearities on the background evolution.
The gravitational non-linearities naturally cut off at very small scales with the main contribution coming from $q_\star \sim k_{\rm NL}$ (in the Einstein-De Sitter universe) or $q_\star \sim k_{\rm eq}$ (in a matter-radiation universe).

\section{Effective Theory of the Perturbed Fluid}
\label{sec:ImperfectFluid}

We have seen that on very large scales, $k \ll {\cal H}$, the effective anisotropic stress induced by second-order scalar fluctuations averages to zero.
On superhorizon scales the effective fluid is therefore well-approximated by a perfect fluid with no dissipation.
However, 
on scales comparable to or smaller than the horizon, $k \gtrsim {\cal H}$, we expect higher-derivative terms to become relevant. For example, we will see that anisotropic stress becomes important for the evolution of long-wavelength perturbations.
The properties of this {\it imperfect fluid} are treated in this section.

In \S\ref{sec:Effective} we introduce the effective theory of the fluid in the presence of long-wavelength perturbations.
The theory is characterized by the UV-IR coupling of short and long modes.
In \S\ref{sec:tidal} we illustrate the main effects for a universe filled with cold dark matter particles. We emphasize the importance of tidal forces from the long modes on the dynamics of the short modes.
We describe, in \S\ref{sec:Viscous}, how the basic parameters of the effective theory are determined in perturbation theory, from matching to numerical simulation, or by fitting to observations.
In \S\ref{sec:StochasticV1} we discuss corrections to the results arising from stochastic fluctuations of the fluid parameters.  Technically these fluctuations arise because quantities in a given realization of the universe are different from their ensemble averages.
Finally, in \S\ref{sec:summary}, we give a brief summary of our results.

\subsection{The Effective Theory}
\label{sec:Effective}

The basic degrees of freedom of the long-wavelength theory are the density contrast $\delta_\ell$ and the velocity divergence\footnote{Below we will also comment on the vorticity ${\boldsymbol{w}}_\ell = \nabla \times {\v}_\ell$. In fact, from a Lagrangian/Hamiltonian viewpoint the mechanical degrees of freedom parameterizing a fluid's configuration space are three scalar fields $\phi^I(\x ,t)$, $I=1,2,3$---the internal coordinates of the fluid element located at $\x$ at time $t$ \cite{Dubovsky:2005xd}. However, at the level of the equations of motion the density and velocity fields suffice.} $\theta_\ell \equiv \nabla \cdot \v_\ell$. 
The (linearized) evolution equation for the velocity divergence is
\beq
\label{equ:EulerL}
\dot \theta_\ell + \H \theta_\ell + \frac{3}{2} \Omega_m \H^2 \delta_\ell = - \frac{1}{\rho_\ell} \nabla_i \nabla_j \langle \tau_{ij} \rangle \, .
\eeq
In this section, we will argue that the non-linear UV-IR coupling of the perturbations leads to the short-wavelength source term $\langle \tau_{ij} \rangle$ being a function of $\delta_\ell$ and $\theta_\ell$. Intuitively, this arises because the presence of long-wavelength modes changes the particle geodesics on small-scales, leading to small changes in the stress-energy associated with short-wavelength modes. These effects originate from tidal forces and  are hence proportional to derivatives of the long-wavelength gravitational potential (in \S\ref{sec:tidal} we will give an explicit example of how these effects arise in a dark matter universe). This leads us to introduce an effective approach, where we expand the source term in Eqn.~(\ref{equ:EulerL}) in derivatives of the long-wavelength fields,
\begin{eqnarray}
\label{equ:exp1}
\langle \tau_{ij} \rangle &=&\rho \left[ c_1 \left(\frac{\langle v_s^2\rangle\partial^2}{\H^2}\right)_{ij}+c_2 \left(\frac{\langle v_s^2\rangle\partial^2}{\H^2}\right)^2_{ij} \ +\ \cdots\right]\;\phi_\ell\; \ \ +\\ \nonumber
&&+\ \ \rho \left[\Bigl(d_1^{(n)}\left(\frac{\partial^2}{\Lambda^2}\right)+d_2^{(n)}\left(\frac{\partial^2}{\Lambda^2}\right)^2\ +\ \cdots \Bigr)\Bigl\{v_\ell^2\, ,\, \delta_\ell\phi_\ell\, ,\, \cdots \Bigr\} \right]_{ij} .
\end{eqnarray}
Here, $c_{i}$ and $d_{i}^{(n)}$ are time-dependent coefficients of order one. The index `$n$' on $d_i^{(n)}$ runs over the fields represented in the vector between curly brackets, $\{v_\ell^2,\delta_\ell\phi_\ell,\cdots\}$. The double index~$_{ij}$ in each term comes from the derivative-structure of the individual terms (constructed from combinations of $\partial_i\partial_j,\;\partial_0$ and $\delta_{ij} \partial^2$). Similar expansions hold for $\tau_{00}$ and $\tau_{0i}$.


The terms in the first line of (\ref{equ:exp1}) are a derivative expansion in powers of $k^2/k_{\rm NL}^2$. It is easy to see how this scaling arises if we think about dark matter in terms of particles: the particles feel tidal forces from gradients of  long-wavelength gravitational modes. Since the particles travel a distance of the order of $v_s \H^{-1}\sim k^{-1}_{\rm NL}$ during a Hubble time, this leads to the $k/k_{\rm NL}$ scaling. Alternatively, we may recall the description of dark matter particles via the Boltzmann equation  in \S\ref{sec:Boltzmann}. There we saw that the moments of the distribution function are connected by a hierarchy of equations. Integration of the Boltzmann equation leads to a hierarchy of moments, with higher moments being suppressed by powers of $k/k_{\rm NL}$. These higher moments feed back into the equations for the first two moments: the density and the velocity. The net effect is equations for the first two moments with additional higher-derivative terms. In an effective approach, we simply introduce these higher-derivative terms directly into the equations of motion and parameterize our ignorance of the theory for the short modes by the coefficients of these higher-derivative terms.  In the next subsection we shall show explicitly how these terms arise in a universe filled with only dark matter.

The terms in the second line of (\ref{equ:exp1}) represent additional higher-derivative terms that are suppressed by powers of $k/\Lambda$, with $\Lambda^{-1}$ being the smoothing scale. These terms arise from smoothing out the short modes to obtain equations for the long modes. 
Contrary to the terms arising from tidal forces, these terms need not be proportional only to the gravitational potential, but, at least in principle, can be proportional to any of the fluctuating variables. These fields are represented in the vector $\{v_\ell^2,\delta_\ell\phi_\ell,\cdots\}$. We point out that the terms originating from the smoothing have to start at quadratic order in fluctuations -- see {\it e.g.}~Eqn.~(\ref{eq:higher_derivatives}). Furthermore, since for a non-relativistic system $\delta_\ell$ appears at non-linear level only when accompanied by other fluctuating fields, terms like $\delta_\ell^2,\delta^3_\ell,\cdots$ do not appear. This makes these higher-derivative terms quite subleading.

Notice that all of the corrections to $\langle\tau_{ij}\rangle$ go to zero as $k\rightarrow 0$ outside of the horizon (once the zero mode has been subtracted; see \S\ref{sec:PerfectFluid}). It is convenient to use the linear Einstein equations to express $\phi_\ell$ and its derivatives in terms of $\delta_\ell$ and $\theta_\ell$. Concentrating only on the leading terms arising from two-derivative terms like $\partial_{i} \partial_j \phi_\ell$, we obtain
\beq
\label{equ:exp}
\frac{k_i k_j}{k^2}\frac{\langle \tau_{ij} \rangle}{\bar \rho} \ =\ c_s^2\, \delta_\ell - c_{\rm vis}^2\, \frac{\theta_\ell}{\H}+\cdots\, .
\eeq
We see that the effect of the higher-derivative terms is to induce in the long-distance fluid an effective pressure perturbation, parametrized by $c_s^2$, and an effective anisotropic pressure (or viscosity), parametrized by $c_{\rm vis}^2$  (see \S\ref{sec:Viscous}). Notice that we parametrized the coefficient of the viscosity as $c_{\rm vis}^2$ in dimensional analogy to $c_s^2$, but the viscosity is of course a dissipative term that does not induce propagating sound waves. The following is worth noticing: $c_s^2$ and $c_{\rm vis}^2$ are comparable in size and lead to comparable effects in the Euler equation, since, at linear level, $\theta_\ell/\H\sim \dot\delta_\ell/\H\sim \delta_\ell$. This makes our effective fluid an unusual one. In conventional fluids the pressure is hierarchically larger than the viscous term by an amount of order $k/k_{\rm c}$, where $k_{\rm c}^{-1}$ is the mean free path of the fluid particles. In our case the pressure happens to be very small, making it comparable to the viscosity.\footnote{This is due to the fact that the speed of sound is small, so that the time derivative of $\delta_\ell$ is dominated by the Hubble flow rather than the effect of the gradient terms. Notice, however, that the effective stress-tensor is still that of a fluid. Higher moments of the particle distribution function are still suppressed with respect to the first two moments by factors of $k/k_{\rm NL}$. It is this feature that ensures that the Boltzmann hierarchy can be truncated and that an effective fluid description is possible (see \S\ref{sec:PerfectFluid}).}

For simplicity, vorticity has been neglected in Eqns.~(\ref{equ:EulerL})--(\ref{equ:exp}). This was not strictly necessary. In practice, the fundamental description of our effective fluid is given by Eqn.~(\ref{equ:exp1}), which describes the effective stress-tensor that enters in the Euler and continuity equations, without any assumptions about vorticity. In reality, since $\tau_{\mu\nu}$ contains viscous terms and the fluid is compressible, even if we start with no vorticity at early times, it will be generated during the evolution (see Eqn.~(\ref{equ:vorticity}) below). This is an important difference to standard perturbation theory. In fact, in the standard approach, dark matter is described as a perfect and pressureless fluid, implying that vorticity cannot be generated unless the velocity field is defined through a proper smoothing similar to our definition (see for example \cite{Pichon:1999tk}). Numerical simulations show that vorticity is generated when non-linear structures are formed. Our viscosity parameters are to be matched to simulations and we expect that the resulting fluid should be able to reproduce the generation of vorticity. 

Ultimately, the parameters of the effective fluid in Eqn.~(\ref{equ:exp}) -- or more precisely in Eqn.~(\ref{equ:exp1}) -- can be calibrated with numerical $N$-body simulations (\S\ref{sec:Viscous}), estimated in perturbation theory (Appendix~\ref{sec:estimates}), or even fitted to observations.
However, before discussing this, we would like to understand the effects more intuitively.

\subsection{Tidal Forces and Anisotropic Stress}
\label{sec:tidal}

It is interesting to see how the higher-derivative terms of the effective theory arise in a concrete example. In this subsection, we consider a universe filled with cold dark matter only\footnote{Our approach of adding higher-derivative terms as in Eqn.~(\ref{equ:exp1}) is of course more general than that: a universe filled with dark matter represents only a particular UV completion of our effective fluid.} and study the dynamics on very short scales. We will see explicitly that in this context, the higher-derivative terms arise because the long modes affect the motion of the short modes through tidal forces, and this then backreacts on the motion of the long modes themselves. To see this effect, we find it convenient to revert to thinking of the dark matter as a collection of particles.
 
 Consider the universe as a collection of small cubes of linear size $\Lambda^{-1}$.
We are interested in the effects of long-wavelength modes on the dynamics inside each small cube.
The equations of motion for each particle `$\alpha$' in the cube are 
\beq
\label{equ:cubeeom}
\frac{d^2 \x_\alpha}{d\eta^2} + \H \frac{d \x_\alpha}{d \eta} = - \nabla \phi_s(\x_\alpha) - \nabla \phi_\ell(\x_\alpha)\, ,
\eeq
where we have split the Newtonian potential into short and long modes, $\phi_s$ and $\phi_\ell$, relative to the size of the cube.
We decompose the particle positions into the center of mass of the cube, $\hat \x$, and coordinates relative to the center of mass, $\delta \x_\alpha$, with
\beq
\label{equ:com}
\sum_\alpha \delta \x_\alpha = 0\, .
\eeq
For simplicity, we have assumed that all particles have the same mass, $m_\alpha =m$, but the method generalizes straightforwardly to a distribution of masses.
Summing Eqn.~(\ref{equ:cubeeom}) over all particles in the cube gives the equation of motion for the center of mass
\beq
\label{equ:comeom}
\frac{d^2 \hat \x}{d\eta^2} + \H \frac{d \hat \x}{d \eta} =- \nabla \phi_\ell(\hat \x)\, ,
\eeq
where we have used the fact that the sum over the short-wavelength modes averages to zero,
\beq
\label{equ:com2}
\sum_\alpha \nabla \phi_s(\x_\alpha) \approx 0\, .
\eeq
Eqn.~(\ref{equ:com2}) essentially represents the fact  the internal forces taken into account in $\phi_s$ do not accelerate the center of mass.
Subtracting Eqn.~(\ref{equ:comeom}) from Eqn.~(\ref{equ:cubeeom}) then gives an equation for the particle motion relative to the center of mass,
\beq
\label{equ:eomX}
\frac{d^2 \delta x_\alpha^i}{d \eta^2} + \H \frac{d \, \delta x_\alpha^i}{d \eta} = - \nabla_i \phi_s - \underline{\underline{\frac{\partial^2 \phi_\ell}{\partial x_i \partial x_j}}} \delta x^j_\alpha \ +\  \cdots\ ,
\eeq
where we have expanded the long-wavelength potential in a power series around $\hat \x$.
We see that
the effect of the long-wavelength mode $\phi_\ell$ on the internal dynamics of particles in the cube is captured by the tidal tensor, $\partial_i \partial_j \phi_{\ell}$.

Even in the absence of the long-wavelength potential the geodesic equation~(\ref{equ:eomX}) typically cannot be solved exactly. In the Zel'dovich approximation $\nabla_i \phi_s(\x_\alpha)$ is evaluated at some initial time $\eta_{0}$, when the particles have initial positions $\x^\alpha_{0}$ \cite{Peebles}.
The lowest-order approximation to the particle deflections then is
\beq
\delta x_\alpha^i(\eta) \approx - b(\eta) \nabla_i \phi_s(\x^\alpha_{0})\, , \qquad {\rm where} \quad b(\eta) \sim \H^{-2}\, .
\eeq
The tidal force term in Eqn.~(\ref{equ:eomX}) is then included perturbatively by considering the source term
\beq
\label{equ:source}
-b(\eta) \frac{\partial^2 \phi_\ell}{\partial x^i \partial x^j} \partial_j \phi_s\, .
\eeq
Eqn.~(\ref{equ:source}) reveals explicitly the UV-IR coupling between long-wavelength modes and short-wavelength modes.
In principle, the Newtonian potential can be a function of time, so the solution of Eqn.~(\ref{equ:eomX}) with the source term (\ref{equ:source}) can involve an integral over time.
However, during the matter era $\phi$ is constant and the solution of Eqn.~(\ref{equ:eomX}) takes the form
\beq
\delta x^i_\alpha(\eta) \approx - b(\eta) \partial_i \phi_s -g(\eta)  \frac{\partial^2 \phi_\ell}{\partial x^i \partial x^j} \partial_j \phi_s\, , \qquad {\rm where} \quad g(\eta) \sim \H^{-4}\, .
\eeq
 The associated three-velocities of the particles then are
 \beq
 v^i_\alpha \approx c_1 \frac{\partial_i \phi_s}{\H} + c_2 \frac{\partial_i \partial_j \phi_\ell \partial_j \phi_s}{\H^3}\, , \qquad {\rm where} \quad c_1, c_2 \sim {\cal O}(1)\, .
 \eeq
For purposes of illustration, we now consider the term  $\rho v_i v_j \subset \tau_{ij}$ and average over short-wavelength fluctuations,
 \bea
 \label{equ:vivj}
\langle  \rho v_i v_j \rangle &\sim& \delta_{ij} \langle \rho v^2_s \rangle + \frac{\partial_i \partial_j \phi_\ell}{\H^2} \langle \rho v^2_s \rangle\, , \\
&\sim&  \delta_{ij} \langle \rho v^2_s \rangle + \frac{v_{(i, j)}^\ell}{\H}  \langle \rho v^2_s \rangle\, ,
 \eea
 where we used
 \beq
 \langle \partial_i \phi_s \partial_j \phi_s \rangle \sim \H^2 \delta_{ij} \langle  v^2_s \rangle \qquad {\rm and} \qquad \partial_{i} \partial_j \phi_\ell \sim \H v_{(i,j)}^\ell \, .
 \eeq
 Here, we have gone back to the continuum description of the dark matter and dropped the particle indices $\alpha$.
 The first term in Eqn.~(\ref{equ:vivj}) describes isotropic pressure and arises even in the absence of the long-wavelength mode, while the second term gives anisotropic stress from the breaking of isotropy induced by the long-wavelength mode. It is easy to estimate that the next term from tidal forces scales as $\langle v_s^4 \rangle \partial^4\phi_\ell/\H^4$, in agreement with what we anticipated in Eqn.~(\ref{equ:exp1}). As expected, the tidal effects are suppressed on superhorizon scales, where the Newtonian potential $\phi_\ell$ is constant.
However, the above analysis shows that on subhorizon scales tidal forces induce an effective anisotropic pressure of the form
 \beq
 (\Sigma_{ij})_{\rm eff} \sim \frac{\langle \rho v_s^2 \rangle}{\H} \sigma_{ij}^\ell\, ,
 \eeq
 where $\sigma_{ij}$ is the shear tensor defined as
 \beq
 \label{equ:sigmaij}
 \sigma_{ij} \equiv v_{(i,j)} - \frac{1}{3} \delta_{ij} \, v_{k,k}\, .
 \eeq
 In addition, the trace of Eqn.~(\ref{equ:vivj}) leads to an isotropic pressure perturbation proportional to $\delta_\ell$ and to a bulk viscosity proportional to $\theta_\ell$.
 Our effective theory, by construction, describes all these effects, as well as higher-derivative corrections.
 
It should be emphasized that
the tidal effects we described in this section are automatically captured in the results of $N$-body simulations.
Those numerical works solve the particle geodesic equations exactly (and in the presence of long-wavelength perturbations) and hence include tidal effects and anisotropic pressure.
It should also be emphasized that this is not a general relativistic effect, but is completely captured by the Newtonian dynamics if the Newtonian equations are solved exactly.
Nothing is therefore lost when the parameters of the effective theory are matched to the results of $N$-body simulations.
 
\subsection{UV Matching}
\label{sec:Viscous}

We now describe the UV matching of the coefficients of the effective theory to either numerical simulations or directly to observations.
For purposes of illustration, let us return to Eqn.~(\ref{equ:exp}), the expansion of the source term in the Euler equation to leading order in
$\delta_\ell$ and $\Theta_\ell \equiv \frac{\theta_\ell}{\H}$ (the dimensionless velocity dispersion),
\beq
\label{equ:A}
{\cal A} \equiv \frac{k_i k_j }{k^2}  \frac{\langle \tau_{ij} \rangle}{\bar \rho}  = c_s^2 \delta_\ell - c_{\rm vis}^2 \Theta_\ell\, .
\eeq
To extract the speeds of sound from Eqn.~(\ref{equ:A}) we correlate ${\cal A}$ with long-wavelength perturbations $\delta_\ell$ and $\Theta_\ell$,
\begin{align}
A_\delta \equiv \langle \delta_\ell {\cal A} \rangle &= c_s^2 P_{\delta \delta} - c_{\rm vis}^2 P_{\delta \theta} \, ,\\
A_\theta \equiv \langle \Theta_\ell {\cal A} \rangle &= c_s^2 P_{\delta \theta} - c_{\rm vis}^2 P_{\theta \theta}\, . 
\end{align}
In linear theory, $\delta^{(1)} = - \Theta^{(1)}$ and $P_{\delta \delta} = P_{\theta \theta} = |P_{\delta \theta}|$, implying that this  approach would only allow us to obtain a linear combination of the sound speeds, $c_s^2 + c_{\rm vis}^2$. However, non-linear corrections will break this degeneracy, $P_{\delta \delta} \ne P_{\theta \theta} \ne |P_{\delta \theta}|$, and allow us to measure (or calculate) $c_s$ and $c_{\rm vis}$ separately,
\begin{align}
\label{equ:csM}
c_{s}^2 &= \frac{P_{\theta \theta} \, A_\delta - P_{\delta \theta} \, A_\theta}{P_{\delta \delta} P_{\theta \theta} - P_{\delta \theta}^2}\, , \\
c_{\rm vis}^2 &= \frac{P_{\delta \theta} \, A_\delta - P_{\delta \delta} \, A_\theta}{P_{\delta \delta} P_{\theta \theta} - P_{\delta \theta}^2}\, .
\label{equ:cvisM}
\end{align}

The parameter $c_{\rm vis}$ may be related to the standard phenomenological parameterization 
of imperfect fluids whose stress tensor may be written as
\beq
\tau_{ij} = \rho u_i u_j + (p-\zeta\theta) \gamma_{ij} + \Sigma_{ij}\, ,
\eeq

The bulk viscosity $\zeta$ and $\Sigma_{ij}$ account for viscous terms.
At leading order in an expansion in spatial derivatives, an ansatz for $\Sigma_{ij}$ that is consistent with the second law of thermodynamics is~\cite{Weinberg}
\beq
\label{equ:shearM}
\Sigma_{ij} = -  \eta \sigma_{ij} \, .
\eeq
Here $\eta > 0$ and $\zeta > 0$ are the coefficients of shear and bulk viscosity, respectively and $\sigma_{ij}$ was defined in Eqn.~(\ref{equ:sigmaij}). With Eqn.~(\ref{equ:shearM}) the Euler equation becomes the non-relativistic Navier-Stokes equation.
From Eqn.~(\ref{equ:shearM}) we see that the viscous speed of sound is related to a linear combination of the shear viscosity $\eta$ and the bulk viscosity $\zeta$,
\beq
\label{equ:speed}
c_{\rm vis}^2 \equiv \Bigl(\frac{2}{3}\eta +\zeta\Bigr) \frac{\H}{\bar \rho}\, .
\eeq

To extract the shear viscosity of the fluid (and hence via Eqn.~(\ref{equ:speed}) also the bulk viscosity) we consider
\beq
{\cal B} \equiv  \frac{(k_i k_j - \frac{1}{3} \delta_{ij}k^2)}{k^2}  \frac{\langle \tau_{ij} \rangle}{\bar \rho} = - \frac{2}{3} \eta \frac{\H}{\bar \rho} \, \Theta_\ell\, .
\eeq
Correlating ${\cal B}$ with $\Theta_\ell$ gives the shear viscosity $\eta$,
\beq
B_\theta \equiv \langle \Theta_\ell {\cal B}\rangle = - \frac{2}{3} \eta \frac{\H}{\bar \rho}\, P_{\theta \theta}\, .
\eeq
We remark in passing that viscosity is a source term in the generation of vorticity.
We see this from the linearized equation of motion for  ${\boldsymbol{w}}_\ell = \nabla \times {\v}_\ell$, which, at leading order in derivatives, is
\bea\label{equ:vorticity}
\dot {w}^l_\ell + \H {w}^l_\ell &=& \frac{1}{2} \frac{\eta}{\rho_\ell} \nabla^2 {w}^l_\ell  \\ \nonumber
&&-\ \ \frac{1}{\rho_\ell}\epsilon^{lik}\left(\eta_{,k}\sigma^\ell_{ij,j}+\eta_{,j}\sigma^\ell_{ij,k}+\eta_{,jk}\sigma^\ell_{ij}-\frac{\rho_{\ell,k}}{\rho_\ell}\left(\eta \sigma^\ell_{ij}\right)_{,j}\right) +\epsilon^{lik}\frac{\rho_{\ell,k}}{\rho_\ell^2}\left(\zeta \theta_\ell\right)_{,i} \\ \nonumber 
&&+\ \ {\cal O}(\delta_\ell^2,\v_\ell^2,\ldots)\, .
\eea
The terms in the second line represent sources for vorticity. Notice that they vanish if the fluid is incompressible.

Finally, we may also  consider correlations with the trace of $\tau_{ij}$,
\beq
{\cal C} \equiv \frac{ \langle \tau^i_{\ i} \rangle}{3 \bar \rho} =  c_s^2 \delta_\ell - \zeta \frac{\H}{\bar \rho} \Theta_\ell\, ,
\eeq
which contains information on isotropic pressure and bulk viscosity.

\vskip 8pt
\noindent
{\sl One-loop perturbation theory.}  \hskip 8pt
As we observed above, a linear (or tree-level) calculation in general does {\it not} allow an extraction of all fluid parameters, unless it is {\it assumed} that the contribution from the bulk viscosity is parametrically smaller (see Appendix~\ref{sec:estimates})
\beq
\label{equ:zeta}
\zeta \frac{\H}{\bar \rho}\ \ll \ c_s^2 \, ,\, c_{\rm vis}^2 \, .
\eeq
If we don't want to make the assumption (\ref{equ:zeta}) we have to go beyond linear order in perturbation theory to compute $c_s$ and $c_{\rm vis}$ separately via Eqns.~(\ref{equ:csM}) and (\ref{equ:cvisM}),\footnote{One may ask how non-linear perturbation theory knows about dissipative effects. Even if the initial conditions are purely in the growing mode, non-linear interactions create both growing and decaying modes~\cite{Bernardeau:2001qr}. The decaying modes running in loops of the perturbation expansion carry the information on dissipative effects.}
where
\bea
P_{\delta \delta} &=& P_{\rm L}(k) + P^{(2,2)}_{\delta \delta} + P^{(1,3)}_{\delta \delta}\, , \\
P_{\theta \theta} &=& P_{\rm L}(k) + P^{(2,2)}_{\theta \theta} + P^{(1,3)}_{\theta \theta}\, , \\
P_{\delta \theta} &=& P_{\rm L}(k) + P^{(2,2)}_{\delta \theta} + P^{(1,3)}_{\delta \theta}\, . 
\eea
Here, $P_{\rm L}(k)$ is the linear power spectrum and $P^{(i,j)}$ are higher-order (one-loop) corrections.
Explicit expressions for the one-loop corrections $P^{(i,j)}$ may be found {\it e.g.}~in Appendix A of Ref.~\cite{Carlson:2009it}. From those expressions one may infer that the velocity fields are more sensitive to tidal fields than the density fields \cite{Pueblas:2008uv}. In fact, the non-linear growth of $\theta$ is {\it smaller} than the linear prediction, {\it i.e.}~on the relevant scales $P_{\theta \theta}(k)$ is suppressed relative to the linear $P(k)$.
The one-loop calculation is feasible, but presenting it here would not be very illuminating.
Furthermore, as Ref.~\cite{Carlson:2009it} recently demonstrated, on small scales one-loop perturbation theory becomes increasingly unreliable in capturing even the correct qualitative features of the velocity-density cross-correlation.
Instead, we now describe how the parameters of the effective fluid should be determined from numerical simulations or directly from observations.

\vskip 8pt
\noindent
{\sl $N$-body simulations.}  \hskip 8pt
An $N$-body simulation of moderate box size allows one to measure the statistics of small-scale fluctuations and construct the stress tensor $\tau_{ij}$. Furthermore, at the {\it matching scale} $\Lambda$ the following correlations can be measured: $P_{\delta \delta}$, $P_{\delta \theta}$, $P_{\theta \theta}$. Measuring the correlations of $k_i k_j\tau_{ij}$ with $\delta_\ell$ and $\theta_\ell$ then provides a measurement of $c_s$ and $c_{\rm vis}$ via Eqns.~(\ref{equ:csM}) and (\ref{equ:cvisM}). 
The basic idea is that a relatively small simulation suffices to normalize the fluid parameters and this fluid is then used in a perturbative treatment of long-wavelength fluctuations.

\vskip 8pt
\noindent
{\sl Fitting to observations.}  \hskip 8pt
Alternatively, the fluid parameters may simply be retained as free parameters of the effective theory. Computations are then performed with these arbitrary coefficients. Finally, the parameters are determined by comparing the results for certain observables to the corresponding measurements of the matter fluctuations.
Notice that this may allow one to take into account effects not normally included in simulations (such as the effects arising from the presence of baryons), by fitting observations to a general one-component (or multi-component) effective fluid for the {\it total} matter density with free parameters as in Eqn.~(\ref{equ:exp1}).\footnote{This proposal is limited to sufficiently large scales for which relativistic effects, such as those due to the epoch of reionization, are small.} 

\subsection{Corrections from Statistical Fluctuations}
\label{sec:StochasticV1}

We have used ensemble averaging and the ergodic theorem to estimate average quantities in domains of size $\Lambda^{-1}$.
For any specific realization of the universe there will be a random statistical error  in that estimate.
In Appendix~\ref{sec:estimates} we estimate the size of that effect in perturbation theory.
Here, we discuss the basic conclusions of that analysis.

Consider for instance the effective pressure
\beq
p_{\rm eff} = \bar p_{\rm eff} + \delta p_{\rm eff}^{\ell} + \delta p_{\rm eff}^{\rm stat}\, .
\eeq
Here, $\bar p_{\rm eff} \sim c_s^2 \bar \rho$ denotes the renormalized  background pressure (\S\ref{sec:super}), $\delta p_{\rm eff}^\ell\sim c_s^2\bar\rho\delta_\ell$ is the perturbation induced by long-wavelength modes $\delta_\ell$ (\S\ref{sec:tidal}), and $\delta p_{\rm eff}^{\rm stat}$ is the perturbation induced by random statistical fluctuations, which enters in the equations for the long-wavelength mode as an external source.  The statistical properties of this random source could be computed in perturbation theory (when this is a good approximation), or measured from simulations.  In \cite{Senatore:2009cf} the treatment and inclusion of these kind of terms has been developed in a different, but related context.
In  Appendix~\ref{sec:estimates}  we establish the following two facts:
\begin{enumerate}
\item Both   $\delta p_{\rm eff}^\ell$ and $\delta p_{\rm eff}^{\rm stat}$ are much smaller than $\bar p_{\rm eff}$ if the momentum integral defining $p_{\rm eff}$ is UV-dominated ({\it i.e.}~if the integral is dominated by scales with $q_\star \gg \Lambda$, where, as before, $q_\star$ is the typical scale of non-linearities: $q_\star \sim k_{\rm NL}$ in Einstein-de Sitter or $q_\star \sim k_{\rm eq}$ in our universe). 
Quantitatively, the stochastic fluctuations and the background pressure are related as follows
\beq
\delta p_{\rm eff}^{\rm stat} = \alpha\,  \bar p_{\rm eff}\, ,
\eeq
where $\alpha$ is a random variable with variance
\beq
\Delta_\alpha^2 \equiv  \frac{\Lambda^3}{q_\star^3}\, .
\eeq
For $\Lambda < q_\star$ stochastic fluctuations are therefore suppressed. This result may be understood as the standard $\frac{1}{\sqrt{N}}$ suppression of random fluctuations, where $N$ is the number of space domains sampled. 

\item At the scale dominating gravitational non-linearities, $q_\star$, the two contributions to pressure fluctuations, $\delta p_{\rm eff}^\ell$ and $\delta p_{\rm eff}^{\rm stat}$, are of the same size.
On larger scales,  $\delta p_{\rm eff}^{\rm stat}$, in fact, dominates over  $\delta p_{\rm eff}^\ell$---{\it i.e.}~$\delta p_{\rm eff}^{\rm stat} \gtrsim \delta p_{\rm eff}^\ell$---with equality holding at $q_\star$.
 However, at least at leading order (see \cite{Senatore:2009cf}), the fluctuations $\delta p_{\rm eff}^{\rm stat}$ are uncorrelated with the long-wavelength density fluctuations $\delta_\ell$, so their effect on the power spectrum of density fluctuations is subdominant for scales with $k \ll q_\star$. 
On all scales, $\Delta P_\delta^\ell > \Delta P_\delta^{\rm stat}$. Nevertheless, for specific applications of the effective theory (see \S\ref{sec:Application}) the corrections from stochastic noise can be significant and will have to be included.
\end{enumerate}
We emphasize that stochastic fluctuations in our fluid are much larger than the corresponding effect in conventional fluids. In practice, the correlations of stochastic noise contributions therefore have to be measured in the simulations or must be included phenomenologically as external sources.

\subsection{Summary}
\label{sec:summary}

Let us briefly summarize our conclusions so far:

We have argued that the result of integrating out small-scale non-linear fluctuations can be represented at long wavelengths by an {\it effective fluid}.
Studying the Boltzmann hierarchy for dark matter particles, we have shown that the fluid is somewhat unconventional. The hierarchy is truncated at the second moment not because there are sizable interactions, but  because the universe has finite age and particles only move a small distance (of order the non-linear scale) over a Hubble time. Higher moments therefore haven't had enough time to develop rather than being suppressed by interactions. 
Also unlike a conventional fluid, our effective fluid can have large stochastic fluctuations. 

In \S\ref{sec:PerfectFluid} we studied the renormalization of the background density and pressure induced by kinetic and potential contributions from small-scale structures. We proved that the contribution to the effective pressure is always positive and vanishes for viralized structures.

\begin{figure}[h!]
    \centering
        \includegraphics[width=0.55\textwidth]{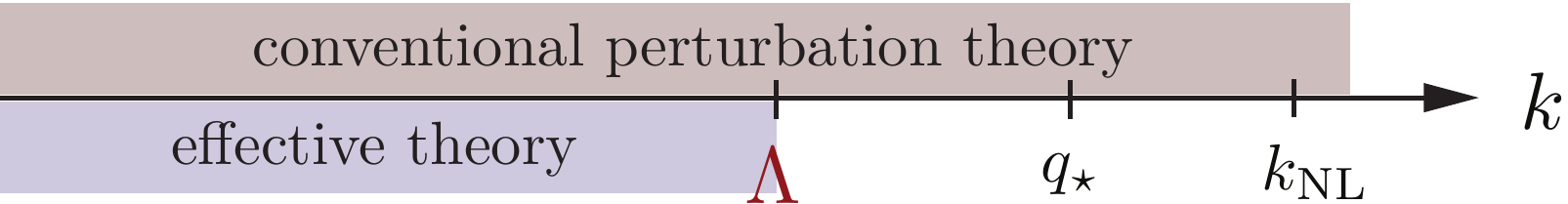}
    \caption{\sl Hierarchy of scales and perturbative expansions. The effective theory  only contains modes with $k < \Lambda$, while conventional perturbation theory contains modes with $k \sim k_{\rm NL}$ where the perturbative expansion is known to break down.}
    \label{fig:scales}
\end{figure}

In \S\ref{sec:ImperfectFluid} we then studied the dynamics of long-wavelength fluctuations of the effective fluid.
We have shown that dissipative effects, such as an effective viscosity, are crucial to understanding the evolution of long-wavelength modes.
We furthermore introduced an {\it effective theory for the perturbed fluid}. Its key elements are:

{\sl Hierarchy of scales.} \hskip 4pt
In the simplest application of the effective theory we have in mind the following hierarchy of scales (see Fig.~\ref{fig:scales})
\beq
\label{equ:hierarchy}
q_\star \ > \ \Lambda \ > \ k \ ,
\eeq
{\it i.e.}~the characteristic scale of non-linearities $q_\star$ is much smaller that the smoothing scale $\Lambda$, which in turn is smaller than the scales $k$ to which we apply the theory.
It is the large hierarchy between $q_\star$ and $k$ that allows for a perturbative treatment and a fluid description.
The hierarchy between $k$ and $\Lambda$ ensures that higher-derivative corrections (see Appendix~\ref{sec:Euler}) are negligible. The hierarchy between $q_\star$ and $\Lambda$ suppresses the effect of stochastic fluctuations (see Appendix~\ref{sec:estimates}).

\vskip 6pt
{\sl UV matching.} \hskip 4pt
Above the momentum scale $\Lambda$, and especially near the scale of non-linearities $q_\star$, the perturbative treatment of fluctuations breaks down. We therefore propose to numerically compute the effect of these non-linear scales. The cummulative effect of these small scales defines the parameters of the effective fluid ($w_{\rm eff}$, $c_s$, $c_{\rm vis}$, etc.)  at the quasi-linear scale $\Lambda$: {\it e.g.}~the parameters $c_s$ and $c_{\rm vis}$ appear as time-dependent expansion parameters in the source term for the Euler equation
\beq
\label{equ:exp2}
\frac{k_i k_j}{k^2}\frac{\langle \tau_{ij} \rangle}{\bar \rho} \ =\ c_s^2\, \delta_\ell - c_{\rm vis}^2\, \frac{\theta_\ell}{\H}\, .
\eeq
The more precise version of the effective theory given in Eqn.~(\ref{equ:exp1}) contains further fluid parameters.
In \S\ref{sec:Viscous} we described how these parameters can be extracted in $N$-body simulations.
Alternatively, the fluid parameters may simply be retained as free parameters to be measured by fitting perturbation theory calculations in the effective theory to the data.

\vskip 6pt
\noindent
{\sl Controlled expansion.} \hskip 4pt
After this matching calculation, the theory below the momentum scale $\Lambda$ has well-defined small perturbation variables
$\delta_\ell(k \ll \Lambda)$ and $\theta_\ell(k \ll \Lambda)$.
On scales with $k \ll \Lambda$ we therefore define the theory as an expansion in $\delta_\ell$ and $\theta_\ell$.
The perturbation theory in $\delta_\ell$ and $\theta_\ell$, evolving in a background characterized by $w_{\rm eff}$, $c_s$ and $c_{\rm vis}$, is well-defined and under control.  This is in contrast to other perturbative schemes, where the definition of small expansion parameters isn't as clear (see Ref.~\cite{Carlson:2009it} for a discussion).

\vskip 6pt
\noindent
{\sl RG flow.} \hskip 4pt
It is expected that the parameters of the effective theory like the speeds of sound $c_s(\Lambda)$ and $c_{\rm vis}(\Lambda)$ 
depend on the cut-off scale $\Lambda$ at which the matching to numerical simulations is performed.
In effective field theories this dependence of parameters on the matching scale is described by the renormalization group (RG) equations \cite{Goldberger:2007hy, Burgess:2007pt, Pich:1998xt}.
In our case, RG flow type equations 
will describe how the fluid quantities change as the cut-off is changed from $\Lambda$ to $\Lambda'$.
The corresponding beta-functions 
can be computed perturbatively in the effective theory.

\vskip 6pt
\noindent
{\sl Loops and higher-derivatives.} \hskip 4pt In principle, perturbation theory in our effective theory allows for arbitrary precision by going to higher and higher order in $\delta_\ell$ and $\theta_\ell$. This is different from the way in which most conventional cosmological perturbation theories are organized.
In practice, at least the following two complications have to be considered:
\begin{enumerate}
\item[i)] Going to higher order in perturbation theory requires also that the matching calculation in Eqn.~(\ref{equ:exp2}) is performed to higher order in $\delta_\ell$ and $\theta_\ell$. This is not a problem in principle, but it complicates the effective theory by adding extra fluid parameters---see Eqn.~(\ref{equ:exp1}).
\item[ii)] At some order, the higher-derivative corrections from the smoothing may not be negligible anymore and should hence be included in the effective $\tau_{\mu \nu}$---{\it cf.}~the second line in Eqn.~(\ref{equ:exp1}).
At what point exactly these higher-derivative corrections become important will depend on the specific application and the hierarchy of scales it involves.
\end{enumerate}

\section{Applications of the Effective Theory}
  \label{sec:Application}

In this section we suggest possible applications of the general formalism we introduced in this paper. Our discussion will be rather schematic, leaving further calculations and numerical investigations to future work~\cite{Future}.

\subsection{An Alternative to Conventional Perturbation Theory}

In typical applications of effective theories one is interested in calculating the effects of some short-distance (high-energy) physics, characterized by a scale $q_\star$, on the dynamics at a long-distance (low-energy) scale $k \ll q_\star$.
If the full theory is known, then these effects can in principle be calculated explicitly. If the full theory is not known (or the full theory is known, but not computable, {\it e.g.}~strongly coupled) then high-energy effects are parameterized as $k/q_\star$ corrections. Allowing all corrections consistent with a given set of symmetries accounts for the UV physics in a completely model-independent way. The parameters of the effective theory are then constrained either from numerical simulations or measured by experiments.

In this work we have applied the effective field theory philosophy to cosmological perturbations. 
The full theory (General Relativity coupled to matter) is known, so in principle all effects of short-scale non-linearities can be computed explicitly.
The parameters of our effective theory are determined either from $N$-body simulations, or matched to observations.
After the UV matching, the effective theory at long-wavelengths is that of an imperfect fluid characterized by only a few physical parameters like the equation of state $w_{\rm eff}$, the speed of sound $c_s$, and the viscosity $c_{\rm vis}$.
On large scales the effective fluid has small perturbation variables ({\it e.g.}~$\delta_\ell < 1$), allowing for a controlled perturbative expansion at low momenta. Conceptually, our approach therefore offers a well-defined and controlled treatment of the effects short-distance non-linearities on the long-wavelength universe.
This is to be contrasted with the breakdown of most conventional cosmological perturbations theory techniques at small scales (see Carlson {\it et al.}~\cite{Carlson:2009it}).
Often the conventional approaches are formulated as an expansion in $\delta$, with loop integrals running over modes with $\delta \gg 1$. Furthermore, the equations that are solved are the equations of a pressureless fluid (but see \cite{McDonald:2009hs}).  Since both the equations and the perturbative expansion is known to break down at the non-linear scale, errors are expected. However, the size of those errors is often hard to estimate {\it a priori}.
In contrast, at low momenta our effective theory has well-defined small expansion parameters, which solve equations that are corrected by the properties of the effective fluid.
Furthermore, like in all effective theories, the errors can be estimated and are under control.

\vskip 6pt
Our examples for applications of the effective theory will work their way from inside the horizon to superhorizon scales:
In \S\ref{sec:BAO} and \ref{sec:preheating} we describe how the fluctuations measured in baryon acoustic oscillations (BAO) and the perturbations arising in models of preheating are affected by the viscosity of the effective fluid.
Then, in \S\ref{sec:evolution} and \S\ref{sec:backreaction}, we explain how the superhorizon evolution of curvature perturbations and the backreaction of non-linearities on the background FRW evolution are understood in the effective approach.

\subsection{Baryon Acoustic Oscillations}
\label{sec:BAO}

\noindent
{\sl BAOs as a probe of dark energy.} \hskip 8pt
Oscillations in the baryon-photon fluid, frozen into the matter distribution at decoupling, provide a standard ruler to constrain the expansion history of the universe (for a review see \cite{Bassett:2009mm}).
These sound waves imprint an almost harmonic series of peaks in the matter power spectrum $P_\delta(k)$, corresponding to a localized feature in the two-point correlation function $\xi_\delta(r)$ at about 100 $h^{-1}$Mpc \cite{Eisenstein:1997ik}.
The position and shape of the BAO peak are powerful probes of dark energy.
While the evolution of density perturbations in the early universe is linear and therefore easily calculated, at late times the non-linear evolution of matter makes the interpretation of the BAO peak more challenging.
The non-linear clustering erases the oscillations on small scales, which shifts and broadens the peak~\cite{Eisenstein:2006nj,Crocce:2007dt}.
An accurate modeling of the effects of non-linearities on the BAO feature is essential in light of the high precision aimed for in future BAO surveys; {\it e.g.}~a shift in the BAO scale of one percent generates systematics
in the deduced dark energy equation of state
parameter $w_{de}$ of about five percent \cite{Angulo:2007fw}, which is not
negligible compared to the expected statistical errors in
the next generation of galaxy surveys.

\vskip 6pt
\noindent
{\sl Damping by viscosity.} \hskip 8pt 
It is suggestive to explain the non-linear broadening of the BAO peak by the viscous damping of density fluctuations in an imperfect fluid (see \S\ref{sec:ImperfectFluid}).
In Fourier space, the broadening of the BAO peak is described by the following phenomenological ansatz~\cite{Eisenstein:2006nj,Crocce:2007dt}
\beq
\label{equ:Pbao}
P_{\rm obs}(k) = e^{-\frac{1}{2}k^2 \Sigma^2} P_{\rm L}(k) + P_{\rm mc}(k)\, ,
\eeq
where $P_{\rm L}(k)$ is the linear power spectrum and $P_{\rm mc}(k)$ is a mode-coupling term.\footnote{The mode-coupling term $P_{\rm mc}(k)$ relates to the stochastic small-scale noise of \S\ref{sec:StochasticV1}.}
The exponential damping of the linear power spectrum in (\ref{equ:Pbao}) is precisely the effect we expect from the viscosity of the effective fluid, with the parameter $\Sigma^2$ being related to a time integral of a viscosity parameter. For instance for a viscous fluid in the absence of gravity with bulk viscosity $\zeta$, sound waves get damped as $\sim e^{-\zeta k^2 t }$.
Related observations have been made in perturbation theory in Ref.~\cite{Crocce:2007dt}, where the exponential factor in (\ref{equ:Pbao}) arises from a renormalization of the propagator for the evolution of density fluctuations. Thanks to the effective theory approach, we conclude that measuring the broadening of the peak corresponds to measuring the viscosity of the effective long-distance fluid. We find this to be quite a powerful statement.

\vskip 6pt
\noindent
{\sl Real-world complications.} \hskip 8pt
Although we defer a detailed analysis to future work~\cite{Future}, we here point out a few of the expected challenges in this specific application of the effective theory. 
First, we notice that the BAO scale $k_{\rm BAO}$ lies between the non-linear scale $k_{\rm NL}$ and the scale of matter-radiation equality $k_{\rm eq}$,
\beq
k_{\rm eq} \ < \ k_{\rm BAO}\ < \ k_{\rm NL}\, .
\eeq
Since the dominant contribution to the gravitational fluid comes from scales with $q_\star \sim k_{\rm eq}$, the hierarchy of scales described in \S\ref{sec:summary} -- see Eqn.~(\ref{equ:hierarchy}) -- cannot be exploited in the BAO application.
Specifically, to study fluctuations at the BAO scale one has to choose the matching scale in between the BAO scale and the non-linear scale: $k_{\rm BAO} < \Lambda < k_{\rm NL}$. This implies that the integrals defining the fluid parameters at the scale $\Lambda$ are IR-dominated ({\it i.e.}~dominated by modes with $k \sim \Lambda$), and therefore the stochastic pressure discussed in Appendix~\ref{sec:estimates} is large and has to be included. Furthermore, the non-linear scale and the BAO scale are not very different, and the expansion parameter in the long-wavelength effective theory is not very small (although it is still smaller than one, making the expansion well-defined). These complications will make the application of the effective theory to the BAO example a bit non-trivial. 

\subsection{Preheating}
\label{sec:preheating}

For dark matter perturbations in the late universe, the large hierarchy between the scale of non-linearities and the horizon scale makes the effects we discussed in this paper small.
It would therefore be interesting to study effects in the early universe where this hierarchy is larger and the UV-IR coupling is potentially more significant.
It is well known that at the end of inflation a dynamical instability called preheating can convert the energy of the homogeneous inflaton field into the excitations of other fields \cite{Mukhanov:2005sc}.
This process is highly inhomogeneous and non-linear and therefore mostly studied numerically in lattice simulations  
\cite{Felder:2000hq}--\cite{GarciaBellido:2007af}.
The scale of non-linearities during preheating is found to be as large as $0.01 \H^{-1}$ \cite{Frolov:2008hy}, compared to $10^{-4} \H^{-1}$, the scale of dark matter non-linearities today.
An effective description of smoothed perturbation fields during preheating may therefore be of interest.

\subsection{Non-Linear Superhorizon Evolution of $\zeta$}
\label{sec:evolution}

Inflation \cite{Guth:1980zm, Linde:1981mu, Albrecht:1982wi} famously produces a nearly scale-invariant spectrum of scalar perturbations \cite{Mukhanov:1990me}.
These perturbations are commonly described by the perturbation to the spatial three-curvature ${\zeta}$, where $g_{ij} = a^2(\eta) \exp[{\zeta}(\eta, \x)] \delta_{ij}$. To relate observations of the cosmic microwave background and the large-scale structure to the primordial value of ${\zeta}$, any evolution of ${\zeta}$ after it is created during inflation has to be taken into account \cite{Baumann:2009ds, Baumann:2008aq}.
One can prove \cite{Cheung:2007sv}, that ${\zeta}$ remains constant outside of the horizon if:
i) all modes that contribute to the non-linear definition of ${\zeta}$ are outside the horizon, and ii) the pressure is a unique function of the energy density $p(\rho)$ (adiabatic condition).
Each Hubble patch then evolves like a separate FRW universe and ${\zeta}$ doesn't evolve in time.
However, short-wavelength fluctuations with wavelengths much smaller than the Hubble scale evade this theorem and can in principle allow a non-linear evolution of ${\zeta}$.
This was recently emphasized in Ref.~\cite{Boubekeur:2008kn}.
In this section we show that this non-linear superhorizon evolution of curvature perturbations is completely consistent with the `separate universe' picture.
We explain that short-wavelength fluctuations induce a non-adiabatic pressure in the effective fluid describing the long-wavelength dynamics.

\vskip 6pt
\noindent
{\sl Superhorizon evolution in comoving gauge.} \hskip 8pt
Ref.~\cite{Boubekeur:2008kn}
derived the 
second-order curvature perturbation in comoving gauge ($T^i_{\ 0}= 0$) as a function of its initial condition ${\zeta}_{0}$,
\beq
\label{equ:NLR}
{\zeta} = {\zeta}_0 - \frac{1}{5} \frac{1}{\H^2} \partial^{-2} \partial_i \partial_j (\partial_i {\zeta}_0 \partial_j {\zeta}_0)\, .
\eeq
This implies a time evolution of ${\zeta}$ of the form
\beq
\dot {\zeta} = - \frac{4}{5} \frac{1}{\H}  \partial^{-2} \partial_i \partial_j (\partial_i {\zeta}_0 \partial_j {\zeta}_0)\, ,
\eeq
where the source term on the r.h.s.~can be written in Fourier space as
\beq
\label{equ:PaoloSource}
{\cal S} \equiv \int_\q \frac{(\k \cdot \q) (k^2 - \k \cdot \q)}{k^2} \, {\zeta}_0(\q) {\zeta}_0(\k - \q)\, .
\eeq
Crucially, Eqn.~(\ref{equ:PaoloSource}) does {\it not} vanish in the $k \to 0$ limit,
\beq
\lim_{k \to 0} {\cal S} = - \int_0^{q_{\rm max}} dq \, d \Omega_{\q} \, (\hat \k \cdot \hat \q)^2\,  \frac{\Delta_{\zeta}^2(q)}{2\pi^2 q}\, ,
\eeq
and is dominated by the UV cutoff at large $q$. Notice that in principle we could imagine a universe in which density fluctuations are never large, the comoving gauge is well-defined on all scales, and the integral above can be taken to $q_{\rm max}=+\infty$. There is then no way out: ${\zeta}$ seems to be evolving. We will now argue that the apparent evolution of ${\zeta}$ is due to the renormalization of the background density and pressure induced by the short-scale non-linearities. Once the background has been redefined to take this into account, the curvature perturbation ${\zeta}$ defined with respect to this background is indeed constant on superhorizon scales.

\vskip 6pt
\noindent
{\sl Superhorizon evolution in Poisson gauge.} \hskip 8pt
We first reformulate the problem in the context of our effective theory in Poisson gauge.
In this way we will see that the apparent superhorizon evolution of ${\zeta}$ has a simple physical interpretation.

Smoothing the $i$--$i$ Einstein equation  (\ref{equ:NLT_0}) we find
\beq
\label{equ:dotRell}
\dot {\zeta}_\ell = \frac{\H}{\bar \rho} \left( \frac{1}{3} \left[ \tau^k_{\ k} \right]_{\Lambda} + \left[\nabla^{-2} \partial_i \partial_j\left( \tau^{ij} - \frac{1}{3} \delta^{ij} \tau^k_{\ k}\right) \right]_\Lambda\right)\, ,
\eeq
where
\beq
\label{equ:Rell}
{\zeta}_\ell \equiv \Phi_\ell + \frac{\H}{\bar \rho + \bar p} \delta q_\ell\, , \qquad {\rm with} \qquad T^{i}_{\ 0} \equiv \partial_i \delta q\, .
\eeq
The quantity ${\zeta}_\ell$ is the standard {\it linear} definition of the {\it comoving curvature perturbation}\footnote{In fact, the comoving curvature perturbation is often denoted ${\cal R}$, reserving ${\zeta}$ for the curvature perturbation on uniform density hypersurfaces. On superhorizon scales ${\cal R}$ and $\zeta$ are equal (up to a sign).}~\cite{Baumann:2009ds}.
Using the $0$--$i$ Einstein equation to replace $\delta q_\ell$, it becomes
\beq
{\zeta}_\ell \ \equiv \  \Phi_\ell - \frac{\H (\dot \Phi_\ell + \H \Psi_\ell)}{\dot \H - \H^2 } \ \ \ \stackrel{\rm MD}{\longrightarrow} \ \ \ \Phi_\ell + \frac{2}{3} \frac{\dot \Phi_\ell + \H \Psi_\ell}{\H}\, ,
\eeq
where the limit applies during matter-domination (MD).
On the r.h.s.~of Eqn.~(\ref{equ:dotRell}) we identify the pressure and scalar anisotropic stress of the effective fluid,
\beq
p_{\rm eff} \equiv \frac{1}{3} \left[ \tau^k_{\ k} \right]_{\Lambda} \qquad {\rm and} \qquad \bar \rho \sigma_{\rm eff} \equiv \left[\nabla^{-2} \partial_i \partial_j \left( \tau^{ij} - \frac{1}{3} \delta^{ij} \tau^k_{\ k}\right)\right]_\Lambda\, ,
\eeq
such that
\beq
\dot {\zeta}_\ell = \frac{\H}{\bar \rho} \left( p_{\rm eff} + \bar \rho \sigma_{\rm eff} \right)\, .
\eeq
This is the standard evolution equation for the long-wavelength curvature perturbation ${\zeta}_\ell$ for a fluid with non-adiabatic pressure $p_{\rm eff}$ and anisotropic stress $\sigma_{\rm eff}$~\cite{Baumann:2009ds}.
Both the pressure and the anisotropic stress can be split into a zero mode (\S\ref{sec:PerfectFluid}), perturbations induced by long-wavelength perturbations (\S\ref{sec:ImperfectFluid}) and stochastic terms (\S\ref{sec:ImperfectFluid} and \S\ref{sec:Stochastic}),
\bea
p_{\rm eff} &=& \bar p_{\rm eff} + \delta p_{\rm eff}^\ell + \delta p_{\rm eff}^{\rm stat}\, , \label{equ:isoE}\\
\sigma_{\rm eff} &=& \bar \sigma_{\rm eff} + \delta \sigma_{\rm eff}^\ell + \delta \sigma_{\rm eff}^{\rm stat}\, . \label{equ:anisoE}
\eea
All terms in Eqn.~(\ref{equ:anisoE}) average to zero on superhorizon scales.
Similarly, $\delta p_{\rm eff}^\ell$ and $ \delta p_{\rm eff}^{\rm stat}$ in Eqn.~(\ref{equ:isoE}) go to zero on superhorizon scales. However, the non-adiabatic zero mode pressure $\bar p_{\rm eff}$ remains and sources a superhorizon time evolution of the curvature perturbation,
\beq
\lim_{k \ll \H} \dot {\zeta}_\ell \ \approx \ \frac{\H}{\bar \rho} \bar p_{\rm eff}\, . \label{equ:evo}
\eeq
 However, Eqn.~(\ref{equ:evo}) simply reflects the fact that short-wavelength non-linearities renormalize the background and therefore slightly change the FRW history of separated Hubble patches. The curvature perturbation ${\zeta}$ defined with respect to this new background would indeed be constant on superhorizon scales.
The superhorizon evolution of ${\zeta}$ found in \cite{Boubekeur:2008kn} has therefore been demystified.
 
\vskip 6pt
\noindent
{\sl Comments on the effective spacetime.} \hskip 8pt
As we have just seen, {\it not all gauges are created equal}.  To illustrate the differences in gauges for the purposes of defining an effective theory on large scales, let us consider the long-wavelength limit of the effective spacetime.
We split the covariant components of the metric into long-wavelength fields, $g_{\mu \nu}^\ell$, and short-wavelength fluctuations, $g_{\mu \nu}^s$,
\beq
g_{\mu \nu} = g_{\mu \nu}^\ell + g_{\mu \nu}^s.
\eeq
We then notice that
smoothing the inverse metric $g^{\mu \nu}$ is {\it not} the same as the inverse of the smoothed metric $g_{\mu \nu}^\ell$, with corrections being of order (see Appendix~\ref{sec:Einstein})
\beq
\bar g_{\alpha \beta} \bigl[ g_s^{\mu \alpha} g_s^{\nu \beta}\bigr]_\Lambda\, .
\label{equ:correction}
\eeq
Importantly, Eqn.~(\ref{equ:correction}) in general doesn't vanish in the limit $k \ll \Lambda$.
However, as we discuss in detail in \S\ref{sec:gauge}, in a gauge where
\beq
\label{equ:good}
\lim_{k \ll \Lambda} g^{\mu \nu}_\ell \gg \lim_{k \ll \Lambda} \bar g_{\alpha \beta} \bigl[ g_s^{\mu \alpha} g_s^{\nu \beta}\bigr]_\Lambda
\eeq
the smoothed spacetime is well-defined up to negligible corrections. In that case the long-wavelength metric, $g^\ell_{\mu \nu}$, is well-defined and can be used to construct covariant actions such as the action of a scalar field $\varphi$ living in this effective spacetime
\beq
\sqrt{-g_\ell} \, g^{\mu \nu}_\ell \partial_\mu \varphi \partial_\nu \varphi\, .
\eeq
As we show in \S\ref{sec:gauge}, the limit (\ref{equ:good}) is satisfied in Poisson gauge, but is violated in comoving gauge.
In comoving gauge the notion of an effective spacetime after integrating out small scales is therefore ill-defined.
This explains why the backreaction of small-scale fluctuations on ${\zeta}$ is easier to interpret in Poisson gauge (see Appendix~\ref{sec:Einstein} for further details).

\subsection{(Non-)Acceleration from Backreaction}
\label{sec:backreaction}

It has been suggested (see {\it e.g.}~Refs.~
\cite{Buchert:1999er}--\cite{Brown:2009tg}) 
that non-linear backreaction of short-wavelength fluctuations on the background evolution could explain the observed acceleration of the universe without the need for dark energy.
In our work we see no indication of such a large effect.
We have shown that small-scale inhomogeneities simply lead to a small renormalization of the FRW background.\footnote{To be clear, cosmological observations do not observe the background directly. Besides renormalization of the background, inhomogeneities in the matter affect light propagation ({\it e.g.}~via lensing) and hence also the measured cosmological observables. For a discussion of these distinct effects see~\cite{Brouzakis:2007zi,Marra:2007pm, Biswas:2006ub,Barausse:2005nf}.}
In fact, we were able to prove that very small scales (in virialized equilibrium) decouple completely from the background dynamics (see \S\ref{sec:PerfectFluid}). This means that the effects of small scales are naturally cut off even in a pure matter universe, in which the effects of non-linearities can naively be expected to be large. 
As we stressed in the Introduction, this decoupling of small scales is more properly a non-renormalization theorem, and goes beyond standard effective field theory decoupling~\cite{Appelquist:1974tg, Goldberger:2007hy, Burgess:2007pt, Pich:1998xt}.
Furthermore, we found that the induced effective pressure is positive, $\bar p_{\rm eff} > 0$, with a magnitude set by the velocity dispersion of small-scale fluctuations, $w_{\rm eff} \sim {\cal O}(v^2) \ll 1$. We therefore do not see how the backreaction of small-scale non-linearities can possibly explain the acceleration of the universe.
To arrive at this conclusion we had to assume that the spacetime can be described by the Poisson metric with small perturbations or equivalently that small scales are well-described by the Newtonian approximation.
This assumption has proven successful in many cosmological applications, but
see Ref.~\cite{Mat} for a discussion of this point. 

Although we did not prove this, we believe that this decoupling of short-wavelength non-linearities should even apply in the extreme case that the universe is filled with a gas of black holes. In this case, our perturbative scheme breaks down. However, there is a large separation of scales between the horizon scale and the scale at which metric perturbation become of order one.\footnote{It is straightforward to show that, even hypothetically, one cannot prepare a {\em dense} gas of black holes. For a large number $N$ of black holes, the number density has to be smaller than $1/N^2$ in Schwarzschild-radius units. For denser configurations, the whole gas collapses to form a big black-hole.}
This means that one can match our effective theory to the effective theory of black holes by Goldberger and Rothstein~\cite{Goldberger:2004jt}. In \cite{Goldberger:2004jt} it was shown that at intermediate scales, between the scale at which curvatures are of order one and the horizon scale, a collection of black holes may be described as point particles with some higher-derivative couplings to the metric. These particles can be taken to be the dark matter particles that we deal with in this paper, and our conclusions are therefore expected to apply straightforwardly even in this extreme example.
Moreover, we proved that decoupling does hold for virialized systems containing black holes. 
We believe this argument makes a large backreaction from gravitational non-linearities even in this case impossible.

\section{Conclusions}
\label{sec:conclusions}

Cosmology has made significant progress by studying {\it linear} perturbations around a homogeneous Friedmann-Robertson-Walker background \cite{Dodelson:2003ft}. 
A crucial simplification of the linear treatment is that large scales (IR) are decoupled from small scales (UV),
{\it i.e.}~at linear order different Fourier modes evolve independently. Furthermore, in linear theory there exists a useful classification of the perturbations into independent scalar, vector and tensor modes.
Beyond linear theory, the Einstein equations couple the UV and the IR, with small-scale fluctuations providing sources for the formation and evolution of large-scale perturbations.
In this paper, we studied this UV-IR coupling of cosmological fluctuations.
By integrating out short-wavelength perturbations we derived an effective theory for the long-wavelength universe.
We observed that on sufficiently large scales the universe can be described by quasi-linear perturbations evolving in the presence  of an effective fluid whose properties are determined by the small-scale structures.
The fluid is somewhat unconventional in the following sense:
higher moments in the Boltzmann hierarchy are small not because of sizable interactions like in a conventional fluid, but because they haven't had sufficient time to develop during one Hubble time ${\cal H}^{-1}$.
In the absence of gravity ${\cal H}^{-1}$ would go to infinity and a macroscopic effective  fluid description would not be applicable at non-zero momentum.
It is in this sense that we refer to the fluid as a `gravitational fluid', highlighting the fundamental importance of gravity.
On superhorizon scales, the only effect of small-scale structures is to renormalize the background density and pressure by terms of order the velocity dispersion. Moreover, we proved that the virial theorem naturally filters out the contributions from very small scales.
On subhorizon scales, dissipative effects like viscosity are induced by the small-scale non-linearities. 
The imperfect fluid is described by a few parameters like the equation of state, the sound speed and a viscosity parameter.
We proposed that this effective description of the long-wavelength universe be used to formulate a
well-defined alternative to conventional perturbation theory.
At scales larger than some smoothing scale $\Lambda^{-1}$ the theory is defined as an 
expansion in small perturbation variables -- $\delta(k \ll \Lambda)$ and $\theta(k \ll \Lambda)$ -- which evolve in a fluid whose physical parameters are determined by a numerical matching calculation.
It remains to be quantified if a simple version of this procedure ({\it e.g.}~a one-loop calculation in the effective theory) will be sufficient to reach the required level of precision 
in a practical application like baryon acoustic oscillations \cite{Future}.

\subsubsection*{Acknowledgments}

We thank Paolo Creminelli, Daniel Green, Lam Hui, Eiichiro Komatsu, Jim Peebles, Massimo Porrati, Kris Sigurdson, Riccardo Sturani, Filippo Vernizzi, David Wands, David Weinberg, Martin White, Simon White, and Amit Yadav for helpful discussions.
A.N.~is especially thankful to Enrico Trincherini for collaboration in the early stages of this project.
D.B.~thanks Rebecca and Sydney Packer in Kingston, JA for their hospitality while this work was completed.
A.N.~would like to thank the Insitute for Advanced Studies for hospitality during this project.

The research of D.B. and M.Z.~is supported by the National Science Foundation under PHY-0855425, AST-0506556 and AST-0907969.
The research of A.N.~is supported by NASA ATP (09-ATP09-0049)
L.S.~is supported in part by the National Science Foundation under PHY-0503584.
M.Z. is supported by the David and Lucile 
Packard Foundation, the Alfred P.~Sloan Foundation and the John D. and Catherine T.~MacArthur Foundation.

\newpage
\clearpage

\appendix
 \section{The Effective Fluid \`a la Newton}
 \label{sec:Euler}
 
 In this appendix we give a simple derivation of the effective stress-energy tensor by smoothing the Newtonian conservation equations.
 
 \subsection{Preliminaries}
 
 
 Consider a collisionless gas of non-relativistic particles.
 We define the single-particle phase space density $f(\x,\p, t)$, such that $f d^3 x d^3 p$ is the number of particles in an infinitesimal phase space volume element. Here, $t,\x$ are the time and spatial coordinates in the Newtonian frame defined in \S\ref{sect:Newton}, and $\p \equiv  m \dot \x$ is the conjugate momentum.
 Phase-space conservation is described by the collisionless Boltzmann equation (or Vlasov equation)~\cite{Peebles},
 \beq
 \label{equ:Boltzmann}
 \frac{D f }{Dt} = \frac{\partial f}{\partial t} + \frac{1}{m} \p \cdot \nabla f -  m \nabla \Phi \cdot \frac{\partial f}{\partial \p} = 0\, ,
 \eeq
 where we have used the particle equation of motion
 \beq
 \frac{d \p }{d t} = - m \nabla \Phi\, .
 \eeq
 Here, $\Phi$ is the Newtonian potential defined as in \S\ref{sect:Newton}.
 Taking moments of the distribution function $f$, we define the particle density, $\rho_m$, the peculiar velocity flow, $\v$, and the stress tensor, $\kappa_{ij}$,
 \bea
 \label{equ:mom1}
m \int_\p\, f(\x, \p, \eta) &\equiv& \rho_m(\x, t)\, ,\\
 m  \int_\p\, \frac{p_i}{m} \,f(\x, \p, t) &\equiv& \rho_m v_i(\x, t)\, ,\\
 m \int_\p\, \frac{p_i p_j}{m^2}\, f(\x, \p, t) &\equiv& \rho_m v_i v_j(\x, t)+ \kappa_{ij}(\x, t)\label{equ:mom3}\, .
 \eea
 Taking moments of the Boltzmann equation (\ref{equ:Boltzmann}) gives equations relating the fields in Eqns.~(\ref{equ:mom1})--(\ref{equ:mom3}).
 The zeroth moment gives the continuity equation (describing mass conservation) 
 \beq
 \partial_t \rho_m + \nabla_i (\rho_m v^i) = 0\, . \label{equ:conV0}
 \eeq
 The first moment gives the Euler equation (describing momentum conservation)
\beq
\rho_m \left[ \partial_t v^i + v^j \nabla_j v^i \right] = - \rho_m \nabla_i \Phi - \nabla_j \kappa_{ij}\, , \label{equ:EulerV0}
\eeq
while the second moment gives an evolution equation for the stress tensor $\kappa_{ij}$.
The hierarchy of moments of the Boltzmann equation is often truncated by  either assuming that the stress tensor is negligible on large enough scales, $\kappa_{ij} \approx 0$, or by postulating a phenomenological ansatz relating $\kappa_{ij}$ to the other fluid quantities, {\it e.g.}
\beq
\kappa_{ij} = - p \delta_{ij} + \eta \Bigl[ \nabla_i v_j + \nabla_j u_i - \frac{2}{3} \delta_{ij} \nabla \cdot \v \Bigr] + \zeta \delta_{ij} \nabla \cdot \v\, , \label{equ:kappaij}
\eeq
where $p$ denotes the pressure and $\eta$ and $\zeta$ are viscosity coefficients (see Appendix~\ref{sec:FluidReview}).
In this appendix we show that even starting with vanishing $\kappa_{ij}$, integrating out short-wavelength fluctuations induces a non-zero effective stress tensor of the form of Eqn.~(\ref{equ:kappaij}) at long-wavelengths.

\subsection{Integrating out Small Scales}
\label{sec:integratingout}

We are interested in the theory at scales $k$ much larger than the scale of non-linearities $q_\star$. To define this effective long-wavelength theory we integrate out short-wavelength modes below a scale $\Lambda \ll q_\star$ starting from Eqns.~(\ref{equ:conV0}) and (\ref{equ:EulerV0}).
In real space, integrating out short-wavelength modes corresponds to a convolution of all fields with a window function $W_\Lambda$,
\begin{align}
\rho_\ell &\equiv  [\rho]_\Lambda(\x) = \int_{\x'} W_\Lambda(|\x-\x'|) \rho_m(\x') \, , \label{equ:smooth1}\\
\phi_\ell &\equiv [\phi]_\Lambda(\x) = \int_{\x'} W_\Lambda(|\x-\x'|) \phi(\x') \, ,\\
\rho_\ell v^i_\ell &\equiv [\rho_m v^i]_\Lambda(\x) = \int_{\x'} W_\Lambda(|\x-\x'|) \rho(\x') v^i(\x') \, . \label{equ:smooth3}
\end{align}
Essentially, this amounts to averaging the fields over domains of size $\Lambda^{-1}$.
The smoothed velocity field $v_\ell^i$ was defined such that $\rho_\ell v_\ell^i$ represents the center of mass of the domain. 
In Fourier space, Eqns.~(\ref{equ:smooth1})--(\ref{equ:smooth3}) become
\begin{align}
\rho_\k^\ell &= W_\Lambda(k) \cdot \rho_\k \, , \\
\phi_\k^\ell &= W_\Lambda(k) \cdot \phi_\k \, , \\
(\rho_\ell v^i_\ell)_\k &=  W_\Lambda(k) \cdot (\rho v^i)_\k\, .
\end{align}
The window function is normalized such that
\beq
\label{equ:norm}
W_\Lambda(k=0) = \int d^3 x \, W_\Lambda(x) \equiv 1\, .
\eeq
For concreteness we sometimes specialize to a Gaussian window function 
\beq
W_\Lambda(k) = \exp \Bigl[ - \frac{1}{2} \frac{k^2}{\Lambda^2}\Bigr]\, .
\eeq
We split all fields into long and short modes
\beq
\label{equ:fields}
\rho_m = \rho_\ell + \rho_s\, , \qquad \phi = \phi_\ell + \phi_s\, , \qquad v^i = v^i_\ell + v^i_s\, ,
\eeq
where in Fourier space the short-wavelength fluctuations are defined by
\begin{align}
\rho_\k^s &= F_\Lambda(k) \cdot \rho_\k \, , \label{equ:rhos}\\
\phi_\k^s &= F_\Lambda(k) \cdot \phi_\k \, , \\
(\rho_m v^i_s)_\k &= F_\Lambda(k) \cdot (\rho_m v^i)_\k\, , \label{equ:vs}
\end{align}
with $F_\Lambda(k) \equiv 1- W_\Lambda(k)$.
For notational simplicity we will often drop the index `$s$' denoting short-wavelength quantities except in cases where this could cause ambiguity.

\subsection{Effective Stress from the Euler Equation}
\label{sec:alt}

To analyze the effects of small-scale non-linearities on the long-wavelength universe, we consider the Euler equation describing momentum conservation in a pressureless fluid
\beq
\label{equ:NLEuler}
\rho_m \left[ \dot v^i + v^j \nabla_j v^i \right] + \rho_m \nabla_i \Phi = 0  \, .
\eeq
We note that the Euler equation is linear in the Newtonian potential $\Phi$ and quadratic in fluid velocities $v^i$.
Notice that we never expand in density perturbations $\delta \rho_m$. The density $\rho_m$ and the Newtonian potential $\Phi$ are related by the Poisson equation
\beq
\label{equ:NLPoisson}
\nabla^2 \Phi = 4\pi G \rho_m \, .
\eeq
Eqns.~(\ref{equ:NLEuler}) and (\ref{equ:NLPoisson}) are valid on subhorizon scales and receive corrections for modes with wavelengths comparable to the horizon.
This is made explicit in the more formal relativistic treatment of the same results in \S\ref{sec:PT}. 

We now show that Eqn.~(\ref{equ:NLEuler}) can be written in the form
\beq
\label{equ:SmoothEuler}
\rho_\ell \left[ \dot v^i_\ell + v^j_\ell \nabla_j v^i_\ell \right] + \rho_\ell \nabla_i \phi_\ell = - \nabla_j \bigl[ \tau^j_{\ i}\bigr]_\Lambda\, ,
\eeq
where 
\beq
\tau^i_{\ j} \equiv \rho_m v_i^s v_j^s - \frac{\phi_{,k}^s \phi_{,k}^s \delta_{ij}-2 \phi_{,i}^s \phi_{,j}^s}{8\pi G}\, .
\eeq
As before
$[...]_\Lambda$ denotes a spatial average over domains of size $\Lambda^{-1}$  and the labels $\ell$ and $s$ denote long-wavelength smoothed fields and small-wavelength fluctuations, respectively. 
We will interpret $\tau_{ij}$ as the effective stress tensor induced by products of small-scale fluctuations. It sources the evolution of the long-wavelength perturbations.

The proof of Eqn.~(\ref{equ:SmoothEuler}) is instructive and shows that the result is valid up to higher-derivative corrections in the long-wavelength fields.

\vspace{0.5cm}
\small
\hrule \vskip 1pt \hrule \vspace{0.3cm}
\noindent 
{\bf Proof}:\vskip 4pt

Consider first the smoothing of the term $\rho_m \dot v^i + \rho_m v^j \nabla_j v_i$,
\beq
\star \ \equiv \ \int_{\x'} W_\Lambda(|\x-\x'|) \left[ \rho_m \partial_t v^i(\x') + \rho_m v^j(\x') \partial_{j'} v^i(\x')\right]\, .
\eeq
Using the continuity equation $\partial_t \rho_m= - \partial_j (\rho_m v^j)$ this may be written as
\beq
\star \ =\ \partial_t (\rho_\ell v^i_\ell) + \int_{\x'} W_\Lambda\, v^i \partial_{j'}(\rho_m v^j) + \int_{\x'} W_\Lambda\, \rho_m v^j \partial_{j'} v^i\, .
\eeq
Integrating the second term by parts and using $\partial_{j'} W_\Lambda = - \partial_j W_\Lambda$, we find
\beq
\label{equ:star}
\star\ =\ \partial_t (\rho_\ell v^i_\ell) + \partial_j \int_{\x'} W_\Lambda\, \rho_m v^i v^j \, . 
\eeq
We split the fluid velocities $v^i$ into long-wavelength modes $v^i_\ell$ and short-wavelength fluctuations $v^i_s$,
\beq
\label{equ:V}
v^i(\x') = v_\ell^i(\x') + v^i_s(\x')\, ,
\eeq
and expand $v_\ell^i(\x')$ in a Taylor series around $\x$,
\beq
\label{equ:Taylor}
v^i_\ell(\x') = v_\ell^i(\x) + \partial_j v^i_\ell(\x) \cdot (\x-\x')^j + \frac{1}{2} \partial_k \partial_j v^i_\ell(\x) \cdot (\x-\x')^j  (\x-\x')^k + \cdots\, .
\eeq
The derivative terms in Eqn.~(\ref{equ:Taylor}) will result in higher-derivative corrections to Eqn.~(\ref{equ:SmoothEuler}).
In particular, cross-terms between $v^i_\ell$ and $v^i_s$ will vanish up to higher-derivative corrections.
For concreteness, we will use a Gaussian window function, such that
 \begin{align}
 \label{equ:Gaussian1}
\partial_{j'} W_\Lambda = - \partial_j W_\Lambda &= \Lambda^2 (\x-\x')^j W_\Lambda \, , \\
\partial_{i'} \partial_{j'} W_\Lambda = \partial_i \partial_j W_\Lambda &= - \Lambda^2 \delta_{ij} W_\Lambda + \Lambda^4 (\x-\x')^i (\x-\x')^j W_\Lambda\, ,  \label{equ:Gaussian2}
\end{align}
but our results will only slightly depend on this example.
Using Eqns.~(\ref{equ:V}), (\ref{equ:Taylor}), (\ref{equ:Gaussian1}) and (\ref{equ:Gaussian2}) we may show from the definition of $v^i_\ell$ that
\bea
\int_{\x'} W_\Lambda \, \rho_m v_s^i &=& - \partial_k v^i_\ell \int_{\x'} W_\Lambda (\x-\x')^k \rho_m - \frac{1}{2} \partial_k \partial_l v^i_\ell \int_{\x'} W_\Lambda (\x-\x')^k (\x-\x')^l \rho_m \, , \\
&=& \frac{\partial_k \rho_\ell \partial_k v^i_\ell}{\Lambda^2} - \frac{1}{2} \rho_\ell \frac{\partial^2 v^i_\ell}{\Lambda^2}\, .
\eea
Using the above derivative expansions we find
\beq
\int_{\x'} W_\Lambda\, \rho_m v^i v^j = \int_{\x'} W_\Lambda\, \rho_m v^i_\ell v^j_\ell + 2 \int_{\x'} W_\Lambda\, \rho_m v_\ell^{(i} v_s^{j)} + \int_{\x'} W_\Lambda\, \rho_m v^i_s v^j_s \, , 
\eeq
where
\beq
 \int_{\x'} W_\Lambda\, \rho_m v^i_\ell v^j_\ell = \rho_\ell v^i_\ell v^j_\ell + {\cal C}_1^{ij} \qquad
{\rm and} \qquad
2 \int_{\x'} W_\Lambda\, \rho_m v_\ell^{(i} v_s^{j)} = {\cal C}_2^{ij}\, .
\eeq
Here, we defined the higher-derivative terms
\bea
\label{equ:C1}
{\cal C}_1^{ij} &\equiv& \rho_\ell \frac{\partial_k v_\ell^i \partial_k v_\ell^j}{\Lambda^2}
 - 2 \frac{\partial_k \rho_\ell \partial_k v_\ell^{(i}}{\Lambda^2} v_\ell^{j)} + \rho_\ell v_\ell^{(i} \frac{\partial^2 v_\ell^{j)}}{\Lambda^2}\, , \\
 {\cal C}_2^{ij} &\equiv& 2  \frac{\partial_k \rho_\ell \partial_k v_\ell^{(i}}{\Lambda^2} v_\ell^{j)} -  \rho_\ell v_\ell^{(i} \frac{ \partial^2 v_\ell^{j)}}{\Lambda^2}\, . \label{equ:C2}
  \eea
Hence, we obtain
\beq
\int_{\x'} W_\Lambda\, \rho_m v^i v^j = \rho_\ell v^i_\ell v^j_\ell + \int_{\x'} W_\Lambda\, \rho_m v^i_s v^j_s + {\cal C}^{ij}\, , 
\eeq
where
\beq
{\cal C}^{ij} \, \equiv\,   \rho_\ell \frac{\nabla v^i_\ell \cdot \nabla v^j_\ell}{\Lambda^2} \, .
\eeq
Notice that many of the higher-derivative corrections in Eqns.~(\ref{equ:C1}) and (\ref{equ:C2}) have canceled.
Eqn.~(\ref{equ:star}) becomes
\beq
\star\ =\ \partial_t(\rho_\ell v^i_\ell) + \partial_j (\rho_\ell v^i_\ell v^j_\ell) + \partial_j [\rho_m v^i_s v^j_s]_\Lambda \, + \, \partial_j {\cal C}^{ij} \, .
\eeq
Expanding the time derivative using the smoothed continuity equation,
\beq
\int_{\x'} W_\Lambda [\partial_t \rho_m + \partial_{i'}(\rho_m v^i)] = \partial_t \rho_\ell + \partial_i (\rho_\ell v^i_\ell) = 0,
\eeq
we find
\beq
\label{equ:star2}
\star\ =\ \rho_\ell \dot v^i_\ell + \rho_\ell v^j_\ell \partial_j v^i_\ell+ \partial_j  [\rho_m v^i_s v^j_s]_\Lambda \, + \, \partial_j {\cal C}^{ij} \, .
\eeq

Next, we consider smoothing of the term $\rho_m \partial_i \Phi$,
\beq
\label{equ:diamond}
\diamond \ =\ \int_{\x'}W_\Lambda(|\x-\x'|) \rho_m(\x') \partial_{i'} \Phi(\x')\, .
\eeq
As before, we define a split into smoothed long-wavelength modes and short-wavelength fluctuations,
\begin{align}
\rho_m(\x') &= \rho_\ell(\x') + \rho_s(\x')\, , \\
\Phi(\x') &= \phi_\ell(\x') + \phi_s(\x')\, ,
\end{align}
where 
\beq
\label{equ:zero2}
\int_{\x'} W_\Lambda \, \rho_s \approx - \frac{1}{2} \frac{\partial^2 \rho_\ell}{\Lambda^2}\qquad {\rm and} \qquad  \int_{\x'} W_\Lambda\, \phi_s \approx - \frac{1}{2}\frac{\partial^2 \phi_\ell}{\Lambda^2}\, .
\eeq
Smoothing of the Poisson equation implies
\beq
\nabla^2 \phi_\ell = 4\pi G\, \rho_\ell \qquad {\rm and} \qquad \nabla^2 \phi_s = 4\pi G\, \rho_s\, .
\eeq
As before, Eqn.~(\ref{equ:zero2}) enforces the vanishing of cross-terms in Eqn.~(\ref{equ:diamond}) up to higher-derivative corrections, so that we get 
\beq
\diamond \ =\ \rho_\ell \partial_i \phi_\ell +  \int_{\x'} W_\Lambda\, \rho_s\,  \partial_{i'} \phi_s\, + \, {\cal D}^i\, ,
\eeq
where
\beq
\label{equ:hd}
{\cal D}^i \equiv \frac{\partial_k \partial_i \phi_\ell \partial_k \rho_\ell}{\Lambda^2}
\eeq
stands for all higher-derivative corrections. 
As before, there have been cancellations of some higher-derivative terms.
Using the Poisson equation, $\nabla^2 \phi_s = 4\pi G \, \rho_s$, we get
\beq
\diamond \ =\  \rho_\ell \partial_i \psi_\ell + \int_{\x'} W_\Lambda \frac{\partial^2_{j'} \psi_s }{4\pi G} \partial_{i'} \psi_s \, + \, {\cal D}^i\, .
\eeq
Integrating by parts and using $\partial_{j'}W_\Lambda = - \partial_j W_\Lambda$, this becomes
\beq
\diamond \ =\  \rho_\ell \partial_i \phi_\ell + \partial_j \int_{\x'} W_\Lambda \frac{\partial_{j'} \phi_s \partial_{i'} \phi_s}{4\pi Ga^2} - \int_{\x'} W_\Lambda \frac{\partial_{j'} \phi_s \partial_{j'} \partial_{i'} \phi_s}{4\pi G} \, + \, {\cal D}^i\, .
\eeq
After using the fact
\beq
\int_{\x'} W_\Lambda \frac{\partial_{j'} \phi_s \partial_{j'} \partial_{i'} \phi_s}{4\pi G} = \frac{1}{2} \partial_i \int_{\x'} W_\Lambda \frac{\partial_{j'} \phi_s \partial_{j'} \phi_s}{4\pi G} \, ,
\eeq
this reduces to
\beq
\label{equ:diamond2}
\diamond \ = \ \rho_\ell \partial_i \phi_\ell + \partial_j \Bigl[\frac{\phi_{,k}^s \phi_{,k}^s \delta^j_i - 2 \phi_{,i}^s \phi_{,j}^s}{8\pi G} \Bigr]_\Lambda \, + \, {\cal D}^i\, .
\eeq
Using the Poisson equation, $\nabla^2 \phi_\ell = 4\pi G \, \rho_\ell$, the higher-derivative term (\ref{equ:hd}) may be written as
\beq
{\cal D}^i = \nabla_j {\cal D}^{ij}\, , \qquad {\rm where} \qquad {\cal D}^{ij} \equiv \frac{\nabla \phi_{,k}^\ell \cdot \nabla \phi_{,k}^\ell \, \delta^j_i - 2 \nabla \phi_{,i}^\ell\cdot  \nabla \phi_{,j}^\ell}{8\pi G \cdot \Lambda^2}\, .
\eeq
Adding the results of Eqns.~(\ref{equ:star2}) ($\star$) and (\ref{equ:diamond2}) ($\diamond$) gives 
\beq
\label{equ:linearEuler}
\rho_\ell \left[ \dot v^i_\ell + v^j_\ell \nabla_j v^i_\ell \right] + \rho_\ell \nabla_i \phi_\ell = - \nabla_j \bigl[ \tau^j_{\ i}\bigr]_\Lambda - \nabla_j [\tau^j_{\ i}]^{\partial^2}\, ,
\eeq
where
\beq
\label{equ:TauEff}
\tau^i_{\ j} = \rho_m  v_{i}^s v_{j}^s - \frac{\phi_{,k}^s \phi_{,k}^s \delta^i_j - 2 \phi_{,i}^s \phi_{,j}^s}{8\pi G}\, ,
\eeq
and
\beq
\label{equ:TauHD}
 [\tau^i_{\ j}]^{\partial^2} \equiv {\cal C}^{ij} + {\cal D}^{ij} = \rho_\ell \frac{\nabla v^i_\ell \cdot \nabla v^j_\ell}{\Lambda^2}  + \frac{\nabla \phi_{,k}^\ell \cdot \nabla \phi_{,k}^\ell \, \delta^i_j - 2 \nabla \phi_{,i}^\ell\cdot  \nabla \phi_{,j}^\ell}{8\pi G \cdot \Lambda^2}\, .
\eeq
Up to the higher-derivative corrections in Eqn.~(\ref{equ:TauHD}) 
we have hence arrived at the desired answer, Eqn.~(\ref{equ:SmoothEuler}).
 \hfill QED $\blacksquare$ 
\vspace{0.2cm} \hrule \vskip 1pt  \hrule
 \vspace{0.7cm}

\normalsize

\subsection{Renormalized Pressure and Density}

From the stress tensor (\ref{equ:TauEff}) we may read of the effective pressure induced at order~$v^2$,
\beq
3 p_{\rm eff} = [\tau^k_{\ k}]_\Lambda = [\rho v_k^s v_k^s]_\Lambda - \frac{[\phi_{,k}^s \phi_{,k}^s]_\Lambda}{8\pi G}\, .
\eeq
Similarly, the background density receives corrections at order $v^2$.
In \S\ref{sec:PerfectFluid} we computed this effective density $\rho_{\rm eff}$ directly from the second-order Einstein equations:
\beq
\rho_{\rm eff} = [\gamma(v_s) \rho_m]_\Lambda + \frac{1}{2}[\rho_m \phi^s]_\Lambda\, , \qquad  \rho_m \equiv \gamma(v_s) e^{-3\phi_s} \rho\, , \qquad \gamma(v_s) \equiv \frac{1}{\sqrt{1-v_s^2}}\, ,
\eeq
or 
\beq
\label{equ:RhoEff}
\rho_{\rm eff} = \left[\rho_m \left(1 + \frac{1}{2} v_s^2 + \frac{1}{2} \phi_s \right)\right]_\Lambda\, ,
\eeq
Alternatively, the corrections to the energy density could be found by considering the first post-Newtonian corrections to the continuity equation.
Eqn.~(\ref{equ:RhoEff}) describes kinetic and gravitational corrections to the total energy density taking into account the relativistic boost to relate the physical number density in the inertial frame where the fluid velocity is $v$ to the number density in the fluid rest frame.
The final answer, Eqn.~(\ref{equ:RhoEff}), is so intuitive that it could have been guessed.
In \S\ref{sect:Newton} we showed that this form of $\rho_{\rm eff}$ is consistent with the conservation of the effective stress-energy tensor, $\partial_\mu \tau^{\mu \nu} = 0$.

\newpage

\section{The Effective Fluid \'a la Einstein}
\label{sec:Einstein}

Our discussion in most of this paper has been formulated in a fixed gauge.
The Poisson gauge was convenient because of its controlled small-scale properties in the Newtonian limit.
In \S\ref{sec:Poisson} we collect standard results from second-order perturbation theory in Poisson gauge.
Second-order gauge transformations can be subtle since the combination of two short-wavelength transformations can induce a long-wavelength change in the metric. We cite the explicit transformations in \S\ref{sec:GT}.
Finally, in \S\ref{sec:gauge} we show that our interpretation of the long-wavelength effective fluid is limited to gauges with controlled small-scale behavior such as the Poisson gauge.
We define the conditions that have to be satisfied in order to define the effective fluid in the long-wavelength limit.

\subsection{Second-Order Equations in Poisson Gauge}
\label{sec:Poisson}

\subsubsection{Metric Perturbations}

The metric in Poisson gauge is given by
\beq
\label{equ:PoissonMetric}
\d s^2 = a^2(\eta) \left[ - e^{2\Psi} \d \eta^2 + 2 \omega_i \d x^i \d \eta + (e^{-2\Phi} \delta_{ij} + \chi_{ij}) \d x^i \d x^j\right]\, ,
\eeq
with $\chi_{ii} = 0$ ({\it i.e.}~the trace is absorbed in $\Phi$) and
\beq
\label{equ:gaugeX}
\omega_{i,i} = \chi_{ij, i} = 0\, .
\eeq
The gauge condition (\ref{equ:gaugeX}) eliminates one scalar degree of freedom from $g_{0i}$ and one scalar and one transverse vector degree of freedom from $g_{ij}$.  Thus, $\omega_i$ is a transverse vector, while $\chi_{ij}$ is a transverse-traceless tensor.

Perturbations are split into first- and second-order terms
\beq
\Psi = \Psi^{(1)} + \frac{1}{2} \Psi^{(2)}\, , \qquad \Phi = \Phi^{(1)} + \frac{1}{2} \Phi^{(2)}\, , \qquad \omega_i = \omega_i^{(2)}\, , \qquad \chi_{ij} = \chi_{ij}^{(2)}\, .
\eeq
We have ignored first-order vector and tensor perturbations as they aren't produced by inflation and in any case decay with the expansion of the universe.
The metric determinant is $\sqrt{-g} = e^{\Psi-3\Phi} a^4$ to second order.

\vskip 6pt
\noindent
{\sl Christoffel symbols.} \hskip 8pt
The second-order Christoffel symbols are
\bea
\Gamma^0_{00} &=& {\cal H} + \dot \Psi \, ,\\
\Gamma^0_{0i} &=& \Psi_{,i} + {\cal H} \omega_i \, ,\\
\Gamma^i_{00} &=&  e^{2\Phi + 2 \Psi} \Psi_{,i} + \dot \omega^i + {\cal H} \omega^i \, ,\\
\Gamma^0_{ij} &=& e^{- 2 \Phi - 2 \Psi} ({\cal H} - \dot \Phi) \delta_{ij} + \frac{1}{2} \dot \chi_{ij} + {\cal H} \chi_{ij}  - \frac{1}{2} (\omega_{j, i} + \omega_{i,j})  \, , \\
\Gamma^i_{0j} &=& ({\cal H} - \dot \Phi) \delta_{ij} + \frac{1}{2} \dot \chi_{ij} + \frac{1}{2} (\omega_{i,j} - \omega_{j,i})\, , \\
\Gamma^i_{jk} &=& - \Phi_{,k} \delta^i_j  - \Phi_{,j} \delta^i_k  + \Phi^{,i} \delta_{jk} - {\cal H} \omega^i \delta_{jk} + \frac{1}{2} \Bigl[(\chi_{ij})_{,k} + (\chi_{ik})_{,j} - (\chi_{jk})_{,i} \Bigr]\, .
\eea
\vskip 6pt
\noindent
{\sl Second-order Einstein tensor.} \hskip 8pt
The Einstein tensor up to second order has the following components
\bea
G^0_{\ 0} &=& - \frac{e^{-2\Psi}}{a^2} \left[ 3 {\cal H}^2 - 6 {\cal H} \dot \Phi +3 \dot \Phi^2 - e^{2\Phi + 2\Psi} (\Phi_{,i} \Phi_{,i} - 2 \Phi_{,kk})\right] \, , \label{equ:G00}\\
G^i_{\ 0} &= & 2 \frac{e^{2\Phi}}{a^2} \left[{\dot \Phi}_{,i} + ({\cal H} - \dot \Phi) \Psi_{,i}\right] - \frac{1}{2a^2} \omega^i_{,kk} + 2 ({\cal H}^2 - \dot {\cal H}) \frac{\omega^i}{a^2} \, , \label{equ:Gi0}\\
G^i_{\ j} &=& \frac{1}{a^2} \Bigl[ e^{-2\Psi} \Bigl( -({\cal H}^2 + 2 \dot{\cal H}) - 2 \dot \Phi \dot \Psi - 3 \dot \Phi^2 + 2 {\cal H} (2\dot \Phi + \dot \Psi) + 2 \ddot \Phi \Bigr) \nonumber \\
&&\ \ \ \ \ \ + \ e^{2\Phi} (\Psi_{,k} \Psi_{,k} + \Psi_{,kk} - \Phi_{,kk}) \Bigr]\delta^i_j \nonumber \\
&& \ \ \ \    + \ \frac{e^{2\Phi}}{a^2} 
( \Phi_{,ij} - \Psi_{,ij} + \Phi_{,i} \Phi_{,j}  -\Psi_{,i} \Psi_{,j}  - \Psi_{,i} \Phi_{,j}  - \Phi_{,i} \Psi_{,j}) \nonumber \\
&& \ \ \ \   - \ \frac{1}{2a^2} \Bigl[ (\dot \omega^i_{,j} + \dot \omega^j_{,i}) + 2 {\cal H} (\omega^i_{,j} + \omega^j_{,i}) \Bigr]
\ + \ \frac{1}{2a^2} \Bigl[\ddot  \chi^i_j + 2 {\cal H} \dot \chi^i_j - (\chi^i_j)_{,kk} \Bigr]\, .
\eea

We split these tensor components into $G^\mu_{\ \nu}=\bar G^\mu_{\ \nu} + (G^\mu_{\ \nu})^{\rm L}+ (G^\mu_{\ \nu})^{\rm NL}$. The background components are
\beq
- a^2 \bar G^0_{\ 0} =  3 {\cal H}^2 \, , \qquad
\bar G^i_{\ 0} =  0 \, , \qquad 
- a^2 \bar G^i_{\ j} = {\cal H}^2 + 2 \dot{\cal H}\, .
\eeq
The linear parts are
\bea
- \frac{a^2}{2}(G^0_{\ 0})^{\rm L} &=& \nabla^2 \Phi - 3 \H (\dot \Phi + \Psi)\, , \\
+ \frac{a^2}{2} (G^i_{\ 0})^{\rm L} &= &[\dot \Phi + \H \Psi]_{,i} \ + \  \frac{1}{4} \nabla^2 \omega^i + (\H^2 - \dot \H) \omega^i \, ,\\
+ \frac{a^2}{2} (G^i_{\ j})^{\rm L} &=&
  \Bigl[  ({\cal H}^2 + 2 \dot{\cal H}) \Psi  +  {\cal H} (2\dot \Phi + \dot \Psi) +  \ddot \Phi - \frac{1}{2} \nabla^2(\Phi - \Psi) \Bigr]\delta^i_j  + \frac{1}{2} 
[\Phi - \Psi]_{,ij} \nonumber \\
&& \    - \ \frac{1}{4} \Bigl[ (\dot \omega^i_{,j} + \dot \omega^j_{,i}) + 2 {\cal H} (\omega^i_{,j} + \omega^j_{,i}) \Bigr]
\ + \ \frac{1}{4} \Bigl[\ddot  \chi^i_j + 2 {\cal H} \dot \chi^i_j - (\chi^i_j)_{,kk} \Bigr]\, .
\eea
Finally, the quadratic parts are
\bea
- a^2 (G^0_{\ 0})^{\rm NL} &=& 12 \H^2 \Psi^2 + 12 \H \dot \Phi \Psi + 3 \dot \Phi^2 - \Phi_{,k} \Phi_{,k} + 4 \Phi \Phi_{,kk}\, ,\\
+ \frac{a^2}{2} (G^i_{\ 0})^{\rm NL} &=& 2\Phi [\dot \Phi + \H \Psi]_{,i} - \dot \Phi \Psi_{,i} \, , \\
+ a^2(G^i_{\ j})^{\rm NL} &=&  \Bigl[  -4 ({\cal H}^2 + 2 \dot{\cal H}) \Psi^2 - 2 \dot \Phi \dot \Psi - 3 \dot \Phi^2 - 4 {\cal H} (2\dot \Phi + \dot \Psi) \Psi - 4 \Psi \ddot \Phi \nonumber \\
&&\ \ \ \ \ \ + \ \Psi_{,k} \Psi_{,k} - 2 \Phi [\Phi - \Psi]_{,kk}  \Bigr]\delta^i_j \nonumber \\
&& \     + \
2 \Phi [\Phi - \Psi]_{,ij} + \Phi_{,i} \Phi_{,j}  -\Psi_{,i} \Psi_{,j}  - \Psi_{,i} \Phi_{,j}  - \Phi_{,i} \Psi_{,j} \, .
\eea

In the absence of anisotropic stress $\phi \equiv \Phi = \Psi$ and using $\phi_{,k} \gg {\cal H} \phi \sim \dot \phi$ on small scales we find
\bea
- a^2 (G^0_{\ 0})^{\rm NL} &\approx&  - \phi_{,k} \phi_{,k} + 4 \phi \phi_{,kk}\, , \label{equ:g1}\\
+ a^2(G^i_{\ j})^{\rm NL} &\approx&  \phi_{,k} \phi_{,k}\,  \delta^i_j -  2 \phi_{,i} \phi_{,j} \, . \label{equ:g2}
\eea
Notice that in this gradient expansion Eqns.~(\ref{equ:g1}) and (\ref{equ:g2}) hold independent of the background equation of state.

\subsubsection{Matter Perturbations}
\label{sec:matterA}

We now cite results for the parameterization of perturbations in the dark matter energy and momentum. We first give results for dark matter in the fluid approximation and then present the energy-momentum tensor for point particles.

\vskip 6pt
\noindent
{\sl Dark matter in the fluid approximation.} \hskip 8pt
We consider the stress-energy tensor of a fluid in the covariant form
\beq
\label{equ:Tmunu}
T^{\mu}_{\ \nu} = (\rho + p ) u^\mu u_\nu + p\, \delta^\mu_\nu  +\Sigma^\mu_{\ \nu}\, ,
\eeq
where
$g_{\mu \nu} u^\mu u^\nu = -1$
and
\beq
\label{equ:SigmaMuNu}
\Sigma^\mu_{\ \mu} = \Sigma^\mu_{\ \nu} u^\nu =0 \, . 
\eeq
Eqn.~(\ref{equ:SigmaMuNu}) implies that only the spatial components of $\Sigma^\mu_{\ \nu}$ are non-zero, {\it i.e.}~$\Sigma^0_{\ 0} = \Sigma^i_{\ 0} = 0$.
In this paper we will often restrict to a background without pressure and anisotropic stress, {\it i.e.}~$p=\Sigma^i_{\ j} =0$, but our results are easily extended to more general backgrounds.
At second order, fluid velocities and gradients of the gravitational potential induce both pressure and anisotropic stress. A characterization of the second-order corrections to the effective fluid is one of the main objectives of this paper.

To second order in metric perturbations the four-velocity is
\begin{align}
u^0 &= a^{-1} e^{-\Psi} \gamma(v) \, , \qquad u^i = a^{-1} e^\Phi v^i\, , \label{equ:umu}\\
u_0 &= - a e^{\Psi} \gamma(v) \, , \qquad \ \ \, u_i = a ( \omega_i + e^{-\Phi} v^i)\, ,
\end{align}
where we defined the Lorentz factor
\beq
\gamma(v) \equiv \frac{1}{\sqrt{1-v^2}} \approx 1 + \frac{1}{2} v^2\, .
\eeq
Eqn.~(\ref{equ:Tmunu}) becomes
\bea
T^0_{\ 0} &=& - \gamma^2 \rho = - \rho(1 + v^2)\, ,\\
T^i_{\ 0} &=& -  e^{\Psi + \Phi} \rho v^i \, ,\\
T^0_{\ i} &=& \rho  (\omega_i + e^{-\Psi - \Phi} v_i)\, , \\
T^i_{\ j} &=& \rho  v^i v_j\, .
\eea

Notice that the conservation equation, $\nabla_\mu T^{\mu \nu}=0$, may be written as
\beq
\label{equ:nablaT}
u_\nu \nabla_\mu T^{\mu \nu} = u_\nu \nabla_\mu (\rho u^\mu u^\nu)  = - \nabla_\mu (\rho u^\mu) = - \partial_\mu (\sqrt{-g} \rho u^\mu) = 0\, .
\eeq
This motivates introducing
\beq
\label{equ:rhoSTAR}
 \rho_m \equiv \frac{\sqrt{-g} \rho u^0}{a^3}  = \gamma(v) e^{-3\Phi} \rho \approx \rho \left(1+\frac{v^2}{2} - 3\Phi \right)\, ,
\eeq
such that Eqn.~(\ref{equ:nablaT}) implies
\beq
\frac{d}{d\eta} \int d^3 x\, a^3 \rho_m = 0\, .
\eeq
The physical role of $\rho_m$ becomes clear when thinking about dark matter in terms of particles and their conserved number density, $n \propto \rho_m$. A relativistic boost relates the physical number density in the intertial frame where the fluid velocity is $v$ to the number density in the fluid rest frame, $n_{\rm rest} \propto \rho$. The factor $\gamma(v) e^{-3\Phi}$ in Eqn.~(\ref{equ:rhoSTAR}) clearly relates the volume element in the rest frame to the physical volume in the moving frame.

The conserved energy of a localized system is ({\it cf.}~Exercise 20.5 in Misner, Thorne and Wheeler~\cite{MTW})
\bea
&& \int \Bigl[ \rho \underbrace{(1-v^2)^{-1/2}}_{\rm Lorentz} + \underbrace{\frac{1}{2} \rho v^2}_{\rm KE} + \underbrace{\frac{1}{2}\rho \Phi}_{\rm PE} \Bigr] \underbrace{(g_{xx} g_{yy} g_{zz})^{1/2} dx dy dz}_{\rm proper \; volume}  \\
&=& -  \int dx^3\, a^3 \Bigl[ T^{0}_{\ 0} - \frac{(G^0_{\ 0})^{\rm NL}}{8\pi G} \Bigr]  \\
&=& \int dx^3\, a^3 \rho_m \Bigl(1+ \frac{1}{2} v^2 + \frac{1}{2} \Phi \Bigr)\, .
\eea
More generally, the conserved four-momentum of a localized system is 
\beq
P^\mu \equiv - \int dx^3\, a^3 \Bigl[ T^{\mu}_{\ 0} - \frac{(G^\mu_{\ 0})^{\rm NL}}{8\pi G} \Bigr]  \equiv - \int d^3 x\, a^3 \tau^\mu_{\ 0}\, .
\eeq
For an early discussion of conserved energy and momentum in General Relativity see~\cite{Tolman1, Tolman2}.

\vskip 6pt
\noindent
{\sl Dark matter as point particles.} \hskip 8pt
When relating our approach to $N$-body simulations a particle description of dark matter is essential.
A collection of particles `${\rm a}$' is described by the following stress-energy tensor~\cite{Weinberg},
\beq
T^{\mu \nu} = \sum_{\rm a} m_\a \frac{u^\mu_\a u^\nu_\a}{u_\a^0} \frac{\delta_{\rm D}(\x-\x_\a(\eta))}{\sqrt{-g}}\, .
\eeq
To second order, its components are
\bea
 T^{0}_{\ 0} &=& - a^{-3} e^{3\Phi}\, \sum_\a m_\a  \left(1+ \frac{v^2_\a}{2}\right)  \delta_{\rm D}(\x-\x_\a(\eta))\, ,\\
T^{i}_{\ 0} &=& a^{-3} e^{3\Phi} \,e^{\Phi+\Psi} \sum_\a m_\a v^i_\a \,  \delta_{\rm D}(\x-\x_\a(\eta))  \, ,\\
T^i_{\ j} &=&  a^{-3} e^{3\Phi}\, \sum_\a m_\a v_\a^i v^j_\a \, \delta_{\rm D}(\x-\x_\a(\eta)) \, .
\eea
The spatial integral of the zero-zero component of the effective stress-energy tensor gives the total energy of the particles
\bea
E \equiv - \int d^3 x \, a^3 \tau^0_{\ 0} &=& - \int d^3 x \, a^3 \left( T^0_{\ 0} - \frac{(G^0_{\ 0})^{\rm NL}}{8\pi G}\right) \\
&=& \sum_\a m_\a \left(1 + \frac{v_\a^2}{2} + \frac{1}{2} \Phi(\x_\a) \right)\, .
\eea
Particles evolve according to the geodesic equation, which at second order is
\beq
\frac{d^2 \x_\a}{d\eta^2} + (\H-2 \dot \Phi-\dot \Psi) \frac{d \x_\a}{d \eta} = - e^{2(\Phi + \Psi)} \boldsymbol{\nabla} \Psi(\x_\a)-\dot{\boldsymbol \omega}-\H{\boldsymbol\omega}\, .
\eeq

\subsubsection{Einstein Equations}

Metric and matter perturbations are related by the Einstein equations 
\beq
G^\mu_{\ \nu} = 8\pi G \, T^\mu_{\ \nu}\, .
\eeq
In the following we present the first-order and second-order equations.

\vskip 6pt
\noindent
{\sl First-order solutions.} \hskip 8pt
In linear perturbation theory, the Einstein tensor is linearized in the metric perturbations $\phi \equiv \Phi^{(1)}$ and $\psi \equiv \Psi^{(1)}$
\bea
\nabla^2 \phi - 3 \H (\dot \phi + \H \psi) & =& - 4 \pi G a^2 \, (T^0_{\ 0} - \bar T^0_{\ 0})\, , \label{equ:lin1}\\
\bigl[ \dot \phi + \H \psi \bigr]_{,i}  & =&  4 \pi G a^2 \,T^i_{\ 0} \, , \label{equ:mom}\\
\ddot \phi + \H (2\dot \phi + \dot \psi) + (\H^2 +2 \dot \H) \psi - \frac{2}{3} \nabla^2 (\phi-\psi) &= & \frac{4\pi G a^2}{3}\, (T^i_{\ i} - \bar T^i_{\ i})\, ,  \label{equ:lin11}\\
\partial_i \partial_j \left[ (\phi - \psi)_{,ij} - \frac{1}{3} \delta_{ij} \nabla^2 (\phi - \psi)\right] & =& 8\pi G a^2 \,\partial_i \partial_j \left[ T^i_{\ j} - \frac{1}{3} \delta^i_j T^k_{\ k} \right]\, , \label{equ:lin2}
\eea
where we have subtracted the background
\beq
{\cal H}^2 = - \frac{8\pi G a^2}{3} \bar T^0_{\ 0}\, , \qquad {\cal H}^2 + 2 \dot {\cal H} = - \frac{8\pi G a^2}{3} \bar T^i_{\ i}\, .
\eeq
In addition, the energy-momentum tensor may be linearized in the fluid three-velocities $v^i$ and the density contrast $\delta$. The Einstein equations then simplify considerably: {\it e.g.}~for a background with constant equation of state $w \equiv \bar p/\bar \rho$, and in the absence of anisotropic stress ($\Sigma_{ij}=0\ \ \Rightarrow \ \ \phi=\psi$) the Newtonian potential obeys the following evolution equation
\beq
\label{equ:phi1st}
\ddot \phi + 3 {\cal H} (1+ w) \dot \phi - w \nabla^2 \phi = 0\, ,
\eeq
where we have used Eqns.~(\ref{equ:lin1}) and (\ref{equ:lin11}) and $\delta p = c_s^2 \, \delta \rho$ with $c_{s}^2 = w$.
In Fourier space Eqn.~(\ref{equ:phi1st}) becomes
\beq
\ddot \phi_{\bf k} + \frac{6(1+w)}{1+ 3w} \frac{1}{\eta} \dot \phi_{\bf k} + w k^2 \phi_{\bf k} = 0\, , \label{equ:phi1st2}
\eeq
which has the following exact solution
\beq
\phi_{\bf k}(\eta) = y^\alpha \left[ C_1(k) J_\alpha(y) + C_2(k) Y_\alpha(y)\right]\, , \qquad y \equiv \sqrt{w} k \eta \, , \quad \alpha \equiv \frac{1}{2} \left( \frac{5+3w}{1+3w} \right)\, ,
\eeq
where $J_\alpha$ and $Y_\alpha$ are Bessel functions of order $\alpha$.
During the matter-dominated era, $w=0$, this becomes
\beq
\phi_{\bf k}(\eta) = C_1(k) + \frac{C_2(k)}{y^5}\, ,
\eeq
whereas during the radiation-dominated era, $w=\frac{1}{3}$, we find
\beq
\phi_{\bf k}(\eta) = \frac{1}{y^2} \left[C_1(k) \Bigl( \frac{\sin y}{y} - \cos y\Bigr) + C_2(k)\Bigl( \frac{\cos y}{y} + \sin y \Bigr)\right]\, .
\eeq
In both cases the decaying mode may be dropped by setting $C_2(k) \equiv 0$.
For a radiation-dominated background, 
the Newtonian potential is time-independent on superhorizon scales, $\lim_{k\eta \to 0} \phi_{\bf k}(\eta) = C_1(k)$, but decays on subhorizon scales. In contrast, during matter-domination
the growing mode linear gravitational potential is time-independent on all scales, with a spatial profile $\phi(\x)$ given by the Poisson equation 
\beq
\label{equ:PoissonX}
\nabla^2 \phi - 3 {\cal H}^2 \phi = 4 \pi G a^2 \bar \rho \delta\, ,
\eeq
or 
\beq
- k^2 \phi_\k = \frac{3}{2} \H^2 \delta_\k + 3 \H^2 \phi_\k\, .
\eeq
 The linear velocity is determined by the gradient of the potential, see Eqn.~(\ref{equ:mom}),
\beq
\label{equ:linvel}
v_i = - \frac{2}{3{\cal H}} \phi_{,i}\, .
\eeq
Using the scalar velocity potential, $v_i = \partial_i \tilde v$, Eqns.~(\ref{equ:PoissonX}) and (\ref{equ:linvel}) may be combined into
\beq
\label{equ:PoissonXX}
-k^2 \phi_\k = \frac{3}{2} \H^2 d_\k\, , 
\eeq
where
\beq
d_\k \equiv \delta_\k -  3 \H \tilde v_\k
\eeq
is the gauge-invariant linear density perturbation in the fluid rest frame.
Eqn.~(\ref{equ:PoissonXX}) is valid on all scales, while the Newtonian approximation, $- k^2 \phi_\k  \approx \frac{3}{2} \H^2 \delta_\k$, is only valid for scales that are much smaller than the horizon.

\vskip 6pt
\noindent
{\sl Second-order theory.} \hskip 8pt
The non-linear Einstein equations can be written in a form that is very similar to the linear Eqns.~(\ref{equ:lin1})--(\ref{equ:lin2}) if we replace the stress-energy tensor $T^\mu_{\ \nu}$ by the stress-energy pseudo-tensor $\tau^\mu_{\ \nu}$:
\beq
T^\mu_{\ \nu} \ \ \ \Rightarrow  \ \ \ \tau^\mu_{\ \nu} \ \equiv \ T^\mu_{\ \nu} - \frac{(G^\mu_{\ \nu})^{\rm NL}}{8\pi G}\, .
\eeq
The Einstein equations for
$\Phi \equiv \Phi^{(1)} + \frac{1}{2} \Phi^{(2)}$ and $\Psi \equiv \Psi^{(1)} + \frac{1}{2} \Psi^{(2)}$ then become, 
\beq
(G^\mu_{\ \nu})^{\rm L} = 8\pi G (\tau^\mu_{\ \nu} - \bar \tau^\mu_{\ \nu})\, ,
\eeq
or explicitly,
\bea
\nabla^2 \Phi - 3 \H (\dot \Phi + \H \Psi) & =& - 4 \pi G a^2 \, (\tau^0_{\ 0} - \bar \tau^0_{\ 0})\, , \label{equ:NL1}\\
\bigl[ \dot \Phi + \H \Psi \bigr]_{,i}  & = & 4 \pi G a^2 \,\tau^i_{\ 0} \, , \label{equ:tau0i}\\
\ddot \Phi + \H (2\dot \Phi + \dot \Psi) + (\H^2 +2 \dot \H) \Psi - \frac{2}{3} \nabla^2 (\Phi-\Psi) &=&  \frac{4\pi G a^2}{3}\, (\tau^i_{\ i} - \bar \tau^i_{\ i})\, , \label{equ:NLT}\\
\partial_i \partial_j \left[ (\Phi - \Psi)_{,ij} - \frac{1}{3} \delta_{ij} \nabla^2 (\Phi - \Psi)\right] & =& 8\pi G a^2 \,\partial_i \partial_j \left[ \tau^i_{\ j} - \frac{1}{3} \delta^i_j \tau^k_{\ k} \right]\, . \label{equ:NL2}
\eea
This defines the dynamics of long-wavelength scalar fluctuations sourced by products of short-wavelength fluctuations.
All gravitational non-linearities have been moved to the r.h.s.~of the Einstein equations.
Eqns.~(\ref{equ:NL1}) and (\ref{equ:tau0i}) can be combined into the Poisson equation,
\beq
\nabla^2 \Phi = 4\pi G a^2 \left[ 3 \H \nabla^{-2} \partial_i \tau^i_{\ 0} - (\tau^0_{\ 0} - \bar \tau^0_{\ 0})\right]\, .
\eeq 
At long wavelengths, $\tau^\mu_{\ \nu}$ is conserved by virtue of the linearized Bianchi identity
\beq
\bar \nabla_\mu (G^\mu_{\ \nu})^{\rm L} + \nabla_\mu^{\rm L} \, \bar G^\mu_{\ \nu} = 0\, .
\eeq
Using $(G^\mu_{\ \nu})^{\rm L} = 8\pi G (\tau^\mu_{\ \nu} - \bar \tau^\mu_{\ \nu})$ and computing the linear covariant derivative action on the background $\bar G^\mu_{\ \nu}$, this is
 \beq
\label{equ:BFRW3}
\bar \nabla_\mu  \tau^\mu_{\ \nu}  =- \bar \rho \left[ 3 \dot \Phi \, \delta_{\nu, 0} + \Psi_{,i} \, \delta_{\nu, i}\right] \, .
\eeq
Notice that if one wished to obtain this equation from the Newtonian stress tensor we found in \S\ref{sect:Newton}, it would be necessary to keep subleading terms in $x^i$ in the expression for $\tau_{\mu\nu}$.

\subsubsection{Conservation Equations}

Finally, we give the velocity expansion of the continuity equations, $\nabla_\mu T^\mu_{\ \nu} = 0$.
At order $v^2$ the mass conservation equation and the Euler equation are
\begin{equation}
\label{equ:N11}
\frac{1}{a^3} \frac{d}{d \eta} (a^3 \rho_m) + \frac{1}{a} \nabla_i (\rho_m v^i) = 0\, , \end{equation} and
\begin{equation}
\label{equ:N12}
\frac{d}{d\eta} (a v_i) + v^j \nabla_j v_i = - \nabla_i \Psi\, ,
\end{equation}
where $\rho_m$ was defined in Eqn.~(\ref{equ:rhoSTAR}).
A common way to organize the post-Newtonian expansion around Minkowski space is as an expansion of the field equations in powers of $c^{-1}$ (see {\it e.g.}~Chandrasekhar \cite{Chandrasekhar:1990uf}).
For cosmology this post-Newtonian expansion at order 1PN ($c^{-2}$) was performed in \cite{Hwang:2005mg}.
Our velocity expansion at order $v^2$ effectively keeps 1PN order in the continuity equation, but Newtonian 0PN in the Euler equation.

\subsection{Second-Order Gauge Transformations}
\label{sec:GT}

To relate the results obtained in Poisson gauge to other gauges we need to consider
second-order gauge transformations.
As a reference for the reader we here cite those transformations for metric perturbations.
More details may be found in the fantastically useful review by Malik and Wands~\cite{Malik:2008im}.

The most general set of metric perturbations can be written as
\beq
\delta g_{00} = - 2 a^2 \psi \, , \qquad \delta g_{0i} = a^2 B_i\, , \qquad \delta g_{ij} = 2 a^2 C_{ij}\, ,
\eeq
where
\begin{align}
B_i &= B_{,i} + \omega_i\, , \\
C_{ij} &= - \phi \delta_{ij} + E_{,ij} + F_{(i,j)} + \frac{1}{2} \chi_{ij}\, .
\end{align}
Here, $\psi$, $B$, $\phi$, and $E$ are scalar perturbations, $\omega_i$ and $F_i$ are vector perturbations, and $\chi_{ij}$ is a tensor perturbation.
In this parameterization the Poisson gauge is defined by
$B=E=F_i =0$.

Consider the second-order coordinate transformation
\beq
\tilde x^\mu = x^\mu + \xi^\mu_{(1)} + \frac{1}{2} (\xi^\mu_{(1),\nu} \xi^\nu_{(1)} + \xi^\mu_{(2)} )\, .
\eeq
At every order the four-vectors $\xi_{(r)}^\mu$, ($r=1,2$), can be decomposed into scalar and vector parts
\beq
\xi^0_{(r)} = \alpha_{(r)}\, , \qquad \xi^{i}_{(r)} = \partial^i \beta_{(r)} + d^i_{(r)}\, , \qquad {\rm with} \quad \partial_i d^i_{(r)} = 0\, .
\eeq
Second-order gauge transformations of tensors are written in a compact form by the use of Lie derivatives.
Any tensor ({\it e.g.}~the metric $g$) transforms as
\beq
\tilde g = g + {\cal L}_{\xi_{(1)}} g + \frac{1}{2} ({\cal L}^2_{\xi_{(1)}} + {\cal L}_{\xi_{(2)}}) g\, ,
\eeq
where we have suppressed indices.\footnote{Explicitly, the Lie derivative of the metric tensor is $$
{\cal L}_\xi g_{\mu \nu} = g_{\mu \nu, \lambda} \xi^\lambda + g_{\mu \lambda} \xi^\lambda_{\ , \nu} + g_{\nu \lambda} \xi^\lambda_{\ , \mu}\, . 
$$
From this follows straightforwardly an expression for ${\cal L}_\xi^2 g_{\mu \nu} = {\cal L}_\xi ({\cal L}_\xi g_{\mu \nu})$.}
Using
\beq
g = \bar g + \delta g_{(1)} + \frac{1}{2} \delta g_{(2)}\, ,
\eeq
this implies
\bea
\delta \tilde g_{(1)} &=& \delta g_{(1)} + {\cal L}_{\xi_{(1)}} \bar g\, ,  \\
\delta \tilde g_{(2)} &=& \delta g_{(2)} + {\cal L}_{\xi_{(2)}} \bar g +{\cal L}_{\xi_{(1)}}^2 \bar g + 2 {\cal L}_{\xi_{(1)}} \delta g_{(1)}\, . \label{equ:L2x}
\eea

\subsubsection{First-Order Transformations}

We find the following first-order transformations:
\begin{itemize}
\item scalar perturbations
\bea
\tilde \phi_1 &=& \phi_1 - {\cal H} \alpha_1 \, ,\\
\tilde \psi_1 &=& \psi_1 + {\cal H} \alpha_1 + \dot \alpha_1 \, ,\\
\tilde B_1 &=& B_1 - \alpha_1 + \dot \beta_1 \, ,\\
\tilde E_1 &=& E_1 + \beta_1\, ,
\eea
\item vector perturbations
\bea
\tilde \omega_1^i &=& \omega_1^i - \dot d_1^i \, ,\\
\tilde F_1^i &=& F_1^i + \dot d_1^i\, ,
\eea
\item tensor perturbations
\bea
\tilde \chi_{1 ij} &=& \chi_{1 ij}\, .
\eea
\end{itemize}
Furthermore, the scalar shear potential, $\sigma_1 \equiv \dot E_1 - B_1$, and the momentum scalar, $q_1 \equiv v_1 + B_1$ transform as
\bea
\tilde \sigma_1 &=& \sigma_1 + \alpha_1\, ,\\
\tilde q_1 &=& q_1 - \alpha_1\, .
\eea

\subsubsection{Second-Order Transformations}
Second-order gauge transformations are considerably more involved since they receive contributions from products of first-order transformations.

We define 
\bea
\Xi &\equiv& \alpha_1 \left[ \ddot \alpha_1 + 5 {\cal H} \dot \alpha_1 + (\dot {\cal H} + 2 {\cal H}) \alpha_1 + 4 {\cal H} \psi_1 + 2 \dot \psi_1 \right] \nonumber \\
&& + 2 \dot \alpha_1 (\dot \alpha_1 +  2 \psi_1) + \xi_{1k} (\dot \alpha_1 + {\cal H} \alpha_1 + 2 \psi_1)_{,k} + \dot \xi_{1}^k \left[\alpha_{1, k} - 2 B_{1k} - \dot \xi_{1k}\right]\, , \label{equ:1}
\eea
\bea
\Theta_{i} &\equiv& 2 \left[ (2{\cal H} B_{1i} + \dot B_{1i}) \alpha_{1}+ B_{1i, k} \xi^k_1 - 2 \psi_1 \alpha_{1,i} + B_{1k} \xi^k_{1,i} + B_{1i} \dot \alpha_1 + 2 C_{1ik} \dot \xi_1^k \right] \nonumber \\
&& + 4 {\cal H} \alpha_1 (\dot \xi_{1i} - \alpha_{1,i}) + \dot \alpha_1 (\dot \xi_{1i} - 3 \alpha_{1,i}) + \alpha_1 (\ddot \xi_{1i} - \dot \alpha_{1,i}) \nonumber\\
 && + \dot \xi_1^k (\xi_{1i} + 2 \xi_{1k,i}) + \xi^k_1 (\dot \xi_{1i,k} - \alpha_{1,ik}) - \alpha_{1,k} \xi^k_{1,i}\, , \label{equ:2}
\eea
and
\bea
\Upsilon_{ij} &\equiv& 2 \left[ ({\cal H}^2 + 2 \dot {\cal H}) \alpha_1^2 + {\cal H}(\alpha_1 \dot \alpha_1 + \alpha_{1,k} \xi_1^k)\right] \delta_{ij} \nonumber \\
&& - 2 \alpha_{1,i} \alpha_{1,j} + 4 \alpha_1 (\dot C_{1 ij} + 2 {\cal H} C_{1 ij} ) + 4 B_{1 (i} \alpha_{,j)} \nonumber \\
&& 4 \left[C_{1ij,k} \xi_1^k + 2 C_{1 k ( i} \xi^k_{1 j)} \right] + 8 {\cal H} \alpha_1 \xi_{1 (i, j)} + 2 \xi_{1 k, i} \xi_{1,j}^k \nonumber \\
&& + 2\alpha_1 \dot \xi_{1 (i,j)} + 2 \xi_{1 (i, j)k} \xi_1^k  + 2\xi_{1 (i, k} \xi^k_{1, j)}+ 2 \dot \xi_{1 (i} \alpha_{1, j)}  \, . \label{equ:3}
\eea
\noindent
The second-order metric transformations then can be written as follows:
\begin{itemize}
\item scalar perturbations
\bea
\tilde \phi_2 &=& \phi_2 - {\cal H} \alpha_2 - \frac{1}{4} \Upsilon^k_{\ k} + \frac{1}{4} \nabla^{-2} \Upsilon^{ij}_{\ \ , ij}\, , \\
\tilde \psi_2 &=& \psi_2 + {\cal H} \alpha_2 + \dot \alpha_2 + \Xi \, ,\\
\tilde B_2 &=& B_2 - \alpha_2 + \dot \beta_2 + \nabla^{-2} \Theta^k_{\ , k} \, ,\\
\tilde E_2 &=& E_2 + \beta_2 + \frac{3}{4} \nabla^{-2} \nabla^{-2} \Upsilon^{ij}_{\ \ , ij} - \frac{1}{4} \nabla^{-2} \Upsilon^{k}_{\ k}\, ,
\eea
\item vector perturbations
\bea
\tilde \omega_{2i} &=& \omega_{2i} - \dot d_{2i} - \Theta_i + \nabla^{-2} \Theta^k_{\ , ki} \, , \\
\tilde F_{2i} &=& F_{2i} + d_{2i} + \nabla^{-2} \Upsilon_{i\ , k}^{\ k} - \nabla^{-2} \nabla^{-2} \Upsilon^{kl}_{\ \ , kl i}\, ,
\eea
\item tensor perturbations
\bea
\tilde \chi_{2ij} &=& \chi_{2ij} + \Upsilon_{ij} + \frac{1}{2} \left( \nabla^{-2} \Upsilon^{kl}_{\ \ , kl} - \Upsilon^k_{\ k}\right) \delta_{ij} + \frac{1}{2} \nabla^{-2} \nabla^{-2} \Upsilon^{kl}_{\ \ , kl ij}\, , \nonumber \\
&& + \frac{1}{2} \nabla^{-2} \Upsilon^k_{\ k,ij} - \nabla^{-2} \left( \Upsilon_{i \  , kj}^{\ k}+ \Upsilon_{j \ , ki}^{\ k}\right)\, .
\eea
\end{itemize}
In the case of a pure time-shift, $\xi_i=0$, Eqns.~(\ref{equ:1}), (\ref{equ:2}) and (\ref{equ:3}) become
\beq
\Xi = \alpha_1 \left[ \ddot \alpha_1 + 5 {\cal H} \dot \alpha_1 + (\dot {\cal H} + 2 {\cal H}) \alpha_1 + 4 {\cal H} \psi_1 + 2 \dot \psi_1 \right]  + 2 \dot \alpha_1 (\dot \alpha_1 +  2 \psi_1) \, , 
\eeq
\beq
\Theta_{i} = 2 \left[ (2{\cal H} B_{1i} + \dot B_{1i}) \alpha_{1} - 2 \psi_1 \alpha_{1,i}  + B_{1i} \dot \alpha_1  \right]  - 4 {\cal H} \alpha_1  \alpha_{1,i} - 3 \dot \alpha_1 \alpha_{1,i} - \alpha_1 \dot \alpha_{1,i}\, ,
\eeq
and
\beq
\Upsilon_{ij}  = 2 \left[ ({\cal H}^2 + 2 \dot {\cal H}) \alpha_1^2 + {\cal H} \alpha_1 \dot \alpha_1\right] \delta_{ij} 
 - 2 \alpha_{1,i} \alpha_{1,j} + 4 \alpha_1 (\dot C_{1 ij} + 2 {\cal H} C_{1 ij} ) + 4 B_{1 (i} \alpha_{,j)} \, .
\eeq

\subsection{Covariance of the Effective Theory}
\label{sec:gauge}

Next, we discuss the gauge conditions that have to be satisfied in order to define the effective fluid in the long-wavelength limit.

\subsubsection{Long-Wavelength Limit of the Metric}

Let us split the metric into a background $\bar g_{\mu \nu}(k \to 0)$, long-wavelength perturbations $H_{\mu \nu}(k \ll \Lambda)$ and short-wavelength perturbations $h_{\mu \nu}(k \gg \Lambda)$, {\it i.e.}
\beq
g_{\mu \nu} = \bar g_{\mu \nu} + H_{\mu \nu} + h_{\mu \nu} \, . 
\eeq
The long-wavelength limit, $k \to 0$, extracts the $H_{\mu \nu}$ perturbations and projects out the $h_{\mu \nu}$ perturbations
\beq
\tilde g_{\mu \nu} \equiv \lim_{k \ll \Lambda} g_{\mu \nu} = \bar g_{\mu \nu} + H_{\mu \nu}\, .
\label{equ:metricL}
\eeq
At linear order in $H_{\mu \nu}$ and quadratic order in $h_{\mu \nu}$ we find
\bea
\sqrt{-g} &=& \sqrt{- \bar g} \left( 1 + \frac{1}{2} H \right) \left( 1 + \frac{1}{2} h + \frac{1}{8} h^2 - \frac{1}{4} h^\alpha_\beta h^\beta_\alpha \right) \, ,\\
g^{\mu \nu} &=& \bar g^{\mu \nu} - H^{\mu \nu} - h^{\mu \nu} + h^{\mu \alpha} h^\nu_\alpha\, ,
\eea
which in the long-wavelength limit becomes
\bea
\lim_{k \ll \Lambda} \sqrt{-g} &=& \sqrt{- \bar g} \left( 1 + \frac{1}{2} H  + \lim_{k \ll \Lambda} \left[\frac{1}{8} h^2 - \frac{1}{4} h^\alpha_\beta h^\beta_\alpha \right]\right)\, , \\
\lim_{k\ll \Lambda} g^{\mu \nu} &=& \bar g^{\mu \nu} - H^{\mu \nu}  + \lim_{k \ll \Lambda}(h^{\mu \alpha} h^\nu_\alpha)\, . \label{equ:metricL2}
\eea
Notice that the long-wavelength limit of the covariant metric component $g_{\mu \nu}$, Eqn.~(\ref{equ:metricL}), is {\it not} the inverse of the  long-wavelength limit of the contravariant metric component $g^{\mu \nu}$, Eqn.~(\ref{equ:metricL2}), unless the long-wavelength limit of the product of short-wavelength perturbations is sufficiently small. 
In other words, to define a {\it covariant effective theory in the long-wavelength limit} we require the following condition to hold
\beq
\label{equ:metriccon}
\lim_{k \ll \Lambda} |H^{\mu \nu}_{(2)}| \gg \lim_{k \ll \Lambda} h^{\mu \alpha}_{(1)} h^{\nu \beta}_{(1)} \, \bar g_{\alpha \beta} \, , 
\eeq
{\it i.e.}~it needs to be the case that the $k \to 0$ limit of the second-order piece, $H_{(2)}$ is bigger than the $k \to 0$ limit of the square of the first-order piece, $h^2_{(1)}$.
In this case the long-wavelength metric is well-defined and can be used to construct covariant actions such as the action of a scalar field $\varphi$ living in this effective spacetime $\tilde g_{\mu \nu}$,
\beq
\sqrt{-\tilde g} \tilde g^{\mu \nu} \partial_\mu \varphi \partial_\nu \varphi\, .
\eeq
We will now discuss in which gauges the condition (\ref{equ:metriccon}) is satisfied.

\subsubsection{Poisson Gauge}

Let us first demonstrate that Poisson gauge is a `good gauge' in the sense defined above.
In Poisson gauge the scalar perturbations of the metric are
\bea
\delta g_{00} &=& - 2 a^2 \left[\Psi^{(1)}_{\rm P} + \frac{1}{2} \Psi^{(2)}_{\rm P} \right] \, ,\\
\delta g_{ij} &=&  - 2 a^2 \left[\Phi^{(1)}_{\rm P} + \frac{1}{2} \Phi^{(2)}_{\rm P} \right]  \delta_{ij} \, ,\\
\delta g_{0i} &=& 0\, ,
\eea
where during pure matter-dominance we find (see also Refs.~\cite{Boubekeur:2008kn,matta})
\beq
 \Psi^{(1)}_{\rm P} = \Phi^{(1)}_{\rm P} \equiv \phi = const.\, ,
\eeq
and
\beq
\Psi^{(2)}_{\rm P} = \frac{2}{7} \frac{1}{{\cal H}^2} {\cal S} + {\cal C}_\psi \, , \qquad
\Phi^{(2)}_{\rm P} = \frac{2}{7} \frac{1}{{\cal H}^2} {\cal S} + {\cal C}_\phi\, .
\eeq
Here, we have defined
\beq
{\cal S} \equiv \frac{1}{3} \nabla^{-2} \left[2 (\partial_i \partial_j \phi)^2 + 5 (\nabla^2 \phi)^2 + 7 \partial_i \phi \partial_i \nabla^2 \phi \right] \, ,
\eeq
and the constants ${\cal C}_\psi$ and ${\cal C}_\phi$ are set by initial conditions \cite{Boubekeur:2008kn,matta}.
For the following estimates we will use
\beq
\Psi^{(2)}_{\rm P} \sim \Phi^{(2)}_{\rm P} \sim \frac{\phi \nabla^2 \phi}{{\cal H}^2} \sim  \frac{|\partial_i \phi|^2}{{\cal H}^2} \, .
\eeq
It is then easy to check that the condition (\ref{equ:metriccon})
implies
\beq
|\partial_i \phi|^2  \gg {\cal H}^2 \phi^2\, .
\label{equ:Pcon}
\eeq
Eqn.~(\ref{equ:Pcon}) is satisfied if we couple two subhorizon perturbations with $k > \H$.
This shows that integrating out subhorizon modes in Poisson gauge gives a well-defined long-wavelength metric.

\subsubsection{Arbitrary Gauges \label{sect:arbitrary_gauges}}

Of course, we don't expect the condition (\ref{equ:metriccon}) to hold in all gauges.
We therefore now exhibit second-order gauge transformations of our Poisson gauge results to define a set of `good gauges' in which a consistent effective fluid description is possible.

\vskip 6pt
\noindent
{\sl Second-order gauge transformations.} \hskip 8pt
Recall the gauge transformations of the metric tensor
\bea
\delta \tilde g_{(1)} &=& \delta g_{(1)} + {\cal L}_{\xi_{(1)}} \bar g\, ,  \\
\delta \tilde g_{(2)} &=& \delta g_{(2)} + {\cal L}_{\xi_{(2)}} \bar g +{\cal L}_{\xi_{(1)}}^2 \bar g + 2 {\cal L}_{\xi_{(1)}} \delta g_{(1)}\, . \label{equ:L2}
\eea
The most general second-order gauge transformations were presented in \S\ref{sec:GT}.
Here, we will consider the special case of a pure time-shift, $\xi^i = 0$, and focus on the 
${\cal L}_{\xi_{(1)}}^2 \bar g$ term in Eqn.~(\ref{equ:L2}) (inspection of the results of \S\ref{sec:GT} shows that this captures the leading effect in an expansion in scale-scale gradients).
For the FRW background
\beq
\bar g_{00} = - a^2\, , \quad \bar g_{0i} = 0\, , \quad \bar g_{ij} = a^2 \delta_{ij}\, ,
\eeq
we may compute  ${\cal L}_{\xi_{(r)}} \bar g$ and ${\cal L}_{\xi_{(1)}}^2 \bar g$, using ${\cal L}_\xi \bar g_{\mu \nu} = \bar g_{\mu \nu, \sigma} \xi^\sigma + \xi^\sigma_{, \mu} \bar g_{\nu \sigma} + \xi^\sigma_{, \nu} \bar g_{\mu \sigma}$.
We find
\bea
{\cal L}_{\xi} \bar g_{00} &=& - 2 a^2 [\dot \alpha + {\cal H} \alpha]\, , \\
{\cal L}_{\xi} \bar g_{0i} &=&  - a^2 \alpha_{,i}\ , \\
{\cal L}_{\xi} \bar g_{ij} &=& + 2 a^2  {\cal H} \alpha \, ,
\eea
and 
\bea
{\cal L}_{\xi_{(1)}}^2 \bar g_{00} &=& - 2 a^2 \left[ \alpha [\ddot \alpha + 5 {\cal H} \dot \alpha + (\dot {\cal H} + {\cal H}^2) \alpha] + 2 \dot \alpha^2\right]\, ,\\
{\cal L}_{\xi_{(1)}}^2 \bar g_{0i} &=& -a^2 \Bigl[ \alpha [ \dot \alpha_{,i} + 4 {\cal H}\alpha_{,i} ]  + 3 \dot \alpha  \alpha_{,i} \Bigr] \, ,\\
{\cal L}_{\xi_{(1)}}^2 \bar g_{ij} &=& - 2 a^2 \Bigl[ - \alpha[ (\dot {\cal H} + {\cal H}^2) \alpha + {\cal H} \dot \alpha] 
  + \frac{1}{3} (\alpha_{,i})^2 
  \Bigr]  \delta_{ij} + \cdots \, .
\eea
We will be interested in first-order time shifts on very small (subhorizon) scales. In that case the dominant terms are the ones with the maximal number of spatial gradients.
In this gradient expansion we find
\beq
{\cal L}_{\xi} \bar g_{00} =  a^2  {\cal O}( {\cal H}\alpha) \, , \quad {\cal L}_{\xi} \bar g_{0i} =  - a^2 \alpha_{,i} \, , \quad
{\cal L}_{\xi} \bar g_{ij} = a^2  {\cal O}({\cal H} \alpha)\, ,
\eeq
and
\bea
{\cal L}_{\xi_{(1)}}^2 \bar g_{00} &=& a^2 {\cal O}({\cal H}^2 \alpha^2)\, , \quad
{\cal L}_{\xi_{(1)}}^2 \bar g_{0i} = a^2 {\cal O}({\cal H} \alpha \alpha_{,i}) \, , \quad \\
{\cal L}_{\xi_{(1)}}^2 \bar g_{ij} &=&- 2 a^2 \Bigl[ 
   \frac{1}{3} (\alpha_{,i})^2  + {\cal O}({\cal H}^2 \alpha^2)
  \Bigr]  \delta_{ij} + \cdots\, .
\eea

\vskip 6pt
\noindent
{\sl Second-order time-shift and `good gauges'.} \hskip 8pt
Let us now go from Poisson gauge to an arbitrary gauge using the second-order gauge transformations discussed above.

We find, {\it e.g.}
\beq
\delta \tilde g_{0i}^{(1)}\subset - a^2 \alpha_{,i} \qquad {\rm and} \qquad
\delta \tilde g_{ij}^{(2)} \subset - 2 a^2 \Bigl[ \Phi^{(2)}_{\rm P} +
   \frac{1}{3} (\alpha_{,i})^2  
  \Bigr]  \delta_{ij} \, .
\eeq
Consider then
\beq
|\tilde H_{i j}^{(2)}| \subset \delta \tilde g_{ij}^{(2)}\, ,
\eeq
and
\beq
\tilde h_{i \alpha}^{(1)} \tilde h_{j \beta}^{(1)} \, \bar g^{\alpha \beta} \subset - a^{-2} \tilde h_{0i}^{(1)} h_{0j}^{(1)} = a^2 \alpha_{,i} \alpha_{,j}\, .
\eeq
The condition
\beq
|\tilde H_{ij}^{(2)}| \gg \tilde h_{i \alpha}^{(1)} \tilde h_{j \beta}^{(1)} \, \bar g^{\alpha \beta}\, ,
\eeq
now implies
\beq
\label{equ:cond}
\Phi^{(2)}_{\rm P} \gg (\alpha_{,i})^2\, .
\eeq
Only gauges that are connected to Poisson gauge in this way have a well-defined long-wavelength limit for the metric after integrating out short-wavelength modes.
This means that we only reach a `good gauge' from Poisson gauge if  $|\partial_i \alpha|^2$ does not give the dominant effect in the new metric perturbation, {\it i.e.}~$ {\cal H}^2 |\partial_i \alpha|^2 \ll |\partial_i \phi|^2$.
A popular gauge that violates the condition (\ref{equ:cond}) is the {\it comoving gauge}, $T^i_{\ 0}=0$.
In that case,
${\cal H} \alpha \sim \phi$, and
\beq
|\tilde H_{ij}^{(2)}| \sim \tilde h_{i \alpha}^{(1)} \tilde h_{j \beta}^{(1)} \, \bar g^{\alpha \beta} \, .
\eeq
In comoving  gauge we therefore cannot define a unique long-wavelength metric
after integrating our short-wavelength fluctuations. This is related to the breakdown of that gauge on small-scales. Analogous considerations apply to the second-order metric in {\it uniform density gauge}.

\subsubsection{Background Field Method}

An elegant alternative to ensure covariance of the theory in the long-wavelength limit is the {\it background field method} (BFM).
The background field method is a technique for quantizing gauge theories without losing explicit gauge invariance of the effective action (see {\it e.g.}~Ref.~\cite{PeskinSchroeder}).
The fields of the classical Lagrangian ${\cal L}$ are decomposed into background fields $\bar \varphi$ (analogous to $\bar g$) and quantum fields $\varphi$ (analogous to $h$).
While the background fields are treated as external sources, only the quantum fields are variables of integration in the functional integral. A gauge-fixing term is added which breaks only the invariance with respect to quantum gauge transformations, but retains the invariance of the functional integral with respect to background field gauge transformations. From the functional integral an effective action $S_{\rm eff}[\bar \varphi]$ for the background fields is derived, which by construction is invariant under gauge transformations of the background fields and thus gauge-invariant.
In this formalism the effective action is manifestly covariant after integrating out the short-wavelength modes. 

The BFM has been applied to perturbative General Relativity by Feynman, DeWitt, 't Hooft and Veltman, {\it e.g.}~Ref.~\cite{'tHooft:1974bx}. As before, the metric is split into a smooth background, $\tilde g_{\mu \nu}$, and short-wavelength perturbations, $h_{\mu \nu}$. Ultimately, we want to derive the effective action for $\tilde g_{\mu \nu}$ by integrating out quadratic fluctuations in $h_{\mu \nu}$.
The key element of the BFM is a gauge-fixing term that is covariant with respect to the long-wavelength metric $\tilde g_{\mu \nu}$,
\beq
\label{gf}
{\cal L}_{\rm gauge} = \widetilde{\Gamma}_\mu  \widetilde{\Gamma}^\mu\, , \qquad \widetilde{\Gamma}_\mu \equiv \widetilde{D}^\nu \Bigl(h_{\mu \nu} - \frac{1}{2} \tilde g_{\mu \nu} h \Bigr)\, ,
\eeq
where $\widetilde{D}_\mu$ is the covariant derivative with respect to the background metric $\tilde g_{\mu \nu}$.
This ensures that the effective action for $\tilde g_{\mu \nu}$ is covariant.
It may be shown that at leading order in our gradient expansion the BFM with the gauge-fixing Lagrangian (\ref{gf}) is equivalent to working in Poisson gauge, $(\widetilde{\Gamma}_\mu)_{\rm P} \approx 0$.

\newpage

\section{Dissipative Fluid Dynamics}
\label{sec:FluidReview}

The theory of imperfect fluids is reviewed comprehensively in the books by Weinberg~\cite{Weinberg} or Landau and Lifshitz~\cite{LL}.
For the benefit of the reader we here collect the main elements of that analysis.
In \S\ref{sec:ImperfectFluid} we showed how this connects to our theory of the gravitational fluid. 

For imperfect fluids the stress tensor may be written as
\beq
\tau_{ij} = \rho u_i u_j + (p-\zeta\theta) \gamma_{ij} + \Sigma_{ij}\, ,
\eeq
where $\Sigma_{ij}$ is the viscous stress tensor describing dissipative dynamics.
At leading order in an expansion in spatial derivatives an ansatz for $\Sigma_{ij}$ that is consistent with the second law of thermodynamics is~\cite{Weinberg}
\beq
\label{equ:shear}
\Sigma_{ij} = -  \eta \sigma_{ij}  \, , 
\eeq
where $\eta > 0$ and $\zeta > 0$ are the coefficients of shear and bulk viscosity, respectively and $\sigma_{ij} \equiv v_{(i,j)} -\frac{1}{3} \delta_{ij} v_{k,k}$. 
With Eqn.~(\ref{equ:shear}) the Euler equation becomes the non-relativistic Navier-Stokes equation. In general, for fluids that are locally at thermal equilibrium one also has terms proportional to the temperature gradient, weighted by the heat conduction. Though we expect to have dissipative terms also in $\tau_{0i}$ that can in principle correspond to temperature gradients, our effective fluid is clearly not in thermal equilibrium, so it is not clear what these terms would correspond to in our case. Fortunately, we do not need such an interpretation: as we saw in the main text, to describe dissipative effects in our case, we just need the higher-derivative expansion of the stress tensor---{\it cf.}~Eqn.~(\ref{equ:exp1}).

In an
adiabatic fluid the pressure is a unique function of the energy density, $p = p(\rho)$,
and the
adiabatic sound speed is  fully determined by the equation of state $w=p/\rho$,
\beq
c_a^2 \equiv \frac{\dot p}{\dot \rho} = w -  \frac{\dot w}{3\H(1+w)}\, .
\eeq
In general, the pressure might not be a unique function of the energy density.
The sound speed is then defined as the ratio of pressure and density perturbations in the frame comoving with the fluid \beq
c_s^2 = \frac{\delta p}{\delta \rho}\, .
\eeq
The speed of sound may be extracted by correlating the pressure with the density contrast of a long-wavelength perturbation $\delta_\ell$,
\beq
\label{equ:csPT}
c_s^2 \equiv \frac{1}{\bar \rho} \frac{\langle \delta_\ell \, p \rangle}{\langle \delta_\ell \delta_\ell\rangle}\, .
\eeq
We define the scalar component $\sigma$ of the anisotropic stress tensor $\Sigma_{ij}$ as
\beq
(\bar \rho + \bar p)  \sigma  = \nabla^{-2}\left(\partial_i \partial_j - \frac{1}{3} \delta_{ij} \nabla^2\right) \Sigma^{ij} \, .
\eeq
We may relate $\sigma$ to the shear viscosity in Eqn.~(\ref{equ:shear}),
\beq
\label{equ:sigmatheta}
\sigma = - \frac{2 \eta}{3 \bar \rho} \, \theta \equiv \tilde c_{\rm vis}^2 \frac{\theta}{\H}\, ,
\eeq
where we introduced the tilde viscous speed of sound, $\tilde c_{\rm vis}$ (as for $c_{\rm vis}$, we stress that this is a dissipative term that does not induce propagating sound waves). 
We see that the parameter $\tilde c_{\rm vis}$ may be determined by correlating the anisotropic stress with the velocity divergence of a long-wavelength perturbation $\theta_\ell$,
\beq
\label{equ:cvisPT}
\tilde c_{\rm vis}^2 \equiv  \H \frac{\langle \theta_\ell \, \sigma \rangle}{\langle \theta_\ell \theta_\ell\rangle}\, .
\eeq
For $\tilde c_{\rm vis}^2 > 0$ this adds a dissipative term to Euler equation.

\newpage
\section{Estimates in Perturbation Theory}
\label{sec:estimates}

In this appendix we illustrate some of the ideas of this paper by presenting sample calculations in tree-level perturbation theory. In particular, we wish to show how the parameters of the long-wavelength effective theory can be computed if we know the theory that describes the dynamics of the short-wavelength universe. This is the standard procedure of matching the parameters between two effective theories that are valid at different scales. In this appendix, as an example, we will implement this procedure by assuming that perturbation theory is applicable to describe the small-scale dynamics in a dark matter universe.

\subsection{Fluid Parameters at Tree Level}

\vspace{0.5cm}
 \hrule
 \vskip 1pt
 \hrule \vspace{0.3cm}
 \small
\noindent 
{\bf Example}:\vskip 4pt

\noindent
{\sl Perturbation theory.}
\hskip 8pt
During the matter era, the method of perturbation theory is essentially to perform an expansion in the scale factor $a(\eta)$ \cite{Bernardeau:2001qr}\footnote{In a general cosmology, the scale factor expansion is replaced by an expansion in the growth factor $D_+(a)$.}
\bea
\delta(\k, \eta) &=& \sum_{n=1}^\infty a^n(\eta) \delta_n(\k) \, , \\
\theta(\k,\eta) &=& - \H(\eta) \sum_{n=1}^\infty a^n(\eta) \theta_n(\k)\, ,
\eea
where $\delta_n$ and $\theta_n$ are time-independent mode-coupling integrals over $n$ powers of the initial density field; {\it e.g.}~at second order,
\beq
\label{equ:delta2k}
\delta_2(\k) = \int_{\q_1} \int_{\q_2} \delta_{\rm D}(\k - \q_1 - \q_2) \, F_2(\q_1, \q_2) \delta_1(\q_1) \delta_1(\q_2)\, , 
\eeq
where
\beq
F_2(\q_1, \q_2) \equiv \frac{5}{7} + \frac{1}{2} \mu \Bigl(\frac{q_1}{q_2} + \frac{q_2}{q_1} \Bigr) + \frac{2}{7} \mu^2\, , \qquad \mu \equiv \frac{\q_1 \cdot \q_2}{q_1 q_2}\, .
\eeq
The angular integral of the kernel results in a pure number
\beq
\nu_2 \equiv \int \frac{d \Omega_{\q_1}}{2\pi} F_2(\q_1, \q_2) = \int_{-1}^1 d\mu \, F_2 = \frac{34}{21}\, .
\eeq
We furthermore define
\beq
\delta_\k^{(1)}(\eta) \equiv a(\eta) \delta_1(\k)\, , \qquad \delta_\k^{(2)}(\eta) \equiv a^2(\eta) \delta_2(\k)\, ,
\eeq
such that
\beq
\delta_\k = \delta_\k^{(1)} + \delta_\k^{(2)}\, .
\eeq

The convolution in Eqn.~(\ref{equ:delta2k}) captures the UV-IR coupling between long-wavelength modes and short-wavelength fluctuations.
Using these results we may evaluate the correlation functions in the definitions of $\bar p_{\rm eff}$, $c_s$ and $\tilde c_{\rm vis}$,
\beq
\bar p_{\rm eff} = \frac{1}{3} \lim_{k \to 0} \langle \tau \rangle\, , \qquad c_s^2 = \frac{1}{\bar \rho} \frac{\langle \delta_\ell p_{\rm eff }\rangle}{\langle \delta_\ell \delta_\ell \rangle}\, , \qquad \tilde  c_{\rm vis}^2 = \frac{\langle \delta_\ell \sigma_{\rm eff }\rangle}{\langle \delta_\ell \delta_\ell \rangle}\, ,
\eeq
where 
\beq
\sigma_{\rm eff} =  \frac{1}{\bar \rho} \frac{\partial_i \partial_j}{\partial^2} \langle [\hat \tau^i_{\ j}]^\Lambda \rangle\ .
\eeq
These formulas assume negligible bulk viscosity, which otherwise would contribute to
the trace of the stress tensor.
As we remarked in the main text, at tree level in perturbation theory it is impossible to distinguish between bulk and shear viscosity. In this appendix we only consider tree-level computations because our purpose is simply to explain how the various terms of the long-wavelength effective theory can be extracted from perturbation theory in principle. 

\begin{figure}[h!]
    \centering
        \includegraphics[width=0.5\textwidth]{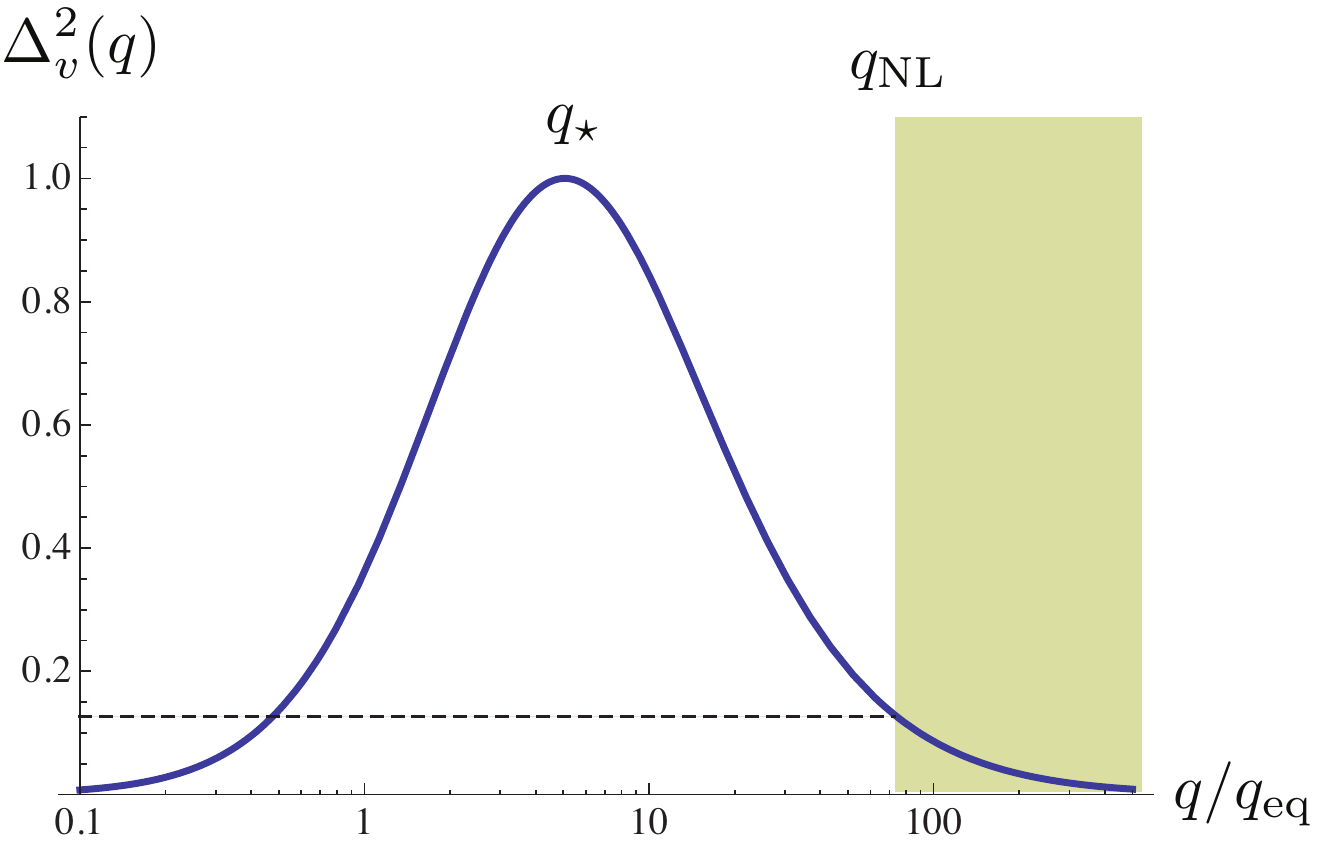}
    \caption{\sl Spectral density of velocity fluctuations, $\Delta_v^2(q)$, computed in linear perturbation theory. Non-linear corrections are significant in the regime $q \gtrsim q_{\rm NL}$. The spectral density $\Delta_v^2(q)$ appears in the kernel of all convolution integrals considered in this paper.}
    \label{fig:kernel}
\end{figure}

\vskip 6pt
\noindent
{\sl Renormalization of the background.} \hskip 8pt
We first consider the renormalization of the background (see \S\ref{sec:PerfectFluid}), {\it i.e.}~the fluid in the limit of long wavelengths.
In perturbation theory we may use the linear relation between the fluid three-velocity and the Newtonian potential, Eqn.~(\ref{equ:linvel}), to write
\beq
\rho v_i v_j \to \frac{4}{3} \frac{\phi_{,i} \phi_{,j}}{8\pi G a^2}\, ,
\eeq
and
\beq
\tau^i_{\ j} \to \frac{\frac{10}{3} \phi_{,i} \phi_{,j} - \phi_{,k} \phi_{,k} \delta_{ij}}{8\pi G a^2}\, , \qquad \tau \equiv \tau^i_{\ i} = \frac{1}{3} \frac{\phi_{,i} \phi_{,i}}{8\pi G a^2} \, .
\eeq
From this we find
\beq
[\tau]^\Lambda_\p = - W_\Lambda(p) \cdot \frac{\H^2 \bar \rho}{4} \int_\q \alpha(\q,\p)\, \delta_\q \delta_{\p-\q}\, , \qquad  \alpha(\q,\p) \equiv \frac{\q\cdot(\p-\q)}{q^2(\p-\q)^2}\, ,
\eeq
and
\beq
\langle [\tau]^\Lambda_\p \rangle = - W_\Lambda(p)\, (2\pi)^3 \delta_{\rm D}(\p) \, \frac{\H^2 \bar \rho}{4} \int_\q  \alpha(\q, \p)\, P_\delta(q)\, .
\eeq
The effective pressure of the background is\footnote{Eqn.~(\ref{equ:peffX}) assumes negligible bulk viscosity which otherwise would also contribute to the trace of $\tau_{ij}$. As we explained in \S\ref{sec:ImperfectFluid} the presence of significant bulk viscosity requires going beyond tree-level perturbation theory and/or numerical simulations to measure the individual viscosity parameters separately.}
\beq
\label{equ:peffX}
\bar p_{\rm eff} \equiv \frac{1}{3} \lim_{p \to 0} \langle [\tau]^\Lambda_\p \rangle = \frac{\bar \rho}{12} \int_\Lambda d \ln q\, \Delta_v^2(q)\, ,
\eeq
where we defined the spectral density of velocity fluctuations (see Fig.~\ref{fig:kernel}) 
\beq
\Delta_v^2(q) = \frac{4}{9} \frac{q^2}{\H^2} \Delta_\phi^2(q) = \frac{\H^2}{q^2} \Delta_\delta^2(q)\, .
\eeq
This shows that the renormalization of the background pressure is always positive and scales with the velocity dispersion of small-scale fluctuations.
We remind the reader that the integral in Eqn.~(\ref{equ:peffX}) is restricted to short-wavelength modes, $q > \Lambda$, or
\beq
\int_\Lambda d \ln q\,  \Delta_v^2(q) \equiv \int d \ln q\,  F_\Lambda^2(q) \Delta_v^2(q)\, ,
\eeq
where $F_\Lambda(q) \equiv 1- W_\Lambda(q)$.

We extract the scalar component of the anisotropic stress as follows
\beq
\sigma_{\rm eff}(\p) =  \frac{1}{\bar \rho} \frac{p_i p_j}{p^2} \langle [\hat \tau^i_{\ j}]^\Lambda_\p \rangle \, ,
\eeq
where
\beq
 \frac{p_i p_j}{p^2} [\hat \tau^i_{\ j}]^\Lambda_\p 
= - W_\Lambda(p) \cdot\frac{5}{2} \H^2\bar \rho \int_\q  \beta(\q,\p)\, \delta_\q \delta_{\p - \q}\, , \qquad \eeq
if we define
\beq
\label{equ:beta}
\beta(\q,\p) \equiv \frac{(\p \cdot \q) \p \cdot (\p-\q) - \frac{1}{3} p^2 \q \cdot (\p - \q)}{p^2 q^2 (\p-\q)^2} \, .
\eeq
Hence, we find
\beq
\sigma_{\rm eff}(\p) = - W_\Lambda(p)\, (2\pi)^3 \delta_{\rm D}(\p) \, \frac{5}{2} \H^2\int_\q \beta(\q, \p) P_\delta(q)\, .
\eeq
Using 
\beq
\lim_{p \to 0} \beta(\q, \p) = - \frac{1}{q^2} \Bigl[\mu^2 - \frac{1}{3} \Bigr]\, ,
\eeq
this implies
\beq
\bar \sigma_{\rm eff} = \lim_{p \to 0} \sigma_{\rm eff}(\p) \propto \int_{-1}^1 d \mu \Bigl[\mu^2 - \frac{1}{3} \Bigr] = 0\, .
\eeq
This confirms our arguments in \S\ref{sec:PerfectFluid} that  the effective anisotropic stress vanishes on very large scales.

\vskip 6pt
\noindent
{\sl Tidal pressure.} \hskip 6pt
We now consider the effects of long-wavelength perturbations on the properties of the effective fluid.
To this end,
we compute the correlation of the effective pressure with a long-wavelength mode $\left. \delta^\ell_\k\, \right|\, k < \Lambda$,
\beq
\langle \delta_\ell\,  p_{\rm eff} \rangle = \frac{1}{3} \langle \delta^\ell_\k \, [\tau]^\Lambda_\p \rangle\, = \, - W_\Lambda(p) \cdot \frac{\H^2 \bar \rho}{12} \int_\q \alpha(\q, \p) \, \langle \delta^\ell_\k \, \delta^s_\q \delta^s_{\p-\q} \rangle\, .
\eeq
The correlation function between a long-wavelength mode, $\delta^\ell_\k = \delta^{(1)}_\k$, and the product of two short-wavelength modes, $\delta_\q^s = \delta_\q^{(1)} + \delta_\q^{(2)}$, is
\bea
 \langle \delta^\ell_\k \, \delta^s_\q \delta^s_{\p-\q} \rangle &=&  \langle \delta^{(1)}_\k \, \delta^{(2)}_\q \delta^{(1)}_{\p-\q} \rangle +  \langle \delta^{(1)}_\k \, \delta^{(1)}_\q \delta^{(2)}_{\p-\q} \rangle\, , \\
 &=& 2 \times  \langle \delta^{(1)}_\k \, \delta^{(2)}_\q \delta^{(1)}_{\p-\q} \rangle\, , \label{equ:3pt}
 \eea
 where we have used the fact that correlation functions with an odd number of $\delta^{(1)}$ vanish if the initial fluctuations are Gaussian. Since $\delta_\q^{(2)}$ is written in Eqn.~(\ref{equ:delta2k}) as a convolution of a long-wavelength mode and a short-wavelength mode, the three-point function in Eqn.~(\ref{equ:3pt}) is related to a four-point function of first-order fluctuations,
 \beq
  \langle \delta^{(1)}_\k \, \delta^{(2)}_\q \delta^{(1)}_{\p-\q} \rangle = \int_{\q_1} \int_{\q_2} \delta_{\rm D}(\q - \q_1 - \q_2) F_2(\q_1, \q_2)\,  \langle \delta_\k^{(1)} \delta_{\q_1}^{(1)} \delta_{\q_2}^{(1)} \delta_{\p-\q}^{(1)}\rangle \, , \label{equ:4pt}
 \eeq
 where
 \bea
\acontraction[1ex] { \delta_\k^{(1)} \delta_{\q_1}^{(1)}}{\delta_{\q_2}^{(1)}}{} {\delta_{{\p}-\q}^{(1)} } 
\acontraction[1ex] {}{ \delta_\k^{(1)}}{} {\delta_{\q_1}^{(1)}} 
\bcontraction[1ex]{}{ \delta_\k^{(1)}} {\delta_{\q_1}^{(1)}} {\delta_{\q_2}^{(1)}}
\bcontraction[1.5ex]{ \delta_\k^{(1)}} {\delta_{\q_1}^{(1)}}{\delta_{\q_2}^{(1)}} {\delta_{{\p}-\q}^{(1)} }
  \langle \delta_\k^{(1)} \delta_{\q_1}^{(1)} \delta_{\q_2}^{(1)} \delta_{{\p}- \q}^{(1)}  \rangle &=& \langle  \delta_\k^{(1)} \delta_{\q_1}^{(1)} \rangle \langle  \delta_{\q_2}^{(1)} \delta_{{\p}- \q}^{(1)} \rangle + \langle  \delta_\k^{(1)} \delta_{\q_2}^{(1)} \rangle \langle  \delta_{\q_1}^{(1)} \delta_{{\p}- \q}^{(1)} \rangle  \\
  &=& 2 \times \langle  \delta_\k^{(1)} \delta_{\q_1}^{(1)} \rangle \langle  \delta_{\q_2}^{(1)} \delta_{{\p}- \q}^{(1)} \rangle \, .
\eea
Using
\bea
\langle \delta_\k^{(1)} \delta^{(1)}_{\q_1} \rangle &=& (2\pi)^3 P_\delta(k) \, \delta_{\rm D}(\k + \q_1)\, , \\
\langle \delta_{\p-\q}^{(1)} \delta_{\q_2}^{(1)} \rangle &=& (2\pi)^3 P_\delta(|\p -\q|)\,  \delta_{\rm D}(\p -\q + \q_2)\, ,
\eea
we perform the $\q_1$ and $\q_2$ integrals in Eqn.~(\ref{equ:4pt}),
 \beq
  \langle \delta^{(1)}_\k \, \delta^{(2)}_\q \delta^{(1)}_{\p-\q} \rangle = \underbrace{(2\pi)^3P_\delta(k) \, \delta_{\rm D}(\k +\p)}_{\langle \delta^\ell_\k \delta^\ell_\p \rangle} \ 2\, (2\pi)^3\, F_2(\k, \k+ \q)\,  P_\delta(|\k + \q|) \, .
 \eeq
We find
\beq
\frac{\langle \delta_\ell \, p_{\rm eff} \rangle}{\langle \delta_\ell \delta_\ell \rangle} =  W_\Lambda(k) \cdot\frac{(2\pi)^3}{3} \H^2 \bar \rho   \int_\q P_\delta(|\k + \q|)\, \alpha(\q, -\k) F_2(\k,\k+\q)\, .
\eeq
Using that $\lim_{k \ll \Lambda} W_\Lambda(k) \approx 1$ and that the integral is dominated by modes with $q \gg k$, this simplifies to
\beq
\frac{\langle \delta_\ell \, p_{\rm eff} \rangle}{\langle \delta_\ell \delta_\ell \rangle} = \frac{(2\pi)^3}{3} \bar \rho   \int_\q P_\delta(q) \frac{\H^2}{q^2} F_2(\k,\q)\, ,
\eeq
or
\beq
\label{equ:csfinal}
\fbox{$\displaystyle
c_{s}^2 =\frac{1}{\bar \rho} \frac{\langle \delta_\ell\, p_{\rm eff} \rangle }{\langle \delta_\ell \delta_\ell \rangle}= \frac{\nu_2}{3}
\int_\Lambda d \ln q \, \Delta^2_v(q) $}\, .
\eeq

\vskip 6pt
\noindent
{\sl Anisotropic stress and viscosity.}
\hskip 8pt
Next, we compute the correlation of the anisotropic stress with a long-wavelength mode $\left. \delta_{\k}^{\ell} = - \theta_\k^\ell/\H  \right| k < \Lambda$,
\beq
\langle \delta_\ell\, \sigma_{\rm eff} \rangle = \frac{1}{\bar \rho}  \frac{p_i p_j}{p^2} \langle \delta^\ell_\k [\hat \tau^i_{\ j}]^\Lambda_\p \rangle = - W_\Lambda(p) \cdot\frac{5}{2} \H^2 \int_\q  \beta(\q,\p) \langle \delta_\k^\ell \delta_\q^s \delta_{\p - \q}^s \rangle \, , \qquad \eeq
where $\beta(\q, \p)$ was defined in Eqn.~(\ref{equ:beta}) and the three-point function was computed above,
\beq
\langle \delta_\k^\ell \delta_\q^s \delta_{\p - \q}^s \rangle = \langle \delta_\k^\ell \delta_\p^\ell \rangle\, 4 \, (2\pi)^3 F_2(\k, \q) P_\delta(q)\, .
\eeq
We find,
\beq
\frac{\langle \delta_\ell \, \sigma_{\rm eff} \rangle}{\langle \delta_\ell \delta_\ell \rangle} = - 10 (2\pi)^3 \H^2   \int_\q  P_\delta(q)\, F_2(\k, \q) \beta(\q, -\k)  
 \, .
\eeq
As before, this simplifies since modes with $q \gg k$ dominate the integral
\beq
\frac{\langle \delta_\ell \, \sigma_{\rm eff} \rangle}{\langle \delta_\ell \delta_\ell \rangle}  = 10 (2\pi)^3 \H^2   \int_\q P_\delta(q) 
F_2(\k,  \q) \frac{(\k \cdot \q)^2  - \frac{1}{3} k^2 \q^2}{k^2 q^4} \, ,
\eeq
or
\beq
\label{equ:cvisfinal}
\fbox{$\displaystyle
\tilde c_{\rm vis}^2 =  \frac{\langle \delta_\ell \, \sigma_{\rm eff} \rangle}{\langle \delta_\ell \delta_\ell \rangle}= 10  \mu_2
\int_\Lambda d \ln q \, \Delta^2_v(q) $}\, ,
\eeq
where
\bea
\mu_2 &\equiv& \int \frac{d \Omega_\q}{2\pi} \, F_2(\hat \k,  \hat \q) \Bigl( (\hat \k \cdot \hat \q)^2 - \frac{1}{3}\Bigr)  = \frac{16}{315}\, .
\eea

The ratio of the sound speed associated with tidal pressure and the sound speed associated with viscosity therefore is
\beq
\frac{c_s^2 }{\tilde c_{\rm vis}^2} = \frac{1}{30} \frac{\nu_2}{\mu_2} = \frac{17}{16}\, .
\eeq
\vspace{0.2cm}
 \hrule
 \vskip 1pt
 \hrule
 \vspace{0.5cm}
\normalsize

\subsection{Random Statistical Fluctuations}
\label{sec:Stochastic}

In the example above we have used ensemble averaging and the ergodic theorem to estimate average quantities in domains of size $\Lambda^{-1}$.
For any specific realization of the universe there will be a random statistical error  in that estimate.
In this section we estimate the size of that effect in perturbation theory.

\subsubsection*{Stochastic Pressure}

As an illustrative example, we consider
the effective pressure 
\beq
p_{\rm eff} = \bar p_{\rm eff} + \delta p_{\rm eff}^{\ell} + \delta p_{\rm eff}^{\rm stat}\, .
\eeq
Here, $\bar p_{\rm eff}$ denotes the renormalization of the background pressure (\S\ref{sec:super}); $\delta p_{\rm eff}^\ell$ is the perturbation induced by long-wavelength modes $\delta_\ell$ (\S\ref{sec:tidal});  $\delta p_{\rm eff}^{\rm stat}$ is the perturbation induced by random statistical fluctuations.
We will now establish the following two facts:
\begin{enumerate}
\item Both   $\delta p_{\rm eff}^\ell$ and $\delta p_{\rm eff}^{\rm stat}$ are much smaller than $\bar p_{\rm eff}$ if the momentum integral defining $p_{\rm eff}$ is UV-dominated ({\it i.e.}~dominated by scales with $q_\star \gg \Lambda$).
\item Typically,  $\delta p_{\rm eff}^{\rm stat} \gtrsim \delta p_{\rm eff}^\ell$ with equality holding at $q_\star \sim k_{\rm NL}$ (in Einstein-de Sitter) or $q_\star \sim k_{\rm eq}$ (in our universe). However, the fluctuations $\delta p_{\rm eff}^{\rm stat}$ are uncorrelated with the long-wavelength density fluctuations $\delta_\ell$ so their effect on the power spectrum of density fluctuations is subdominant for scales with $k \ll q_\star$. 
\end{enumerate}

\vspace{0.5cm}
\small
\hrule \vskip 1pt \hrule \vspace{0.3cm}
\noindent 
{\bf Proof}:\vskip 4pt

In perturbation theory we have
\beq
8\pi Ga^2\,  \tau_{ij} = \frac{10}{3} \phi_{,i} \phi_{,j} - \phi_{,k} \phi_{,k} \delta_{ij}\, .
\eeq
Consider the pressure source term, $\tau \equiv \tau_{ii}$,
\beq
{\cal S} \equiv 8\pi G a^2\, \tau = \frac{1}{3}  \phi_{,i} \phi_{,i} \, .
\eeq
Recall that the
spatial average over a domain of size $\Lambda^{-1}$ is
\beq
[{\cal S}]_\Lambda = \int_{\x'} W_\Lambda(|\x-\x'|) {\cal S}(\x') =\int_{\q_1} \int_{\q_2} e^{-i (\q_1 + \q_2) \cdot \x}\, W_\Lambda(|\q_1+ \q_2|) \, \frac{\q_1 \cdot \q_2}{3}\, \phi_{\q_1} \phi_{\q_2}\, ,
\eeq
and its ensemble average is
\beq
\langle [{\cal S}]_\Lambda \rangle = 
\frac{1}{3} \int d \ln q\ q^2 \Delta_\phi^2(q) \equiv 24\pi G a^2 \, p_{\rm eff}\, .
\eeq
The ensemble average approximates the true average of a given realization with some statistical error
\beq
[{\cal S}]_\Lambda = \langle [{\cal S}]_\Lambda \rangle  \pm \Delta [{\cal S}]_\Lambda\, .
\eeq
How different is $[{\cal S}]_\Lambda$ from its ensemble average $\langle [{\cal S}]_\Lambda \rangle  $ in each cube?
To assess this we compute the variance
\bea
\langle [{\cal S}]_\Lambda [{\cal S}]_\Lambda \rangle &=& \int_{\q_1}  \int_{\q_2}  \int_{\q_3} \int_{\q_4} e^{-i(\q_1+\q_2+\q_3+\q_4)\cdot \x} \ W_\Lambda(|\q_1+ \q_2|) W_\Lambda(|\q_3+ \q_4|)  \,  \nonumber \\
&& \hspace{3cm} \times\ \frac{\q_1 \cdot \q_2}{3} \frac{\q_3 \cdot \q_4}{3}    \, \langle \phi_{\q_1} \phi_{\q_2}
 \phi_{\q_3} \phi_{\q_4} \rangle\, .
\eea
Using Wick's theorem to expand the four-point function for Gaussian fields, we find
\beq
\langle [{\cal S}]_\Lambda [{\cal S}]_\Lambda \rangle - \langle [{\cal S}]_\Lambda \rangle^2 =  2 
\int_{\q_1}  \int_{\q_2}   W_\Lambda(|\q_1+ \q_2|)^2 \,  
\frac{(\q_1 \cdot \q_2)^2}{9}     \, P_\phi(q_1) P_\phi(q_2)\, .
\eeq
{\it Assuming} that the integrals are dominated by modes with $q_1, q_2 \gg \Lambda$, we may make the following substitution, $W_\Lambda(|\q_1 + \q_2|) \approx \delta_{\rm D}(|\q_1 + \q_2|/\Lambda)$. This gives
\beq
\langle [{\cal S}]_\Lambda [{\cal S}]_\Lambda \rangle - \langle [{\cal S}]_\Lambda \rangle^2 \approx  \frac{\Lambda^3}{18 \pi} 
\int d \ln q     \, q \Delta_\phi^4(q)\, .
\eeq
and
\beq
\frac{\langle [{\cal S}]_\Lambda [{\cal S}]_\Lambda \rangle - \langle [{\cal S}]_\Lambda \rangle^2}{\langle [{\cal S}]_\Lambda \rangle^2} \approx  \frac{f}{2 \pi}  \frac{\Lambda^3}{q_\star^3}\, ,
\eeq
where
\beq
f \equiv \frac{\int d \ln x \, x \Delta_\phi^4(x)}{[\int d \ln x \, x^2 \Delta_\phi^2(x)]^2}\, , \quad {\rm with} \quad x \equiv \frac{q}{q_\star}\, .
\eeq
Here, we have factored out the scale $q_\star$ that gives the dominant contribution in the integrals.
In a matter-only universe $q_\star \sim q_{\rm NL}$, while in a universe with matter and radiation $q_\star \sim q_{\rm eq}$.
With this normalization the factor $f$ is of order unity, and
\beq
\label{equ:var}
\frac{\langle [{\cal S}]_\Lambda [{\cal S}]_\Lambda \rangle - \langle [{\cal S}]_\Lambda \rangle^2}{\langle [{\cal S}]_\Lambda \rangle^2}  \sim \frac{\Lambda^3}{q_\star^3} \, .
\eeq
This implies,
\beq
\label{equ:pstat}
\delta p_{\rm eff}^{\rm stat} = \alpha \, p_{\rm eff}\, , 
\eeq
where $\alpha$ is a random variable with variance
\beq
\Delta_\alpha^2 \equiv  \frac{\Lambda^3}{q_\star^3}\, .
\eeq
Eqn.~(\ref{equ:pstat}) characterizes stochastic pressure perturbations.
Since we have assumed $\Lambda \ll q_\star$, this shows that in this case $\delta p_{\rm eff}^{\rm stat} \ll \bar p_{\rm eff}$.
\hfill
QED $\blacksquare$
\vspace{0.2cm}  \hrule \vskip 1pt \hrule
 \vspace{0.5cm}
 \normalsize
 
The $\Lambda/q_\star$ suppression in Eqn.~(\ref{equ:pstat}) may be understood as the standard $\frac{1}{\sqrt{N}}$ suppression of random fluctuations, where $N$ is the number of space domains sampled.

\subsubsection*{Simple Estimates for Einstein-de Sitter}

We now give simple estimates of the relative sizes of the above effects in the case of an Einstein-de Sitter universe for which the transfer function for the Newtonian potential is trivial and $q_\star \sim q_{\rm NL}$. Generalizing this to the case of a universe with radiation would be straightforward.
Specifically, we will show that the fluctuations $\delta p_{\rm eff}^{\rm stat}$ are uncorrelated with the long-wavelength density fluctuations $\delta_\ell$ so their effect on the power spectrum of density fluctuations is subdominant for scales with $k \ll q_\star$. 

\vspace{0.5cm}
\hrule \vskip 1pt \hrule \vspace{0.3cm}
\noindent 
\small
{\bf Proof}:\vskip 4pt

We capture fluctuations at a scale $k$ by taking the smoothing scale $\Lambda \to k$.
First, we compare the relative sizes of the two pressure perturbations
\beq
\delta p_{\rm eff}^\ell \sim c_s^2 \bar \rho\, \delta_\ell \qquad {\rm and} \qquad \delta p_{\rm eff}^{\rm stat} \sim c_s^2 \bar \rho\, \alpha\, .
\eeq
We find
\beq
\frac{|\delta p_{\rm eff}^\ell |^2}{|\delta p_{\rm eff}^{\rm stat}|^2} \sim \frac{\Delta_\delta^2(k)}{\Delta_\alpha^2(k)}\, .
\eeq
We see that the two terms are equal at $k \sim q_{\rm NL}$, where $\Delta_\delta(q_{\rm NL}) = \Delta_\alpha(q_{\rm NL}) =1$.
For $k \ll q_{\rm NL}$ we have
\beq
\frac{|\delta p_{\rm eff}^\ell |^2}{|\delta p_{\rm eff}^{\rm stat}|^2} \sim \Delta_\phi^2(k) \frac{k}{\H} \left(\frac{q_{\rm NL}}{\H}\right)^3 \sim \frac{k}{\H}  \Delta_\phi^{-1/2} \sim 0.01 \frac{k}{\H}   \, .
\eeq
Near the horizon scale $\delta p_{\rm eff}^{\rm stat}$ is therefore larger than
$\delta p_{\rm eff}^\ell$ by an order of magnitude.

However, $\delta p_{\rm eff}^{\rm stat}$ is uncorrelated with long-wavelength density perturbations $\delta_\ell$ so its effect on the density power spectrum is in fact suppressed.
To see this consider the evolution equation for density perturbations 
\bea
\ddot \delta_\ell + \H \dot \delta_\ell + \frac{3}{2}\H^2 \delta_\ell &=& - \frac{\nabla^2 p_{\rm eff}}{\bar \rho} \\
&=& k^2 c_s^2 \left[ \delta_\ell +\alpha\right]\, .
\eea
Using $\hat \delta$ for the solution in the $c_s = 0$ limit, we estimate the pressure-induced corrections
\beq
\delta_\ell \approx \hat \delta_\ell + \frac{k^2}{\H^2} c_s^2 \hat \delta_\ell + \frac{k^2}{\H^2} c_s^2 \alpha\, .
\eeq
The first correction term (arising from tidal pressure) is correlated with $\delta_\ell$, while the second (arising from stochastic pressure) is not.
Hence, the corrections to the power spectrum are
\beq
\Delta_\delta^2  = \Delta_{\hat \delta}^2   +   \frac{k^2}{\H^2} c_s^2   \Delta_{\hat \delta}^2  +  \left(\frac{k^2}{\H^2} c_s^2\right)^2 \Delta_\alpha^2\, ,
\eeq
and
\beq
\frac{\Delta P_\delta^\ell}{\Delta P_\delta^{\rm stat}} \sim   \left(\frac{k^2}{\H^2} c_s^2\right)^{-1} \frac{\Delta_\delta^2}{\Delta_\alpha^2} \, .
\eeq
Using
\beq
c_s^2 \sim \frac{q_{\rm NL}^2}{\H^2} \Delta_\phi^2(q_{\rm NL}) \qquad {\rm and} \qquad \frac{k^2}{\H^2} c_s^2 \sim \left( \frac{k}{q_{\rm NL}}\right)^2 \Delta_\delta^2(q_{\rm NL}) =  \left( \frac{q_{\rm NL}}{k}\right)^2  \Delta_\delta^2(k)\, ,
\eeq
we find
\beq
\frac{\Delta P_\delta^\ell}{\Delta P_\delta^{\rm stat}} \sim   \frac{q_{\rm NL}}{k}  \, .
\eeq
We see that the corrections to the power spectrum are equal at the non-linear scale, but dominated by tidal pressure on larger scales, $k^{-1}  \gg q_{\rm NL}^{-1}$.
\hfill
QED $\blacksquare$
\vspace{0.2cm}  \hrule \vskip 1pt \hrule
 \vspace{0.5cm}
 \normalsize

 \newpage

\vfil
\end{document}